\documentclass[article,floatfix,12pt,showpacs,superscriptaddress,tightenlines,nofootinbib,longbibliography,a4paper]{revtex4-1} 
\pdfoutput=1
\usepackage{tablefootnote}
\usepackage{amsfonts}
\usepackage{amsmath}
\usepackage{xcolor}
\usepackage{graphicx}
\usepackage{hyperref}
\usepackage{multirow}
\usepackage{url}
\usepackage{lineno}
\hypersetup{pdfstartview=FitV,colorlinks=true,linkcolor=blue,citecolor=red,filecolor=black,urlcolor=blue}
\def\beq {\begin{equation}}
\def\eeq {\end{equation}}
\def\bi {\begin{itemize}}
\def\ei {\end{itemize}}
\def\bea {\begin{eqnarray}}
\def\eea {\end{eqnarray}}

\newcommand{\br}{\begin{eqnarray}}
\newcommand{\er}{\end{eqnarray}}
\newcommand{\be}{\begin{equation}}
\newcommand{\ee}{\end{equation}}

\def\lum{{\cal L}}

\newcommand{\wt}{\widetilde}
\def \lsbot{\widetilde{b}_{1}}
\def \mlsbot{m_{\lsbot}}
\def \lstop{\widetilde{t}_{1}}
\def \mlstop{m_{\lstop}}

\def \lspone{\wt\chi_1^0}
\def \mlspone{m_{\lspone}}
\def \lsptwo{\wt\chi_2^0}

\def \lspthree{\wt\chi_3^0}

\usepackage{array}
\newcolumntype{L}[1]{>{\raggedright\let\newline\\\arraybackslash\hspace{0pt}}m{#1}}
\newcolumntype{C}[1]{>{\centering\let\newline\\\arraybackslash\hspace{0pt}}m{#1}}
\newcolumntype{R}[1]{>{\raggedleft\let\newline\\\arraybackslash\hspace{0pt}}m{#1}}

\begin{document}

\title{Current status of MSSM Higgs sector with LHC 13 TeV data}

\author{Rahool Kumar Barman}
\email{rahoolbarman@iisc.ac.in}

\author{Biplob Bhattacherjee}
\email{biplob@iisc.ac.in}

\affiliation{Centre for High Energy Physics, 
Indian Institute of Science, Bangalore 560 012, India}

\author{Arghya Choudhury}
\email{a.choudhury@sheffield.ac.uk}
\affiliation{Department of Physics, Indian Institute of Technology Patna, Bihta, Bihar 801103, India}
\affiliation{Consortium for Fundamental Physics, Department of Physics and Astronomy, University of Sheffield, Sheffield S3 7RH, United Kingdom}
\affiliation{Consortium for Fundamental Physics, Department of Physics and Astronomy, University of Manchester, Manchester, M13 9PL, United Kingdom}

\author{Debtosh Chowdhury}
\email{debtosh.chowdhury@polytechnique.edu}

\affiliation{Istituto Nazionale di Fisica Nucleare, Sezione di Roma, Piazzale Aldo Moro 2, I-00185 Roma, Italy}

\author{Jayita Lahiri}
\email{jayita@chep.iisc.ernet.in}

\affiliation{Centre for High Energy Physics, 
Indian Institute of Science, Bangalore 560 012, India}

\author{Shamayita Ray}
\email{shamayita.ray@gmail.com}

\affiliation{Laboratory for Elementary Particle Physics, Cornell University, Ithaca, NY 14853, USA}
\affiliation{Department of Physics, University of Calcutta, Kolkata 700 009, India}

\date{\today}


\pacs{14.80.Bn, 14.80.Da, 14.80.Ly}

\begin{abstract}

ATLAS and CMS collaborations have reported the results on the Higgs search analyzing $\sim 36$ fb$^{-1}$ data from Run-II of LHC at 13 TeV. In this work, we study the Higgs sector of the phenomenological Minimal Supersymmetric Standard Model, in light of the recent Higgs data, by studying separately the impact of Run-I and Run-II data. One of the major impacts of the new data on the parameter space comes from the direct searches of neutral CP-even and CP-odd heavy Higgses ($H$ and $A$, respectively) in the $H/A \to \tau^{+} \tau^{-}$ channel which disfavours high $\tan\beta$ regions more efficiently than Run-I data. Secondly, we show that the latest result of the rare radiative decay of $B$ meson imposes a slightly stronger constraint on low $\tan \beta$ and low $M_A$ region of the parameter space, as compared to its previous measurement. Further, we find that in a global fit Run-II light Higgs signal strength data is almost comparable in strength with the corresponding Run-I data. Finally, we discuss scenarios with the Heavy Higgs boson decaying into electroweakinos and third generation squarks and sleptons.

\end{abstract}

\maketitle

\tableofcontents

\section{Introduction}
\label{sec:intro}

Discovery of the $125$ GeV Higgs particle during Run-I ($\approx 5$ fb$^{-1}$ data at 7 TeV and 
$\approx 20$ fb$^{-1}$ data at 8 TeV) phase of the LHC by the ATLAS and CMS collaborations \cite{Aad:2012tfa,Chatrchyan:2012xdj}
is indeed a triumph of the Standard Model (SM) of particle physics. It confirms the validity of the 
theory of spontaneous electro-weak symmetry breaking, which in turn produces the masses of the 
weak gauge bosons and also the masses of quarks and leptons 
via Yukawa interactions, within the framework of the SM.

Confirmation of the newly discovered boson as the SM-like Higgs boson by Run-I data, however, does not 
rule out the possibility of the existence of an extended Higgs sector. At the same time, beyond the Standard Model (BSM) 
theories are essential for explaining issues like neutrino masses, existence of dark matter, hierarchy problem. Various BSM theories have been proposed to address these shortcomings of the SM, out of which some contain an extended Higgs sector. One among such theories is the Minimal Supersymmetric extension of the SM, also known as the Minimal Supersymmetric Standard Model (MSSM) \cite{Drees:2004jm,Baer:2006rs,Martin:1997ns}. In this work, we confine ourselves within the MSSM framework.

The MSSM has two Higgs doublets and contains five massive Higgs states after electro-weak symmetry breaking: 
two CP-even Higgs bosons $h$ and $H$, one CP-odd Higgs boson $A$ and the charged Higgs bosons $H^\pm$.
Previous studies of Higgs mass calculation in the framework of MSSM \cite{Ellis:1990nz,Okada:1990vk,Haber:1990aw,Carena:2002es,Heinemeyer:2004gx} have predicted an upper-bound of $\sim 135$ GeV for the lightest CP-even Higgs boson $h$ (for sparticle masses $\lesssim 1$ TeV), indicating the observed Higgs mass to be consistent with and really close to the upper-bound. 
%
%
LHC direct search limits also set a quite high lower limit on the masses of the strongly-interacting supersymmetric (or SUSY) particles like stop and sbottom squarks~\cite{Aaboud:2017vwy,Aaboud:2017hrg,Sirunyan:2017cwe,Sirunyan:2018vjp,Aaboud:2017ayj,
Sirunyan:2017wif,Sirunyan:2017leh}  (with exceptions under some special circumstances\footnote{Degenerate SUSY scenarios are an example of such special circumstances. Phenomenological aspects of such scenarios can be found in \cite{Alves:2010za,LeCompte:2011cn,Drees:2012dd,Dreiner:2012gx,
Bhattacherjee:2012mz,Bhattacherjee:2013wna,Dutta:2015exw,Tobioka:2015vsv,Chowdhury:2016qnz}.}), making them difficult to be observed experimentally.
On the other hand, heavy Higgs bosons \cite{Aad:2014vgg,ATLAS-CONF-2015-061,Aad:2014ioa,Aaboud:2016tru,Aad:2015agg,ATLAS-CONF-2016-021,Aad:2015kna,ATLAS-CONF-2016-016,ATLAS-CONF-2016-012,Aaboud:2016trl,Aad:2015kna,Aad:2015xja,Aaboud:2016xco,Aad:2015wra,ATLAS-CONF-2016-015,Aad:2014kga,Aaboud:2016dig,Aad:2015typ,ATLAS-CONF-2016-056,ATLAS-CONF-2016-082,ATLAS-CONF-2016-079,ATLAS-CONF-2016-074,ATLAS-CONF-2016-062,ATLAS-CONF-2016-059,ATLAS-CONF-2016-085,ATLAS-CONF-2016-049,ATLAS-CONF-2016-088,ATLAS-CONF-2016-089,CMS-PAS-HIG-14-029,CMS-PAS-HIG-16-006,
Khachatryan:2016hje,Khachatryan:2015tra,Khachatryan:2015cwa,
Khachatryan:2016sey,Khachatryan:2015yea,Khachatryan:2015tha,Khachatryan:2015tha,
Khachatryan:2015lba,Khachatryan:2015qxa,CMS-PAS-HIG-16-023,CMS-PAS-EXO-16-027,CMS-PAS-HIG-16-037,CMS-PAS-HIG-16-025} and electroweak sparticles \cite{Sirunyan:2018ubx,Aaboud:2017nhr} are still allowed to be light with masses of a few hundreds of GeV.

Extensive studies have already been performed where the allowed ranges of masses and other MSSM parameters have been obtained, in light of the constraints on flavor physics observables and the constraints derived from Run-I data of LHC~\cite{Carena:2011aa,Arbey:2011ab,Baer:2012mv,Arbey:2012dq,Altmannshofer:2012ks,Cheung:2013kla,Chowdhury:2015yja,Bhattacherjee:2015sga,Bechtle:2016kui,Bechtle:2012jw,Djouadi:2013lra,Bechtle:2015pma}. 
Several studies have also analyzed the implications of additionally imposing cosmological constraints on the parameter space~\cite{Buchmueller:2013rsa,Scopel:2013bba,deVries:2015hva}. The MSSM parameter space has also been analyzed in the context of the $\sim 15$ fb$^{-1}$ dataset of Run-II~\cite{Barr:2016sho,Kowalska:2016ent,Han:2016xet,Buckley:2016kvr,
Zhao:2017qpe,Bagnaschi:2016afc,Bagnaschi:2016xfg}. Global-fit analysis of various GUT-scale SUSY models and MSSM has been performed by the GAMBIT collaboration \cite{Athron:2017qdc,Athron:2017yua}. Recently, the MaterCode collaboration has also performed likelihood analysis with a selection of $\sim 36$ fb$^{-1}$ data taken at 13 TeV for sub-GUT MSSM and pMSSM scenarios \cite{Bagnaschi:2017tru,Costa2018}.


 
In this paper, we review the status of the parameter space of phenomenological MSSM (pMSSM), in light of the latest data from Run-II of LHC (13 TeV, $\sim 36 {\rm fb^{-1}}$), as reported by the CMS and ATLAS collaborations, and concentrate on the Higgs sector. Along with the latest data, we also take into account the LHC Run-I data. Apart from the constraints coming from the mass measurement of the $125$ GeV Higgs boson, 
we discuss the constraints on flavor physics observables and impose them through the bounds on the branching fraction of rare-decays: $B \to X_s \gamma$, $B_s \to \mu^+ \mu^-$, and $B^+ \to \tau^+ \nu_{\tau}$.
We then study the implications of the heavy Higgs searches by ATLAS and CMS collaborations on the already 
constrained parameter space, both with 8 TeV and 13 TeV data, where we consider the channels 
$H/A \to \gamma \gamma, b\bar{b}, t\bar{t}, \tau^{+} \tau^{-}$, $H \to W^{+}W^{-}, ZZ, hh$ and $A \to Zh$. As for the charged Higgs searches, constraints coming from $H^\pm$ decaying to $\tau \nu_\tau$ as well as $t\bar{b}$ final states have been considered, 
as reported by both the collaborations with 8 TeV and 13 TeV data. Finally, we consider the possibilities of the 
decay of neutral heavy Higgses ($H$, $A$) to SUSY particles, within the allowed region 
of the parameter space. We also comment on how the future improvements from various heavy Higgs decay channels would affect the parameter space.

The paper is organized as follows. In section \ref{sec:II}, we discuss the range of different MSSM parameters on which the scan has been performed. We impose the Higgs mass constraint and then perform a global fit analysis by taking into account flavor physics constraints coming from $B$-meson decays and Higgs signal strength constraints coming from 8 TeV and 13 TeV data, and discuss its implication on the MSSM parameter space. We then analyze the impact of the direct search for heavy Higgs (neutral and charged) at 8 TeV and 13 TeV by ATLAS and CMS collaborations in various search channels. Section \ref{sec:III} discusses the impact of all these constraints put together on certain regions of our interest. In section \ref{sec:IV} we discuss the non-standard decay modes of the Higgs i.e. decay of heavy Higgs boson to SUSY particles and the effect of various direct searches on them. In section~\ref{sec:V} we comment on the future projection on the heavy Higgs searches and then summarize the main results of this work.

\section{Parameter space and current bounds}
\label{sec:II}

We begin this section by specifying the region of parameter space scanned, followed by a description of all the experimental bounds imposed to constrain the MSSM parameter space.
\subsection{Parameter space scan}
The parameters relevant to our analysis are the higgsino mass parameter $\mu$, the gaugino mass parameters $M_{1,2,3}$, the ratio of vacuum expectation values (vevs) of the two Higgs doublet $\tan\beta$, the pseudoscalar mass parameter $M_{A}$, the trilinear couplings of the third generation squarks $A_{t,b}$ ($A_{t}$ and $A_{b}$ has been varied independently), and the masses of all three generation of squarks $M_{\tilde{u}_{1},\tilde{d}_{1},\tilde{Q}_{1}},M_{\tilde{u}_{2},\tilde{d}_{2},\tilde{Q}_{2}}~{\rm and}~M_{\tilde{Q}_{3},\tilde{u}_{3},\tilde{d}_{3}}$. Here we would like to note that all the first and third generation squark masses have been independently varied and the second generation squark masses have been set equal to their first generation counterparts. The masses of all three generations of sleptons have been fixed at a value of 2 TeV. We would also like to mention that $M_{1}$ and $M_{2}$ have been restricted above $600~\rm{GeV}$ in order to exclude the possibility of heavy Higgs decaying into electroweak gauginos. These decays will be considered in detail in Section~\ref{sec:IV}. We vary $A_{t}$, squark mass parameters and $\mu$ over a wide range in order to maximize the number of parameter space points, with $M_{h}$ in the correct Higgs mass region (discussed in Sec.~\ref{light_higgs_mass}). Since we also analyzed the effect of direct heavy Higgs searches on our parameter space in the later sections, we have considered $M_A \lesssim 1 ~{\rm TeV}$ as the heavy Higgs production cross-section becomes quite small for heavy Higgs masses $\sim$ 1 TeV, for a $\sqrt{s} =$ 14 TeV collider.

The input parameters are randomly varied over the following ranges:
\begin{eqnarray*}
 &600~ \mathrm{GeV}~<~M_{1}~<~ 5\ \mathrm{TeV},\quad ~ 600~ \mathrm{GeV}~<~M_{2}~<~ 5\ \mathrm{TeV},\quad ~500~{\rm GeV} < M_{3} < 5~{\rm TeV},  \\
& 1 < \tan\beta < 60,\quad 100~{\rm GeV} <  M_{A} < 1~{\rm TeV}, \quad 100~{\rm GeV} < \mu < 5~{\rm TeV}, \quad \\ 
&600~ \mathrm{GeV}~<~M_{\tilde{Q}_{1}}~<~ 5\ \mathrm{TeV},\quad ~600~ \mathrm{GeV}~<~M_{\tilde{u}_{1}}~<~ 5\ \mathrm{TeV}, \\ 
&600~ \mathrm{GeV}~<~M_{\tilde{d}_{1}}~<~ 5\ \mathrm{TeV},~M_{\tilde{Q}_{2}}=M_{\tilde{Q}_{1}},~M_{\tilde{u}_{2}}=M_{\tilde{u}_{1}},~M_{\tilde{d}_{2}}=M_{\tilde{d}_{1}}, \\
&A_{e,\mu,\tau} = A_{u,d,c,s} = 0,\quad -10~{\rm TeV} < A_{b,t} < 10~{\rm TeV},\\
&200~{\rm GeV} < M_{\tilde{Q}_{3},\tilde{u}_{3},\tilde{d}_{3}} < 10~{\rm TeV},\quad M_{\tilde{e}_{1_{L}},\tilde{e}_{1_{R}},\tilde{e}_{2_{L}},\tilde{e}_{2_{R}},\tilde{e}_{3_{L}},\tilde{e}_{3_{R}}}~=~2~{\rm TeV}
\end{eqnarray*}

Apart from the above parameters, we vary the input top pole mass ($m_t^\mathrm{pole}$) in gaussian distribution around a central value of 173.21 GeV and a standard deviation of 0.55 GeV~\cite{Agashe:2014kda}, and $m_t^\mathrm{pole}$ is randomly extracted from the distribution. We sample $\sim 6 \times 10^{8}$ points within the above mentioned ranges. 
The SUSY sparticle spectrum and various production and decay modes of the Higgs sector are generated using FeynHiggs 2.12.0 \cite{Heinemeyer:1998yj,Heinemeyer:1998np,Degrassi:2002fi,Frank:2006yh,Heinemeyer:2007aq,Bahl:2016brp}. 
The neutral Higgs production cross-section through gluon gluon fusion (ggF) \cite{deFlorian:2009hc} and the charged Higgs production \cite{Plehn:2002vy,Berger:2003sm,Dittmaier:2009np} cross-section are calculated using FeynHiggs. 
On the other hand, the neutral Higgs production cross-sections through vector boson fusion ($qqH$), associated production with vector bosons ($VH$) and associated production with $b\bar{b}$ pair ($b\bar{b}H$) are evaluated using the following prescription: we fit the SM-like heavy Higgs production cross-section calculated by the LHC Higgs Cross-section Working Group (HXSWG) \cite{lhchxswg,Dittmaier:2011ti,Dittmaier:2012vm,Heinemeyer:2013tqa} to a function and 
multiply it with the appropriate MSSM/SM ratio of the couplings involved in the 
process. Generically, presence of light SUSY particles in the spectrum could significantly alter these cross-sections and under such circumstances it would not be correct to implement the fitting procedure described above. However, in the context of our analysis, where the gluino, squark and slepton masses have been fixed at a rather high value, the SUSY corrections would impart a very small change, and, under such circumstances, the fitting procedure adopted to calculate the $qqH,~VH$ and $b\bar{b}H$ cross-sections would yield correct approximations~\cite{Anastasiou:2008rm,Harlander:2010wr,Pak:2012xr,
Degrassi:2012vt,Dittmaier:2014sva}\footnote{Detailed work on evaluation of uncertainties and correlation matrix of Higgs cross-sections and partial decay width has been done in \cite{Bechtle:2013xfa,Bechtle:2014ewa,Arbey:2016kqi}.}.
We also evaluated the $ggh/H/A$ and $bbh/H/A$ cross-sections through \texttt{SusHi} \cite{Harlander:2016hcx,Harlander:2012pb}. The $ggh/H/A$ cross-sections obtained from \texttt{SusHi} were compared against the corresponding FeynHiggs cross-section, while the $bbh/H/A$ cross-sections generated by \texttt{SusHi} were compared against the cross-sections obtained through the previously mentioned procedure using fitting functions, for a set of $10^{4}$ randomly chosen points from the entire data-set. It has been observed that the cross-sections deviated at most by $\sim 15\%$ in both cases. We would like to note that \texttt{SusHi} uses the $5$ flavor scheme to compute the $bbh/H/A$ cross-sections while the HXSWG cross-section values, used to derive the fitting function, correspond to $4$ flavor $+$ $5$ flavor Santander matched cross-sections~\cite{Harlander:2011aa,PhysRevD.67.093005}. This difference in the cross-section evaluation scheme can result in a deviation of $\sim 5-15\%$, which partially explains the deviation in the cross-section values computed by \texttt{SusHi} and the fitting functions.


\subsection{Current bounds}

The SUSY spectrum generated from the above variation of input parameters gets initially constrained by the mass of the light CP-even Higgs boson, $M_{h}$, which we assume to be the $125~{\rm GeV}$ Higgs discovered at LHC.
Constraints on Br$(B \rightarrow X_{s}\gamma)$, Br$(B_{s} \rightarrow \mu^{+} \mu^{-})$ and Br$(B^{+} \rightarrow \tau^{+} \nu_{\tau})$ derived from flavor physics experiments, and the existing bounds on signal strength variables derived by CMS and ATLAS from a combined analysis of Run-I (7 TeV and 8 TeV) and Run-II (13 TeV) Higgs data are imposed through a global $\chi^{2}$-fit.
Finally, the minimum value of the $\chi^{2}$ is determined and parameter space points with $\chi^{2}$ value outside the 2$\sigma$ range from the minimum are neglected. 
Further constraints coming from the heavy Higgs searches at Run-I and Run-II ($\sim 3~{\rm fb^{-1}},~15~{\rm fb^{-1}},~36~{\rm fb^{-1}}$) have also been incorporated. We discuss these four constraints in detail in the following subsections.


\subsubsection{Constraints on $M_{h}$}
\label{light_higgs_mass}

The combined experimental measurements of Higgs mass by ATLAS and CMS \cite{Aad:2015zhl} allow for a window of $124.4-125.8$ GeV at $3\sigma$. The available calculation of the Higgs mass in pMSSM is not exact. FeynHiggs 2.12.0 provides dominant 2-loop corrections and partial NNLL resummation \cite{Heinemeyer:1998yj,Heinemeyer:1998np,Degrassi:2002fi,Frank:2006yh,Heinemeyer:2007aq,
Borowka:2015ura,Allanach:2004rh} for the Higgs mass calculation in MSSM. To account for these uncertainties in the Higgs mass calculation, we allow a $\pm 3$ GeV uncertainty in the Higgs mass. Hence, the Higgs mass range considered in the present analysis is $122-128$ GeV~\cite{Allanach:2004rh}.



\subsubsection{Flavor physics constraints}
\label{Sec:Low_energy}
We consider the relevant constraints on the branching fractions of rare $B$-decay channels which are most sensitive to new physics, namely $B \to X_s \gamma$, $B_s \to \mu^+ \mu^-$ and $B^{+} \to \tau^{+} \nu_\tau$.  In the MSSM, without any new source of flavor and CP violations, the generic contribution to $B \to X_s \gamma$ can be characterized as \cite{Misiak:2015xwa,Altmannshofer:2012ks}
\begin{align} \label{bsg1}
R_{bs\gamma}&\equiv{\mathrm{Br}(B \to X_s \gamma) \over \mathrm{Br}(B \to X_s \gamma)_{\text{SM}}} =1-2.45
C_7^{NP}-0.59C_8^{NP},
\end{align}
where $C_{7,8}^{NP}$ are Wilson coefficients which encapsulate the new physics contributions to electromagnetic and chromo-magnetic $b \to s \gamma $ operators. In the MSSM, $C_{7,8}^{NP}$ receive contributions coming from charged Higgs-stop loops, higgsino-stop loops, neutral Higgs-bottom loops and gaugino-squark loops. In the SM, the NNLO prediction for branching ratio is Br$(B \to X_s \gamma)_{\text{SM}} = (3.36 \pm 0.23) \times 10^{-4}$ \cite{Czakon:2015exa,Misiak:2006zs,PhysRevLett.114.221801}, while the present world average of experimental measurements is Br$(B \to X_s \gamma)_{\text{exp.}} = (3.32 \pm 0.16) \times 10^{-4}$ \cite{Amhis:2016xyh}. This leaves room for new physics in $R_{bs\gamma}$, defined in Eq.~\eqref{bsg1}, as
\begin{align} \label{Rbsg}
R_{bs\gamma}= 0.99 \pm 0.08\ .
\end{align}
Here we would like to note that $R_{bs\gamma}$ has been obtained by assuming all statistical and systematic uncertainties in Br$(B \to X_s \gamma)_{\text{SM}}$ and Br$(B \to X_s \gamma)_{\text{exp}}$ to be gaussian, thus allowing them to be combined in quadrature. 
We have used micrOMEGAs 4.3.0 \cite{Belanger:2014vza,Belanger:2001fz,Belanger:2004yn} to compute Br$(B \to X_s \gamma)$ at NLO at the parameter space points. Then $R_{bs\gamma}$ is evaluated using the SM NLO prediction for Br$(B \to X_s \gamma) = (3.28 \pm 0.33) \times 10^{-4} $ \cite{Chetyrkin:1996vx}.

We also consider the new physics contribution to Br$(B_s\to \mu^+\mu^-)$. In the MSSM, at one-loop this process is mediated by heavy neutral and pseudoscalar Higgs ($H/A$) penguin and box diagrams. For penguin diagrams, flavor changing $b \to s$ quark transition is induced through charged Higgs - up-quark and up-squark - chargino loop. In addition, it also receives contribution through up-quark - charged Higgs - neutrino and up-squark - chargino - sneutrino box diagrams. In this work, we consider the latest experimentally measured value of Br$(B_s\to \mu^+\mu^-)$, as measured by LHCb \cite{PhysRevLett.118.191801}, $\text{Br}(B_s \to \mu^+ \mu^-)_{\text{exp.}} = (3.0\pm 0.6^{+0.3}_{-0.2}) \times 10^{-9}$.

A combination of two recent Belle measurements, using hadronic \cite{Adachi:2012mm} and semileptonic \cite{Kronenbitter:2015kls} tagging method and taking into account all correlated systematic  uncertainties, gives the branching fraction for the process  $B^{+} \to \tau^{+} \nu_\tau$ as Br$(B^{+} \to \tau^{+} \nu_\tau) = (0.91 \pm 0.19\, (\mathrm{stat.}) \pm 0.11\, (\mathrm{syst.})) \times 10^{-4}$ \cite{Kronenbitter:2015kls}, while the SM value for the same is Br$(B^{+} \to \tau^{+} \nu_\tau)_{\mathrm{SM}} = (0.828 \pm 0.060) \times 10^{-4}$ \cite{utfit}. In this analysis, we have considered the ratio of the latest experimental measurement of Br$(B^{+} \to \tau^{+} \nu_\tau)_{\mathrm{exp.}} = (1.06 \pm 0.19) \times 10^{-4}$ \cite{Amhis:2016xyh} and its SM value (Ratio = Br$(B^{+} \to \tau^{+} \nu_\tau)_{\mathrm{exp.}}/ Br(B^{+} \to \tau^{+} \nu_\tau)_{\mathrm{SM}} = 1.28019 \pm 0.247511$), in the global $\chi^{2}$ analysis.
As already stated, micrOMEGAs 4.3.0 \cite{Belanger:2014vza,Belanger:2001fz,Belanger:2004yn} is used to calculate the SUSY contributions to all the above three flavor observables\footnote{We did not consider the semi-leptonic decay of $B \to D^{(*)} \tau \bar{\nu}_{\tau}$ \cite{Lees:2012xj,Abdesselam:2016cgx,Aaij:2015yra} in our analysis, which shows some tension with respect to the SM prediction. It is extremely challenging to account for large deviations from the SM rates~\cite{Fajfer:2012vx,Crivellin:2012ye,Celis:2012dk}.}.


\subsubsection{Constraints from Higgs data}

CMS and ATLAS have analysed the LHC data collected at Run-I and Run-II, and derived constraints on the couplings of Higgs boson. 
These studies have been performed for the most significant production modes of Higgs boson at LHC viz. gluon gluon fusion (ggF), vector boson fusion (VBF), associated production with top-quark pairs ($t\bar{t}h$) and associated production with vector bosons (Vh), and for the decay modes $ h \rightarrow  ZZ, W^{+}W^{-}, \gamma \gamma,\tau^{+} \tau^{-}$ and $b \bar{b}$.  Their analysis uses the signal strength formalism, where the signal strength variable $\mu^{f}_{i}$ is defined as: 
\begin{equation}
\mu_{i}^{f}  =  \frac{\sigma_{i}}{\sigma_{i_{SM}}} \cdot \frac{B^{f} }{B^{f}_{SM}} \, .
\end{equation}
Here $i$ stands for production modes of the Higgs and $f$ stands for the Higgs boson decay modes. The parameters  with subscript `SM' represents the corresponding values in the SM.
The experimentally obtained best-fit signal strength values implemented in our analysis are tabulated in Table~\ref{table-mu8} and Table~\ref{table-mu13}. It may be noted that some of these signal strength rates ($ W^{+}W^{-}$ channel : $ 0/1~jet~ggF~tagged$ and $VBF~ tagged$ (CMS), and $\tau^{+}\tau^{-}$ channel : $0~jet~ggF$, $1~jet~ggF$ and $VBF~tagged$ (CMS))  have been obtained after imposition of specific jet vetos. In our analysis, we compare these rates against the signal strength rates of our parameter space points computed by adding the \texttt{FeynHiggs} inclusive cross-section from all contributing channels in the specified fraction. This comparison is motivated from a simplistic point of view that the signal strength rates would incur small changes upon application of exclusive cuts, under the assumption that the signal distribution and the SM distribution gets equally affected.

\begin{table}[htb!]
\renewcommand{\arraystretch}{1.5}
\begin{center}
\begin{tabular}{| c || c | C{2.5cm} || c | C{2.5cm}  |}
\hline \hline
\centering 
\multirow{2}{*}{\shortstack{Decay \\ channel}}  & \multirow{2}{*}{\shortstack{Production \\ mode}} & \multirow{2}{*}{ATLAS} & \multirow{2}{*}{\shortstack{Production \\ mode}} & \multirow{2}{*}{CMS}  \\ 
& & & & \\ \hline \hline
\multirow{5}{*}{$\gamma \gamma$} & $ ggF $ & $\rm 1.32^{+0.38}_{-0.38} $ \cite{Aad:2014eha} & $ ggF $ & $\rm 1.12^{+0.37}_{-0.32} $ \cite{Khachatryan:2014ira} \\
& $VBF$ & $\rm 0.8^{+0.7}_{-0.7} $\cite{Aad:2014eha} & $ VBF$ & $\rm 1.58^{+0.77}_{-0.68} $ \cite{Khachatryan:2014ira} \\
& $Wh$ & $\rm 1.0^{+1.60}_{-1.60} $\cite{Aad:2014eha} & $Wh$ & $\rm -0.16^{+1.16}_{-0.79} $  \cite{Khachatryan:2014ira} \\
& $t\bar{t}h$ & $\rm 1.60^{+2.70}_{-1.80} $\cite{Aad:2014eha}  &  $t\bar{t}h$ & $\rm 2.69^{+2.51}_{-1.81}$\cite{Khachatryan:2014ira} \\
& $Zh$ & $\rm 0.1^{+3.70}_{-0.10} $\cite{Aad:2014eha} &  - & - \\ \hline

\multirow{2}{*}{$ZZ$} & $VBF+Vh$ & $\rm 0.26^{+1.64}_{-0.94} $  \cite{Aad:2014eva} & $VBF+Vh$ & $\rm 1.70^{+2.2}_{-2.1} $ \cite{Chatrchyan:2013mxa} \\
& $ggF+t\bar{t}h+b\bar{b}h $ & $\rm 1.66^{+0.51}_{-0.44} $ \cite{Aad:2014eva} & $ggF+t\bar{t}h $ &  $\rm 0.80^{+0.46}_{-0.36} $ \cite{Chatrchyan:2013mxa} \\ \hline

\multirow{4}{*}{$W^{+}W^{-}$} & $ ggF $ & $\rm 1.02^{+0.29}_{-0.26} $ \cite{ATLAS:2014aga} & \shortstack{$ 0/1~jet$ \\ ($ 97\%~ggF,~3\%~VBF$)}& $\rm 0.74^{+0.22}_{-0.20} $ \cite{Chatrchyan:2013iaa} \\
& $VBF$ & $\rm 1.27^{+0.53}_{-0.45} $ \cite{ATLAS:2014aga} & \shortstack{$ VBF~tagged$ \\ ($ 17\%~ggF,~83\%~VBF$)} & $\rm 0.60^{+0.57}_{-0.46} $ \cite{Chatrchyan:2013iaa} \\
& $Vh$ & $\rm 3.0^{+1.64}_{-1.30} $ \cite{Aad:2015ona} & $Vh~tagged$ & $\rm 0.39^{+1.97}_{-1.87} $ \cite{Chatrchyan:2013iaa} \\
& - & - & $Wh~tagged$ & $\rm 0.56^{+1.27}_{-0.95} $ \cite{Chatrchyan:2013iaa}  \\ \hline 

\multirow{1}{*}{$b\bar{b}$} & $Vh$ & $\rm 0.51^{+0.40}_{-0.37} $ \cite{Aad:2014xzb} & $Vh$ & $\rm 1.0^{+0.5}_{-0.5} $ \cite{Chatrchyan:2013zna} \\ \hline

\multirow{6}{*}{$\tau^{+} \tau^{-}$} & $ggF$ & $\rm 1.93^{+1.45}_{-1.15} $ \cite{ATLAS-CONF-2014-061} & \shortstack{$ 0~jet$ ($ 96.9\%~ggF$,\\$~1\%~VBF,~2.1\%~Vh$)} & $\rm 0.34^{+1.09}_{-1.09} $ \cite{Chatrchyan:2014nva} \\
& $VBF(60\%)+Vh(40\%)$ & $\rm 1.24^{+0.58}_{-0.54} $ \cite{ATLAS-CONF-2014-061} & \shortstack{$ 1~jet$ ($ 75.7\%~ggF$,\\$~14\%~VBF,~10.3\%~Vh$)} & $\rm 1.07^{+0.46}_{-0.46} $ \cite{Chatrchyan:2014nva} \\
& - & - & \shortstack{$ VBF~tagged$ \\ ($ 19.6\%~ggF,~80.4\%~VBF$)} & $\rm 0.94^{+0.41}_{-0.41} $ \cite{Chatrchyan:2014nva} \\
& - & - & \shortstack{$ Vh~tagged$} & $\rm -0.33^{+1.02}_{-1.02} $ \cite{Chatrchyan:2014nva} \\ \hline \hline 

\end{tabular}
\caption{Best-fit value of signal strength variables, along with the associated errors, derived by ATLAS and CMS using LHC Run-I data, which have been implemented to perform the global $\chi^{2}$ analysis to obtain the allowed parameter space.}
\label{table-mu8}
\end{center}
\end{table}

\begin{table}[htb!]
\renewcommand{\arraystretch}{1.5}
\begin{center}
\begin{tabular}{| c || c | C{2.5cm} || c | C{2.5cm}  |}
\hline \hline
\centering 
\multirow{2}{*}{\shortstack{Decay \\ channel}}  & \multirow{2}{*}{\shortstack{Production \\ mode}} & \multirow{2}{*}{ATLAS} & \multirow{2}{*}{\shortstack{Production \\ mode}} & \multirow{2}{*}{CMS}  \\ 
& & & & \\ \hline \hline

\multirow{4}{*}{$\gamma \gamma$}  & $ ggF$ & $\rm 0.80^{+0.19}_{-0.18} $\cite{ATLAS-CONF-2017-047} & $ ggF $ & $\rm 1.11^{+0.19}_{-0.18} $\cite{CMS-PAS-HIG-16-040} \\ 
& $VBF$ & $\rm 2.1^{+0.60}_{-0.60} $\cite{ATLAS-CONF-2017-047} & $ VBF $ & $\rm 0.5^{+0.6}_{-0.5} $\cite{CMS-PAS-HIG-16-040} \\
& $t\bar{t}h$ & $\rm 0.5^{+0.60}_{-0.60} $\cite{ATLAS-CONF-2017-047} & $t\bar{t}h$ & $\rm 2.2^{+0.9}_{-0.8} $\cite{CMS-PAS-HIG-16-040} \\
& $Vh$ & $\rm 0.70^{+0.9}_{-0.8} $\cite{ATLAS-CONF-2017-047} & $Vh$ & $\rm 2.3^{+1.1}_{-1.0} $\cite{CMS-PAS-HIG-16-040} \\ \hline

\multirow{3}{*}{$ ZZ $} & $ ggF$ & $\rm 1.17^{+0.41}_{-0.50} $\cite{ATLAS-CONF-2017-047} & $ ggF $ & $\rm 1.20^{+0.22}_{-0.21} $\cite{Sirunyan:2017exp} \\
& - & - & $ VBF $ & $\rm 0.05^{+1.03}_{-0.05}$\cite{Sirunyan:2017exp} \\
& - & - & $t\bar{t}h$ & $\rm 0.00^{+1.19}_{-0.00} $\cite{Sirunyan:2017exp} \\\hline 

\multirow{3}{*}{$b \bar{b}$} & $VBF $ & $\rm -3.9^{+2.8}_{-2.9} $ \cite{ATLAS-CONF-2016-063} & $ VBF $ & $\rm -3.7^{+2.4}_{-2.5} $ \cite{CMS-PAS-HIG-16-003} \\
& $t\bar{t}h$ & $\rm 2.1^{+1.0}_{-0.9} $ \cite{ATLAS-CONF-2016-080} & $t\bar{t}h$ & $\rm -2.0^{+1.8}_{-1.8} $ \cite{CMS-PAS-HIG-16-004} \\ 
& $ Vh $ & $\rm 0.21^{+0.51}_{-0.50} $ \cite{ATLAS-CONF-2016-091} & - & - \\ \hline 

\multirow{2}{*}{$ \tau^{+}\tau^{-}$} & - & - & $ ggh $ & $\rm 1.05^{+0.49}_{-0.46} $ \cite{CMS-PAS-HIG-16-043} \\
 & - & - & $ q\bar{q}h+Wh+Zh $ & $\rm 1.07^{+0.45}_{-0.43} $ \cite{CMS-PAS-HIG-16-043} \\\hline 

\end{tabular}
\caption{Best-fit value of strength variables, along with the associated errors, derived by ATLAS and CMS using LHC Run-II data, which have been implemented to perform the global $\chi^{2}$ analysis to obtain the allowed parameter space.}
\label{table-mu13}
\end{center}
\end{table}


\subsection{Global fit analysis}

We perform a global $\chi^{2}$ analysis over the entire scanned parameter space taking into account the most relevant flavor physics observables discussed and Higgs signal strength constraints tabulated in Table~\ref{table-mu8} and Table~\ref{table-mu13}. The light Higgs mass constraint has been imposed separately and parameter space points are dropped out if they generate $M_{h}$ outside the range specified in Sec.~\ref{light_higgs_mass}. 


Value of $\chi^{2}$ is computed for all parameter space points, with $\chi^{2}$ is defined as
\begin{eqnarray}
\chi^{2} = \sum_{i} \frac{(\bar{x}_{i} - x_{i})^{2}}{\Delta x^{2}_{i}} \, ,
\label{chi_2}
\end{eqnarray}
where $x_{i}$ corresponds to the experimentally obtained best-fit value of the observable and $\bar{x}_{i}$ is the value of the corresponding observable computed for the parameter space point in MSSM. $\Delta x^{2}_{i}$ represents the error associated with the experimental measurement. 
In case of Higgs signal strength constraints $x_{i}$ corresponds to $\mu_{i}$ where $\mu_{i}$ represents Higgs signal strength corresponding to a certain Higgs boson production mode and Higgs decay channel. 
In case of flavor physics constraints, $x$ corresponds to the branching fraction of the decay channels $B_{s} \to \mu^{+}\mu^{-}$ and $B^{+} \to \tau^{+} \nu_{\tau}$, and to the ratio $Br(b \to s \gamma)/Br(b \to s \gamma)_{SM}$ for the channel $B \to X_s \gamma$. 
The summation over $i$ in Eq.~\eqref{chi_2} represents that $\chi^{2}$ is calculated by summing over all the experimentally obtained signal strength observables tabulated in Table~\ref{table-mu8} and Table~\ref{table-mu13}, corresponding to various production and decay modes of Higgs boson and also the flavor physics observables discussed in Sec.~\ref{Sec:Low_energy}.  Following the global $\chi^{2}$ approach mentioned in Ref.~\cite{Bhattacherjee:2015sga}, we combine 28 observables corresponding to the signal strength measurements from LHC Run-I data (including both CMS and ATLAS analysis as shown in Table.~\ref{table-mu8}), 18 observables from LHC Run-II (including results from $\sim 36$ fb$^{-1}$ as well as $\sim 15$ fb$^{-1}$ luminosity data and from both CMS and ATLAS analysis as shown in Table~\ref{table-mu13}) and $3$ $B$-physics observables. Finally we calculate $\chi^{2}$ for each parameter space point, and determine the minimum $\chi^{2}_{min}$. 


\begin{table}
\begin{tabular}{| c | c | c | c | c |}
\hline
&   & \textbf{8 TeV}(d.o.f) & \textbf{13 TeV}(d.o.f) &  \textbf{Combined}(d.o.f) \\
\hline
\multirow{2}{*}{MSSM} & Without flavor & 15.133(11) & 31.739(1) & 47.046(29) \\ 
& With flavor & 16.686(14) & 33.474(4) & 48.801(32) \\ 
\hline
\multirow{2}{*}{SM} & Without flavor & 15.531(28) & 32.649(18) & 48.179(46) \\ 
& With flavor & 17.766(31) & 34.884(21) & 50.415(49) \\ 
\hline
\end{tabular}
\caption{Values of $\chi_{min}^{2}$ obtained upon combing the constraints from 8 TeV and/or 13 TeV Higgs signal strength observables along with/without the constraints from flavor physics observables.}
\label{chisquaresm}
\end{table}

\begin{figure}[!htb]
\begin{center}
\includegraphics[angle=0, width=0.49\textwidth]{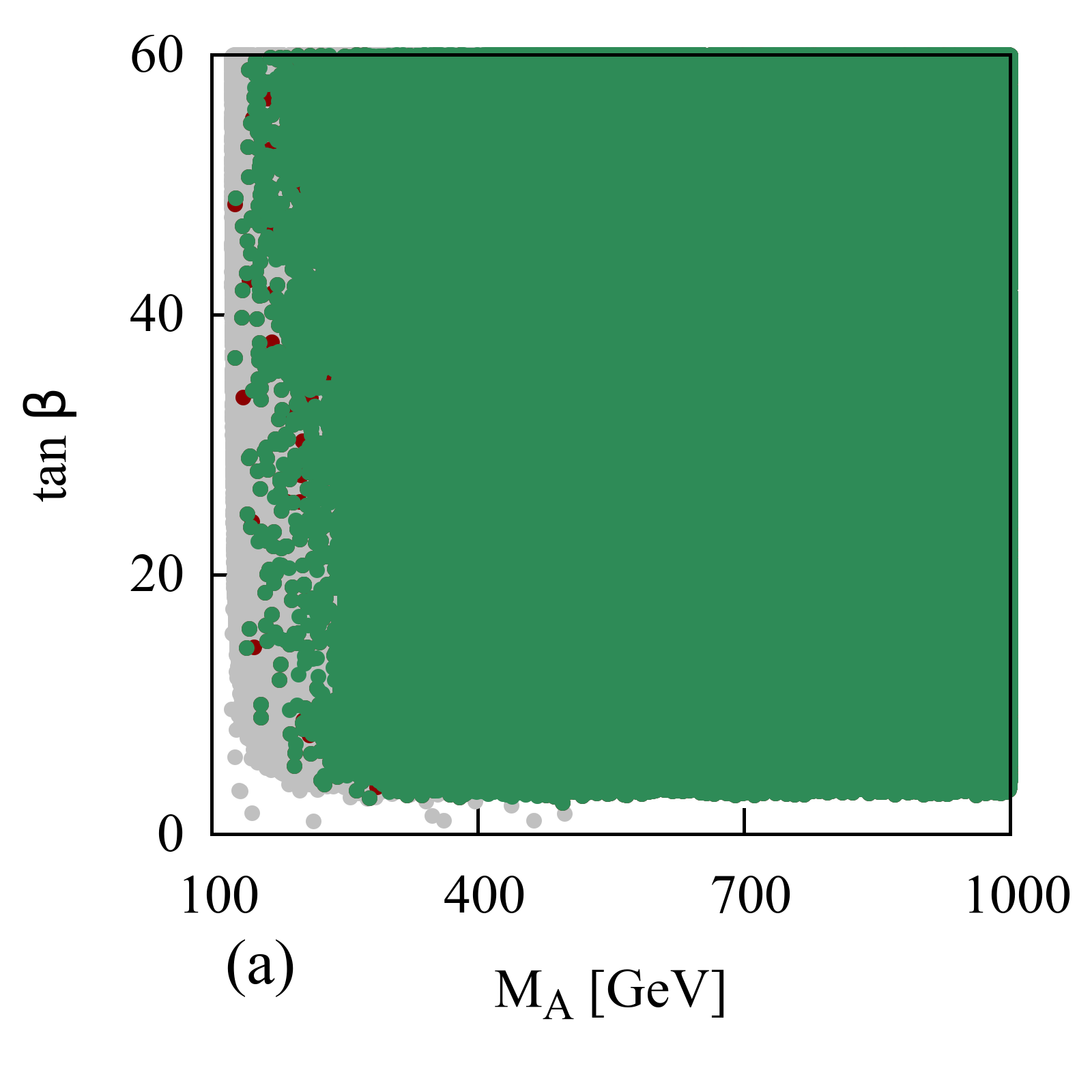} \includegraphics[angle=0, width=0.49\textwidth]{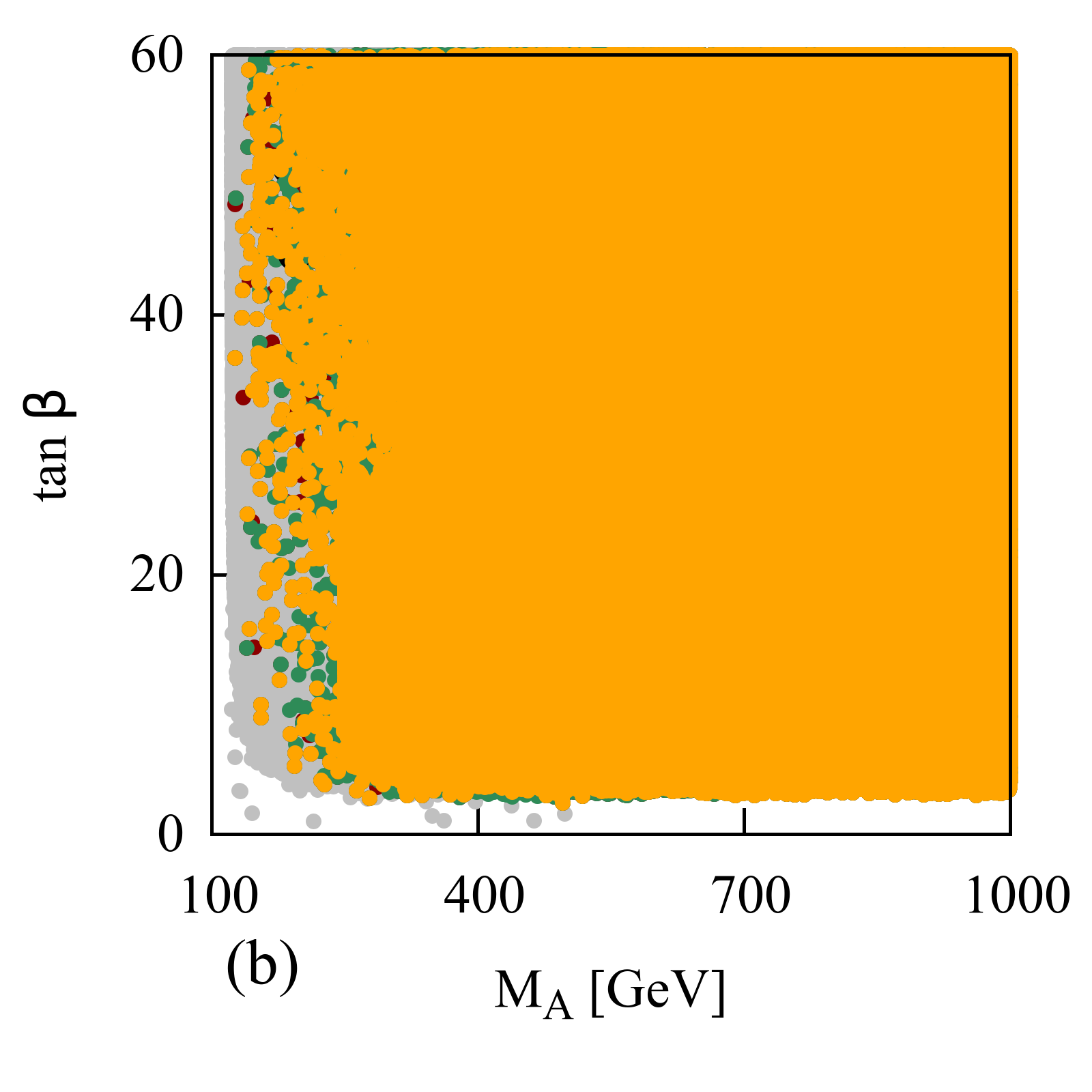}\\
\caption{Scatter plot in the $M_{A} - \tan\beta$ plane, the grey colored points satisfy the light Higgs mass constraint. The red colored points in Fig.~\ref{fig:ma_tb1}(a) and Fig.~\ref{fig:ma_tb1}(b) represent the parameter space points which lie within the $2\sigma$ interval around the $\chi^{2}_{min}$, calculated by combining the Higgs signal strength constraints derived from LHC Run-II data (tabulated in Table.~\ref{table-mu13}). The green colored points in Fig~\ref{fig:ma_tb1}(a) and Fig~\ref{fig:ma_tb1}(b) lie within the $2\sigma$ interval of $\chi^{2}_{min}$ computed by combining the Higgs signal strength constraints derived from LHC Run-I data (tabulated in Table.~\ref{table-mu8}). The yellow colored points in Fig.~\ref{fig:ma_tb1}(b) correspond to the parameter space points which fall within the $2\sigma$ interval of the $\chi^{2}_{min}$ obtained by combining the Higgs signal strength constraints for both Run-I and Run-II.}  
\label{fig:ma_tb1}
\end{center}
\end{figure}

\begin{figure}[!htb]
\begin{center}
\includegraphics[angle=0, width=0.49\textwidth]{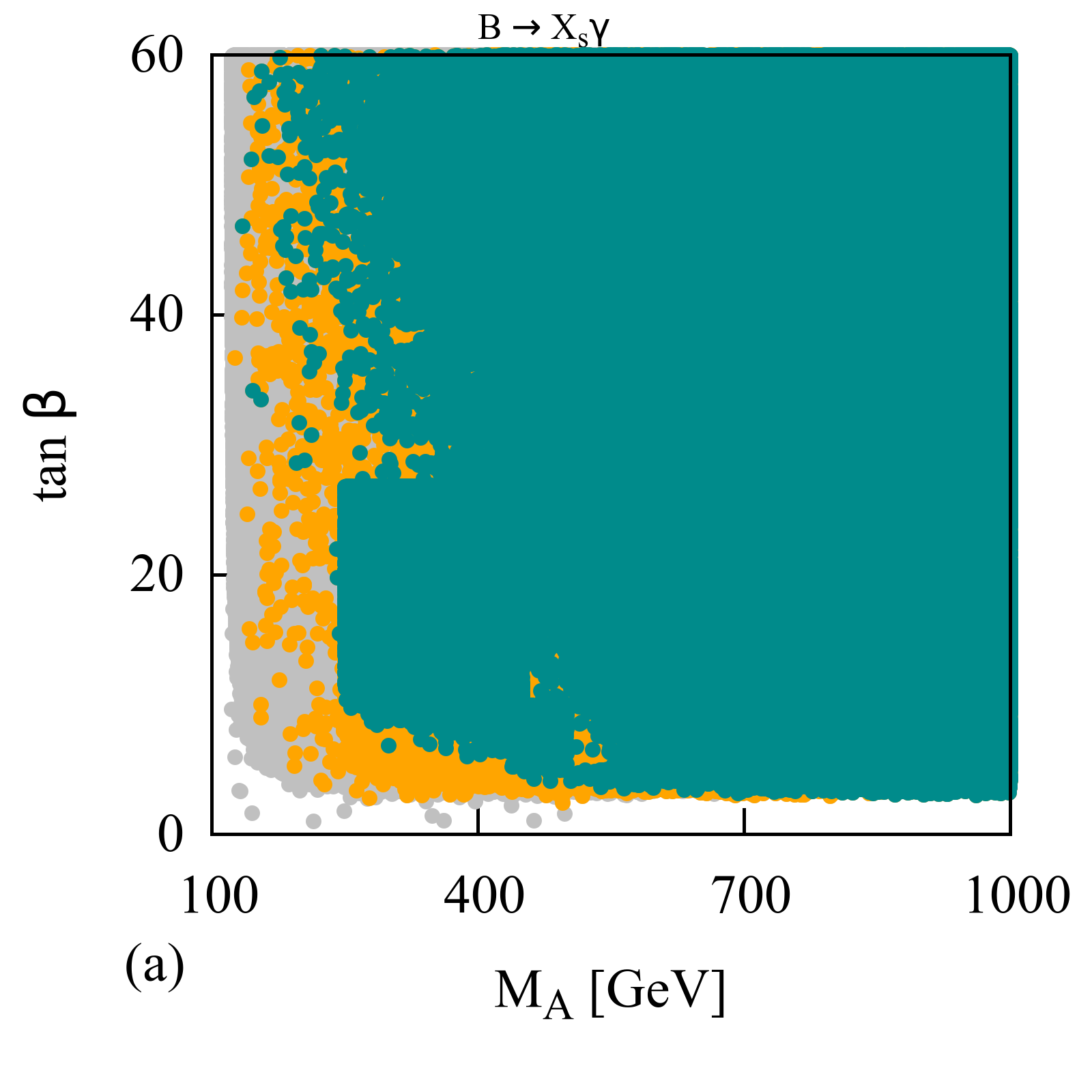} \includegraphics[angle=0, width=0.49\textwidth]{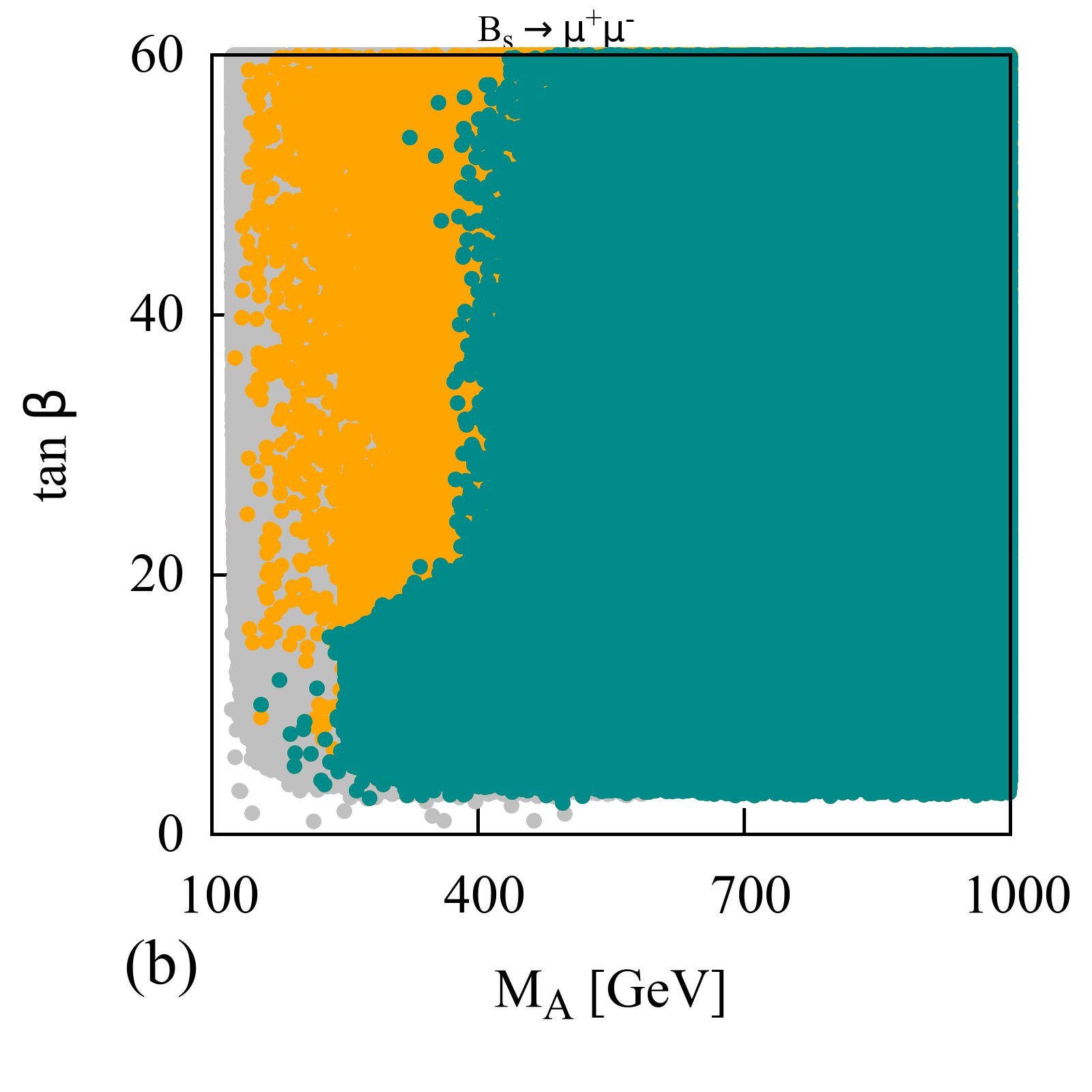}\\
\includegraphics[angle=0, width=0.49\textwidth]{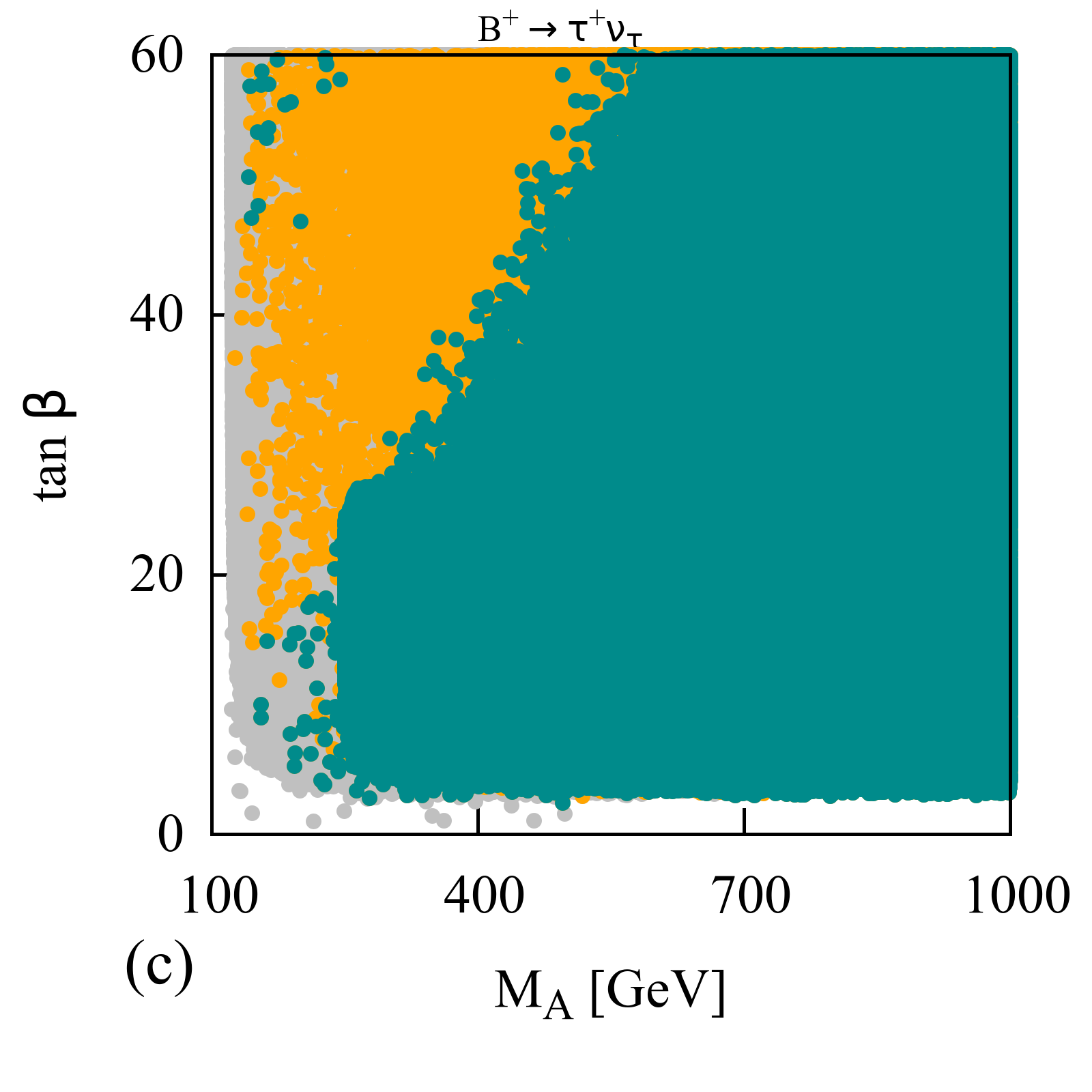} \includegraphics[angle=0, width=0.49\textwidth]{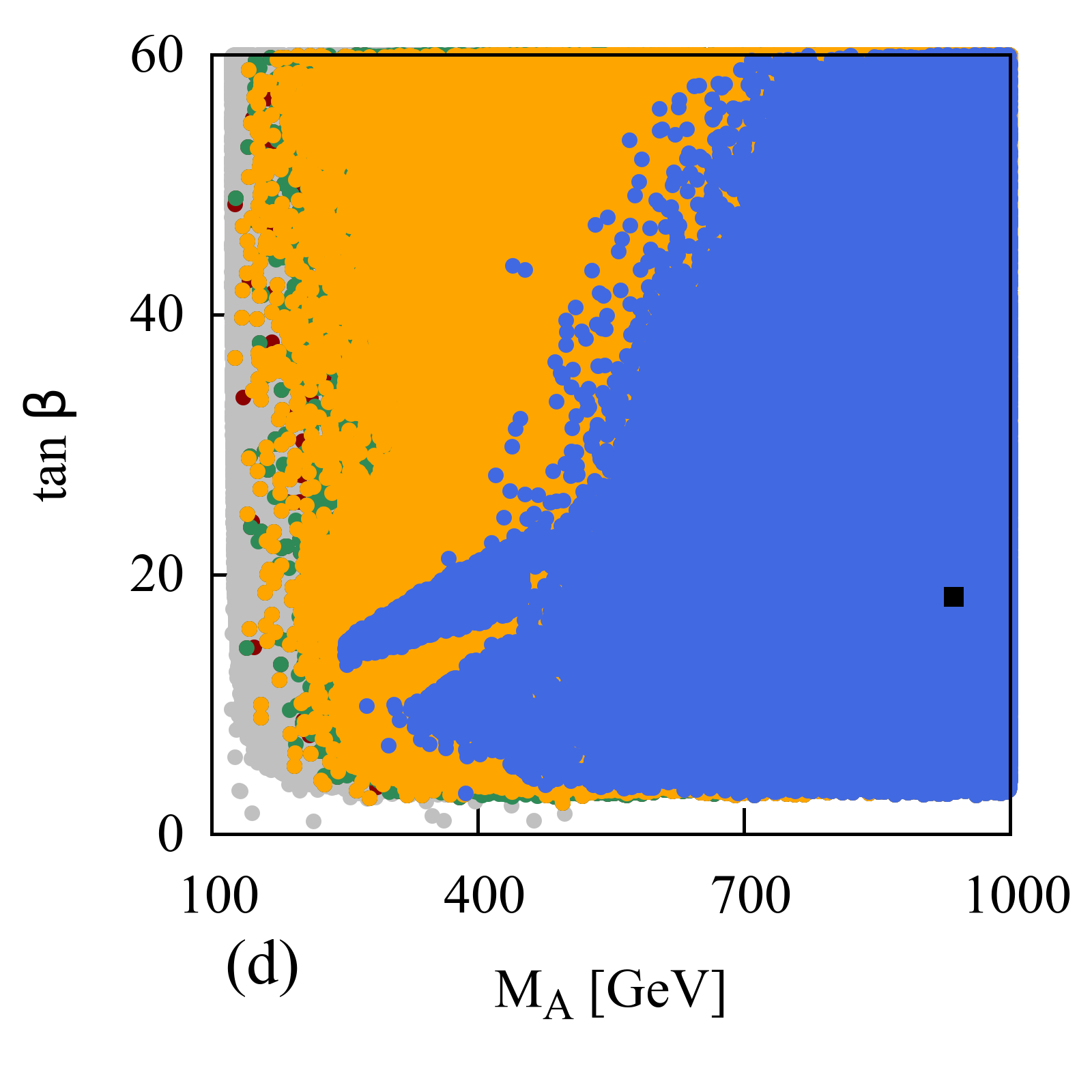}\\
\caption{Scatter plot in the $M_{A} - \tan\beta$ plane, the grey colored points satisfy the light Higgs mass constraint. The light-green colored points in Fig.~\ref{fig:ma_tb2}(a),(b) and (c) represent the parameter space points which lie within the $2\sigma$ interval of $\chi^{2}_{min}$, calculated by combining the Higgs signal strength constraints derived from LHC Run-I data (tabulated in Table.~\ref{table-mu8}), LHC Run-II data (tabulated in Table.~\ref{table-mu13}) along with the constraint from $B \to X_{s} \gamma $, $B_s \rightarrow \mu^+ \mu^-$ and $B^+ \rightarrow \tau^+ \nu_{\tau}$, respectively. The grey and yellow colored points in Fig.~\ref{fig:ma_tb2}(a),(b),(c)correspond to the same color coding as in Fig.~\ref{fig:ma_tb1}. Similarly, in Fig.~\ref{fig:ma_tb2}(d), the grey, green and the yellow colored points correspond to the same color coded points as in Fig.~\ref{fig:ma_tb1}. However, the blue colored points have been obtained by combining all the flavor physics observables along with Run-I and Run-II Higgs signal strength constraints in the $\chi^{2}$ evaluation and then taking points within $2\sigma$ interval of the $\chi^{2}_{min}$. We refer to the blue colored points of Fig.~\ref{fig:ma_tb2}(d) as the allowed parameter space in the remainder of this analysis, unless otherwise stated. In Fig.~\ref{fig:ma_tb2}(d), the
parameter space point which generates the minima ($\chi^{2}_{min}$) has been shown in black and corresponds
to $M_{A}= 936.2~{\rm GeV}$ and $\tan\beta=18.3$.}  
\label{fig:ma_tb2}
\end{center}
\end{figure}

We compute $\chi^2_{min}$ by considering 8 TeV and 13 TeV Higgs signal
 strength data separately and in combination as well, with and without the flavor constraints. We present the values of $\chi^2_{min}$ for all these cases in the MSSM and the SM in Table~\ref{chisquaresm}, 
along with the corresponding degrees of freedom (d.o.f). 
The d.o.f for SM is equivalent to the total number of independent constraints summed together in the evaluation of $\chi^{2}_{min}$, as can be seen in Table~\ref{chisquaresm}.
In case of MSSM, d.o.f is obtained by subtracting the total number of independently varied input parameters relevant to our parameter space from the sum of independent observables used in the evaluation of $\chi^{2}_{min}$. In this analysis, we vary over $17$ input parameters in order to scan over the parameter space of our interest.
We can see that the $\chi^2_{min}$ for MSSM is less than $\chi^2_{min}$ for SM in all the cases. 

 In Fig.~\ref{fig:ma_tb1} and Fig.~\ref{fig:ma_tb2} we show the implications of taking various combinations of observables in the global $\chi^2$ analysis. All the points shown in the plots have $\chi^2$ values within the $2\sigma$ interval of $\chi^2_{min}$ i.e. $\chi^{2} \leq \chi^{2}_{min} + 6.18$, where $6.18$ is the 2$\sigma$ interval associated with two degrees of freedom. 
 The grey colored points in Fig.~\ref{fig:ma_tb1} and \ref{fig:ma_tb2} represent the parameter space points which satisfy light Higgs mass constraints. 
 The red points in Fig.~\ref{fig:ma_tb1}(a), \ref{fig:ma_tb1}(b) and \ref{fig:ma_tb2}(d) 
 represent those evaluated by considering only 13 TeV Higgs signal strength data, while
the green points in Fig.~\ref{fig:ma_tb1}(a), \ref{fig:ma_tb1}(b) and \ref{fig:ma_tb2}(d) represent the points obtained by considering only 8 TeV Higgs signal strength data. The yellow points in Fig.~\ref{fig:ma_tb1}(b), \ref{fig:ma_tb2}(a), \ref{fig:ma_tb2}(b), \ref{fig:ma_tb2}(c) and \ref{fig:ma_tb2}(d) represent the points computed by considering both 8 TeV and 13 TeV data. 
The light-green colored points in Fig.~\ref{fig:ma_tb2}(a), \ref{fig:ma_tb2}(b) and \ref{fig:ma_tb2}(c) represent the parameter space points evaluated by combining 8 TeV and 13 TeV Higgs signal strength data along with the flavor physics constraint from $B \to X_{s} \gamma $, $B_s \rightarrow \mu^+ \mu^-$ and $B^+ \rightarrow \tau^+ \nu_{\tau}$, respectively. 
Finally, the blue points in Fig.~\ref{fig:ma_tb2}(d) represent those which are obtained by considering the combination of 8 TeV and 13 TeV signal strength data along with the flavor physics constraints discussed in Sec.~\ref{Sec:Low_energy}. We represent the parameter space point with the minimum value of $\chi^{2}$ ($\chi^{2}_{min}=48.801$) in black in Fig.~\ref{fig:ma_tb2}(d), and it corresponds to $M_A=936.2~{\rm GeV}$ and $\tan\beta =18.3$. 
We note that the $b \rightarrow s \gamma$ results constrain the low $M_A$ regions, while $B_s \rightarrow \mu^+ \mu^-$ results constrain the low $M_A$ and high $\tan \beta$ regions. $B^+ \rightarrow \tau^+ \nu_{\tau}$ constrains the low $M_A$ and high $\tan \beta$ region even further. We observe that parameter space points with $\tan{\beta}~\gtrsim~55$ for $M_{A}~\lesssim~600~{\rm GeV}$ are excluded after combining all the $B$-physics observables in the $\chi^{2}$ evaluation. Interestingly, a wedge shaped region around $16 \lesssim \tan\beta \lesssim 12$ and $M_{A} \lesssim 450~{\rm GeV}$ also get excluded through the global $\chi^{2}$ analysis, as shown in Fig.~\ref{fig:ma_tb2}(d). This particular exclusion is mainly driven by the constraints from $B_s \rightarrow \mu^+ \mu^-$. For a fixed $M_{A}$, the contribution to $\mathrm{Br}(B_s \rightarrow \mu^+ \mu^-)$ coming from the Higgsino loop interferes destructively with the SM contribution for the case of positive $\mu A_{t}$~\cite{Altmannshofer:2012ks}. Although, we have varied $A_{t}$ across both, positive and negative values, the wedge shaped excluded region corresponds to positive values of $A_{t}$. The destructive interference lowers down the value of $\mathrm{Br}(B_s \rightarrow \mu^+ \mu^-)$ below the SM expectation. In such scenarios, the net contribution from $\mathrm{Br}(B_s \rightarrow \mu^+ \mu^-)$ to the global $\chi^{2}$ increases, thereby, sending it outside the $\chi^{2}_{min}+ 2\sigma$ region. We would like to mention here that the lower limit on $\tan \beta$ is also dependent on $M_S$. By increasing the soft SUSY-breaking scale by few orders of magnitude one can open up the very low $\tan \beta$ region i.e., $\tan \beta < 3$~\cite{Djouadi:2013vqa}. In the rest of this work, we will refer to the blue colored points of Fig.~\ref{fig:ma_tb2}(d) as the allowed parameter space, unless otherwise stated.

\begin{figure}[t!]
\begin{center}
{\includegraphics[angle=0,width=0.60\textwidth]{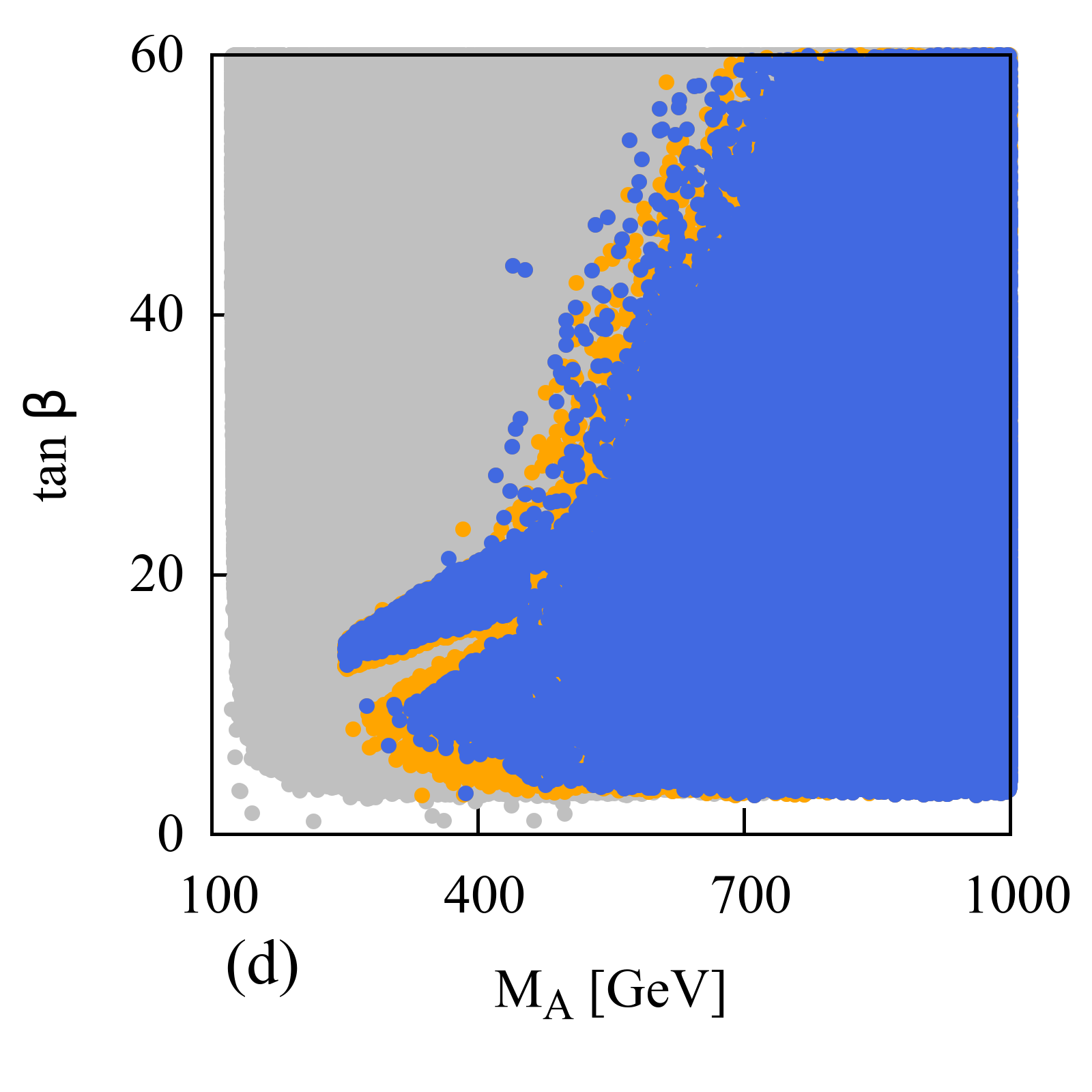}}
\caption{Allowed parameter space points obtained after performing the global $\chi^2$ analysis with the older $b \rightarrow s \gamma$ constraints~\cite{Amhis:2014hma} (shown in yellow color), and with the most recent $b \rightarrow s \gamma$ constraints~\cite{Amhis:2016xyh} (shown as blue colored points). }
\label{fig-bsg}
\end{center}
\end{figure}

%
In an attempt to have a better understanding about the implication of the latest measurement of $Br(B \rightarrow X_{s} \gamma)$ on our parameter space, we performed a global $\chi^2$ analysis by considering the earlier world average of Br($B \to X_s \gamma)_{\text exp} = (3.49 \pm 0.19) \times 10^{-4}$ given in Ref.~\cite{Amhis:2014hma}, which resulted in $R_{bs\gamma}= 1.04 \pm 0.09$, and compared its result against the parameter space obtained from a global $\chi^{2}$ analysis considering the latest world average as given in Eq.~\eqref{Rbsg}. We observe that the new data imposes a slightly stronger constraint on the value of $M_A$ in the low $\tan \beta$ region, as can be seen in Fig.~\ref{fig-bsg}.
                                            
%
%


\subsection{Heavy Higgs searches}

In the previous section we presented the parameter space satisfying flavor physics and Higgs signal strength constraints. 
In this section we apply the most updated bounds on the masses and cross-section times branching ratios of the heavy Higgs bosons, 
and study its effect on the allowed parameter space. 
We consider several direct Heavy Higgs searches by ATLAS and CMS collaborations at 8 TeV and 13 TeV e.g. $H/A \rightarrow \gamma \gamma, b \bar b, \tau^{+} \tau^{-}$, $H \rightarrow W^{+}W^{-}, ZZ, hh$ and $A  \rightarrow Zh$. 
A complete list of the heavy Higgs searches considered here is summarized in Table~\ref{tab:heavysearches} and 
Table~\ref{tab:heavysearchesichep}. 

\begin{table}
\begin{footnotesize}
\begin{tabular}{| c | c | c | c |}
\hline
\textbf{Channel} & \textbf{Experiment} & \textbf{Mass range (GeV)}  &  \textbf{Luminosity} \\
\hline
\multirow{2}{*}{$gg\to H/A \to \tau^{+}\tau^{-}$} & ATLAS 8 TeV~\cite{Aad:2014vgg} & 90-1000 & 19.5-20.3 fb$^{-1}$ \\
& CMS 8 TeV~\cite{CMS-PAS-HIG-14-029} &  90-1000  &19.7 fb$^{-1}$ \\
& ATLAS 13 TeV~\cite{ATLAS-CONF-2015-061} & 200-1200 &3.2 fb$^{-1}$ \\
& CMS 13 TeV~\cite{CMS-PAS-HIG-16-006}&100-3000  & 2.3 fb$^{-1}$ \\
\hline
\multirow{2}{*}{$b\bar{b}\to H/A \to \tau^{+}\tau^{-}$} & ATLAS 8 TeV~\cite{Aad:2014vgg} & 90-1000 & 19.5-20.3 fb$^{-1}$ \\
& CMS 8 TeV \cite{CMS-PAS-HIG-14-029}& 90-1000 & 19.7 fb$^{-1}$  \\
& ATLAS 13 TeV \cite{ATLAS-CONF-2015-061}& 200-1200   & 3.2 fb$^{-1}$ \\
& CMS 13 TeV  \cite{CMS-PAS-HIG-16-006} & 100-3000 & 2.3 fb$^{-1}$ \\
\hline
\multirow{2}{*}{$gg\to H/A \to \gamma\gamma$} & ATLAS 8 TeV \cite{Aad:2014ioa} & 65-600 & 20.3 fb$^{-1}$ \\
& CMS 8+13 TeV \cite{Khachatryan:2016hje}& 500-4000 & 19.7+3.3 fb$^{-1}$ \\
& ATLAS 13 TeV \cite{Aaboud:2016tru}& 200-2000 & 3.2 fb$^{-1}$ \\
\hline
$pp\to b H/A( H/A \to b\bar{b})$ & CMS 8 TeV  \cite{Khachatryan:2015tra} & 100-900 & 19.7 fb$^{-1}$ \\ 
\hline
\multirow{2}{*}{$gg\to H\to W^{+}W^{-}$} & ATLAS 8 TeV  \cite{Aad:2015agg}& 300-1500  &  20.3 fb$^{-1}$\\ 
& ATLAS 13 TeV  \cite{ATLAS-CONF-2016-021} &  500-3000 & 3.2 fb$^{-1}$\\
\hline
\multirow{2}{*}{$W^{+}W^{-}/ZZ \to H\to W^{+}W^{-}$} & ATLAS 8 TeV \cite{Aad:2015agg}&  300-1500 & 20.3 fb$^{-1}$ \\
& ATLAS 13 TeV\cite{ATLAS-CONF-2016-021} &  500-3000   &   3.2 fb$^{-1}$ \\
\hline
$gg\to H\to ZZ$ & ATLAS 8 TeV \cite{Aad:2015kna}& 160-1000 & 20.3 fb$^{-1}$ \\
\hline
$gg\to H\to ZZ \to (\ell \ell)(qq)$  & ATLAS 13 TeV \cite{ATLAS-CONF-2016-016}&  300-1000 & 3.2 fb$^{-1}$ \\
\hline
$gg\to H\to ZZ \to (\ell \ell)(\nu \nu)$  & ATLAS 13 TeV  \cite{ATLAS-CONF-2016-012}  & 300-1000 & 3.2 fb$^{-1}$\\
\hline
$pp \to H\to Z \gamma $ & ATLAS 13 TeV \cite{Aaboud:2016trl} &  250-2750 & 3.2 fb$^{-1}$\\
\hline
$W^{+}W^{-}/ZZ \to H\to ZZ$ & ATLAS 8 TeV  \cite{Aad:2015kna}& 160-1000 & 20.3 fb$^{-1}$\\
\hline
$pp \to H\to ZZ$ & CMS 8 TeV  \cite{Khachatryan:2015cwa}&  150-1000 & 5.1 fb$^{-1}$\\
\hline
$pp\to H\to W^{+}W^{-}$ & CMS 8 TeV \cite{Khachatryan:2015cwa} &  150-1000 & 5.1 fb$^{-1}$ \\
\hline
$gg\to H\to hh$ & ATLAS 8 TeV \cite{Aad:2015xja}& 260-1000 & 20.3 fb$^{-1}$\\
\hline
$pp\to H\to hh \rightarrow (b \bar b) (b \bar b)$ & ATLAS 13 TeV  \cite{Aaboud:2016xco} &  500-3000  & 3.2 fb$^{-1}$\\
\hline
$pp\to H\to hh \to (\gamma \gamma) (b \bar{b})$ & CMS 8 TeV \cite{Khachatryan:2016sey} & 250-1100  & 19.7 fb$^{-1}$\\
\hline
$pp\to H\to hh \to (b\bar{b}) (b\bar{b})$ & CMS 8 TeV \cite{Khachatryan:2015yea}&   270-1100   & 17.9 fb$^{-1}$\\
\hline
$gg\to H\to hh \to (b\bar{b}) (\tau^{+}\tau^{-})$ & CMS 8 TeV \cite{Khachatryan:2015tha}&  260-350 & 19.7 fb$^{-1}$\\
\hline
$gg\to A\to Zh \to (\tau^{+}\tau^{-}) (\ell \ell)$ & CMS 8 TeV \cite{Khachatryan:2015tha}& 220-350 & 19.7 fb$^{-1}$\\
\hline
$gg\to A\to Zh \to (b\bar{b}) (\ell \ell)$ & CMS 8 TeV \cite{Khachatryan:2015lba} & 225-600 &19.7 fb$^{-1}$ \\
\hline
$gg\to A\to Zh \to Z (\tau^{+}\tau^{-}) $ & ATLAS 8 TeV \cite{Aad:2015wra}&220-1000 & 20.3 fb$^{-1}$ \\
\hline
\multirow{2}{*}{$gg\to A\to Zh \to Z (b\bar{b})$} & ATLAS 8 TeV \cite{Aad:2015wra}& 220-1000 & 20.3 fb$^{-1}$  \\
& ATLAS 13 TeV \cite{ATLAS-CONF-2016-015} & 200-2000 & 3.2 fb$^{-1}$\\
\hline
$pp\to A b \bar b\to Zh b \bar b  \to Z (b\bar{b}) (b \bar b) $ & ATLAS 13 TeV \cite{ATLAS-CONF-2016-015} & 200-1000 & 3.2 fb$^{-1}$\\
\hline
$pp\to t H^{\pm} (H^{\pm} \to \tau^{\pm} \nu) + X $ & ATLAS 8 TeV \cite{Aad:2014kga}& 180-1000  & 19.5 fb$^{-1}$\\
\hline
\multirow{2}{*}{$pp\to t b H^{\pm} (H^{\pm} \to \tau^{\pm} \nu) $} & ATLAS 13 TeV \cite{Aaboud:2016dig}& 200-2000 & 3.2 fb$^{-1}$ \\
& CMS 8 TeV \cite{Khachatryan:2015qxa}& 200-600 & $19.7\pm0.5$ fb$^{-1}$\\
\hline
$gb \to t H^{\pm}(H^{\pm} \to tb)$  & ATLAS 8 TeV  \cite{Aad:2015typ} &  200-600  & 20.3 fb$^{-1}$\\
\hline
$q q' \to H^{\pm}(H^{\pm} \to tb) \to (l + \mathrm{jets})$  & ATLAS 8 TeV \cite{Aad:2015typ}& 400-2000 & 20.3 fb$^{-1}$ \\
\hline
$q q' \to H^{\pm}(H^{\pm} \to tb) \to \mathrm{(all\ had.)}$  & ATLAS 8 TeV \cite{Aad:2015typ}& 400-2000 & 20.3 fb$^{-1}$\\
\hline
$pp\to \bar t b H^{\pm} (H^{\pm} \to tb) $ & CMS 8 TeV \cite{Khachatryan:2015qxa} & 200-600 & $19.7\pm0.5$ fb$^{-1}$\\
\hline
\hline
\end{tabular}
\end{footnotesize}
\caption{Listing of the $ H $, $ A $ and $H^{\pm}$ searches considered in this analysis, performed by CMS and ATLAS from Run-I and pre-ICHEP (2016) Run-II LHC data.}
\label{tab:heavysearches}
\end{table}

\begin{table}
\begin{footnotesize}
\begin{tabular}{| c | c | c | c |}
\hline
\textbf{Channel} & \textbf{Experiment} & \textbf{Mass range(GeV)} &  \textbf{Luminosity} \\
\hline
$gg\to H \to ZZ(\ell \ell \nu \nu + \ell \ell \ell \ell)$ & ATLAS 13 TeV~\cite{Aaboud:2017rel} & 200-1200 & $36.1 fb^{-1}$\\
$gg\to H \to ZZ(\ell \ell \nu \nu)$ & ATLAS 13 TeV~\cite{ATLAS-CONF-2016-056} & 300-1000  & 13.3 fb$^{-1}$ \\ 
$gg\to H\to ZZ(\nu \nu qq)$ & ATLAS 13 TeV~\cite{ATLAS-CONF-2016-082} & 500-3000 & 13.2 fb$^{-1}$ \\ 
$gg/VV\to H\to ZZ(\ell \ell qq)$ & ATLAS 13 TeV~\cite{ATLAS-CONF-2016-082} & 500-3000 & 13.2 fb$^{-1}$ \\ 
$gg/VV\to H\to ZZ(4\ell)$ & ATLAS 13 TeV~\cite{ATLAS-CONF-2016-079} & 500-3000 & 14.8 fb$^{-1}$ \\ 
\hline
$gg\to H\to W^{+}W^{-}(e \nu \mu \nu)$ & ATLAS 13 TeV~\cite{Aaboud:2017gsl} & 200-4000 & $36.1 fb^{-1}$\\
$gg/VV\to H\to W^{+}W^{-}(\ell \nu \ell \nu)$ & ATLAS 13 TeV~\cite{ATLAS-CONF-2016-074} & 200-3000 & 13.2 fb$^{-1}$ \\
$gg\to H\to W^{+}W^{-}(\ell \nu qq)$ & ATLAS 13 TeV~\cite{ATLAS-CONF-2016-062} & 500-3000 & 13.2 fb$^{-1}$ \\
\hline
$gg+VV \to H\to W^{+}W^{-}(\ell \nu \ell \nu)$ & CMS 13 TeV~\cite{CMS-PAS-HIG-16-023} &   200-1000 & 2.3 fb$^{-1}$ \\
\hline
$pp \to H\to \gamma \gamma$ & ATLAS 13 TeV~\cite{Aaboud:2017yyg} & 200-2700 & 36.7 fb$^{-1}$ \\
$pp \to H\to \gamma \gamma$ & ATLAS 13 TeV~\cite{ATLAS-CONF-2016-059} & 200-2400  & 15.4 fb$^{-1}$ \\
\hline
$pp \to H\to \gamma \gamma$ & CMS 13 TeV~\cite{CMS-PAS-EXO-16-027} & 500-4000  & 12.9 fb$^{-1}$ \\
\hline
$gg/b\bar{b} \to H\to \tau^{+} \tau^{-}$ & ATLAS 13 TeV~\cite{Aaboud:2017sjh} & 200-2300 & 36.1 fb$^{-1}$ \\
$gg/b\bar{b} \to H\to \tau^{+} \tau^{-}$ & ATLAS 13 TeV~\cite{ATLAS-CONF-2016-085} & 200-1200  &  13.3 fb$^{-1}$ \\
\hline
$gg/b\bar{b} \to H/A\to \tau^{+} \tau^{-}$ & CMS 13 TeV~\cite{CMS:2017epy} & 90-3100  & 35.9 fb$^{-1}$ \\
$gg/b\bar{b} \to H/A\to \tau^{+} \tau^{-}$ & CMS 13 TeV~\cite{CMS-PAS-HIG-16-037} & 90-3200  &  12.9 fb$^{-1}$ \\
\hline
$gg/b\bar{b} \to H\to b\bar{b}$ & CMS 13 TeV~\cite{CMS-PAS-HIG-16-025} & 550-1200  &  2.7 fb$^{-1}$ \\
\hline
$pp \to H\to hh \to b\bar{b}b\bar{b}$ & ATLAS 13 TeV~\cite{ATLAS-CONF-2016-049} & 300-3000  &  13.3 fb$^{-1}$ \\
\hline
$pp \to H\to hh \to b\bar{b} b \bar{b}$ & CMS 13 TeV~\cite{CMS-PAS-HIG-17-009} & 260-1200  &  35.9 fb$^{-1}$ \\
$pp \to H\to hh \to b\bar{b} \gamma \gamma$ & CMS 13 TeV~\cite{CMS:2017ihs} & 250-900  &  35.9 fb$^{-1}$ \\
$pp \to H\to hh \to b\bar{b} \tau^{+} \tau^{-}$ & CMS 13 TeV~\cite{Sirunyan:2017djm} & 250-900  &  35.9 fb$^{-1}$ \\
$pp \to H\to hh \to b\bar{b} \tau^{+} \tau^{-}$ & CMS 13 TeV~\cite{CMS-PAS-HIG-16-029} & 250-900  &  12.9 fb$^{-1}$ \\
\hline
$gg \to A \to Zh,~h \to b\bar{b}$ & ATLAS 13 TeV ~\cite{Aaboud:2017cxo} & 200-2200 &  36.1 fb$^{-1}$ \\
$b\bar{b}A \to Zh,~h \to b\bar{b}$ & ATLAS 13 TeV ~\cite{Aaboud:2017cxo} & 200-2200 &  36.1 fb$^{-1}$ \\
\hline
$pp\to t H^{\pm} (H^{\pm} \to \tau^{\pm} \nu) + X $ & ATLAS 13 TeV~\cite{ATLAS-CONF-2016-088} & 200-2000& 14.7 fb$^{-1}$ \\
\hline
\hline
\end{tabular}
\caption{Listing of the $ H $, $ A $ and $H^{\pm}$ searches considered in this analysis. These correspond to the updated search results presented during and after ICHEP 2016 conference.}
\label{tab:heavysearchesichep}
\end{footnotesize}
\end{table}


\subsubsection{Search for heavy Higgs decaying to $WW$ and $ZZ$ final states}

We calculate the gluon-gluon fusion ($ggH$) production cross-section times the branching ratio of the 
neutral CP-even heavy Higg $H$ into $W^{+}W^{-}$ and $ZZ$, i.e. $\sigma_{ggF} \times  Br(H \rightarrow W^{+}W^{-})$, for the allowed parameter space points (blue colored points in Fig.~\ref{fig:ma_tb1}(d)) and then compare them with the upper bounds obtained on the same quantity 
by both CMS and ATLAS for 8 TeV~\cite{Khachatryan:2015cwa,Aad:2015kna,Aad:2015agg,Aad:2015kna} and 13 TeV~\cite{ATLAS-CONF-2016-021,ATLAS-CONF-2016-012,ATLAS-CONF-2016-016,ATLAS-CONF-2016-074,ATLAS-CONF-2016-062,ATLAS-CONF-2016-079,ATLAS-CONF-2016-056,ATLAS-CONF-2016-082,Aaboud:2017gsl,CMS-PAS-HIG-16-023}.

\begin{figure}[htb!]
\begin{center}
{\includegraphics[angle=0,width=0.48\textwidth]{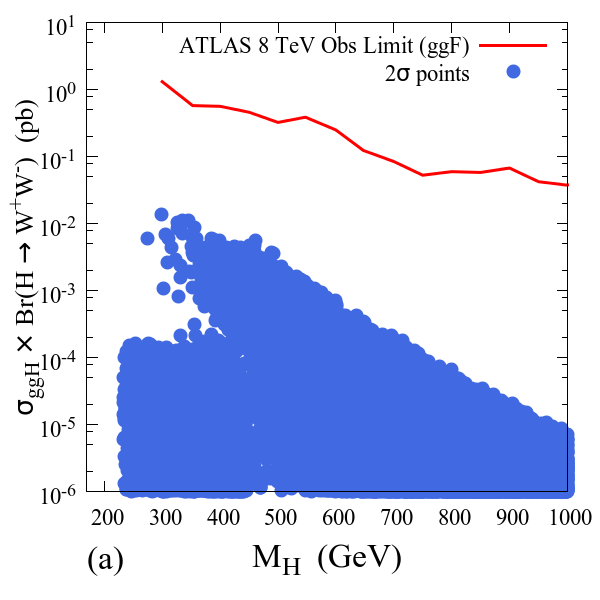}}
{\includegraphics[angle=0,width=0.48\textwidth]{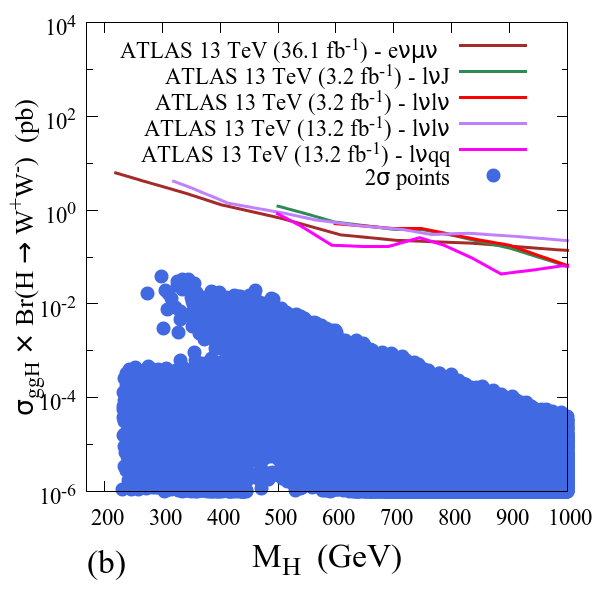}}
\caption{Scatter plot in the $M_H-[\sigma_{ggH} \times$ Br$(H \to W^{+}W^{-}$)] plane, for the allowed parameter space (blue colored points in Fig.~\ref{fig:ma_tb1}(d)). 
Fig.~\ref{figww}(a) (Fig.~\ref{figww}(b)) represents $ggH$ production cross-section times 
branching ratio for 8 (13) TeV. The red solid line in Fig.~\ref{figww}(a) denotes the 95\% C.L. upper limit derived by ATLAS for LHC 8 TeV data~\cite{Aad:2015agg}. In Fig.~\ref{figww}(b), the colored lines represent the upper limit obtained by ATLAS in various final states using 13 TeV data~\cite{ATLAS-CONF-2016-074,ATLAS-CONF-2016-062,ATLAS-CONF-2016-021,Aaboud:2017gsl}. }
\label{figww}
\end{center}
\end{figure}

\begin{figure}[htb!]
\begin{center}
{\includegraphics[angle=0,width=0.48\textwidth]{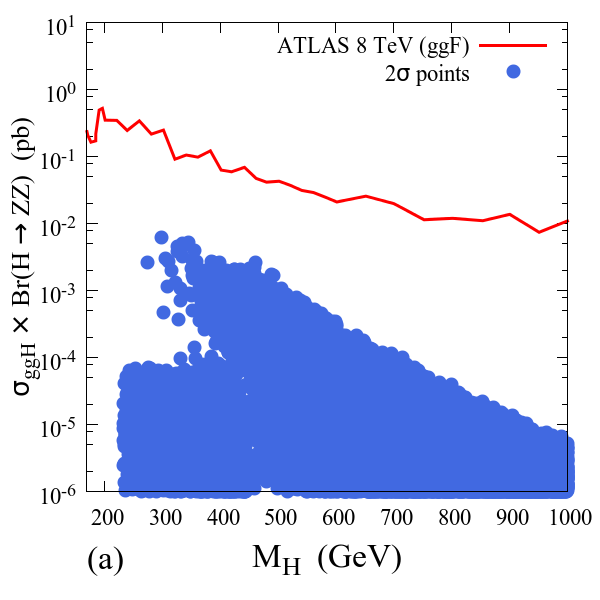}}
{\includegraphics[angle=0,width=0.48\textwidth]{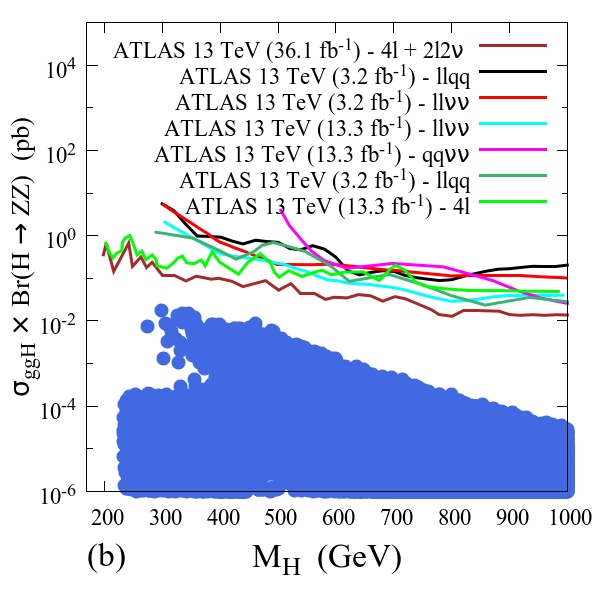}}
\caption {Scatter plot in the $M_H-[\sigma_{ggH} \times$ Br$(H \to ZZ$)] plane, for the allowed parameter space. Fig.~\ref{figzz}(a) (Fig.~\ref{figzz}(b)) represents the 8 (13) TeV gluon fusion cross-section times branching ratio. The red solid line in Fig.~\ref{figzz}(a) represents the 95\% C.L. upper limit on the $ggH$ production cross-section times branching ratio given by ATLAS at 8 TeV~\cite{Aad:2015kna}. In Fig.~\ref{figzz}(b) the colored lines represent the 13 TeV~\cite{ATLAS-CONF-2016-012,ATLAS-CONF-2016-016,ATLAS-CONF-2016-056,ATLAS-CONF-2016-082,ATLAS-CONF-2016-079,Aaboud:2017rel} ATLAS upper limit for different final states. }
\label{figzz}
\end{center}
\end{figure}

In Fig.~\ref{figww} we plot the quantity $\sigma_{ggH} \times $ Br$(H \to W^{+}W^{-})$ for all the allowed parameter space points.
The red solid  line in Fig~\ref{figww}(a) shows the 95\% C.L. upper limit on 
$\sigma_{ggH} \times $ Br$(H \to W^{+}W^{-})$ from 8 TeV ATLAS data~\cite{Aad:2015agg}. The colored
lines in Fig.~\ref{figww}(b) denote the observed limits in various final states considering $H$ to be produced through gluon fusion, from ATLAS 13 TeV data~\cite{ATLAS-CONF-2016-074,ATLAS-CONF-2016-062,ATLAS-CONF-2016-021,Aaboud:2017gsl}. 

In Fig~\ref{figzz} we plot the quantity $\sigma_{ggH} \times $ Br$(H \to ZZ)$ assuming gluon fusion for the production of the CP-even neutral heavy Higgs, for the allowed parameter space points. The red solid line in Fig.~\ref{figzz}(a) shows the 95\% C.L. 
upper limit on  $\sigma_{ggH} \times $ Br$(H \to ZZ)$ by ATLAS collaboration 
using Run-I data~\cite{Aad:2015kna}. 
The colored lines in Fig.~\ref{figzz}(b) denote the observed limits in various final states ($H$ being produced though gluon fusion mode) from 13 TeV ATLAS data~\cite{ATLAS-CONF-2016-012,ATLAS-CONF-2016-016,ATLAS-CONF-2016-056,ATLAS-CONF-2016-082,ATLAS-CONF-2016-079,Aaboud:2017rel}.

We would like to mention that the agreement between Higgs signal strengths and 
their SM expectation values pushes us to the limit where ($\beta - \alpha) \approx \pi/2$, also known as the `decoupling limit'. 
In this limit, the branching ratios of heavy Higgs boson decaying into gauge bosons are highly suppressed 
and attain values of the order of $\sim10^{-2} - 10^{-4}$.
Consequently, the quantities $\sigma_{ggH} \times \mathrm{Br}(H \rightarrow W^{+}W^{-} )$ and  
$\sigma_{ggH} \times \mathrm{Br}(H \rightarrow ZZ )$ are very small.
As a result, as can be seen from Fig~\ref{figww} and Fig~\ref{figzz}, at least three orders of magnitude improvement in the 
upper limit on cross-section measurement is required to probe the 
regions of interest in the pMSSM parameter space. 

We would like to note that the $b\bar{b}H$ production mode takes over the $ggH$ mode of Higgs production at large values of $\tan{\beta}$ because of the enhancement of the bottom Yukawa coupling. Therefore, it is expected that the parameter space points in Fig.~\ref{figww} and Fig.~\ref{figzz} would shift upwards upon inclusion of the $b\bar{b}H$ cross-sections, especially in the high $\tan{\beta}$ regime. In this context, we analyzed the effect of including the $b\bar{b}H$ cross-sections along with the $ggH$ cross-sections, and, have presented the allowed parameter space points in the $\sigma_{ggH + b\bar{b}H} \times Br(H \to WW) - M_{H}$ and $\sigma_{ggH + b\bar{b}H} \times Br(H \to ZZ) - M_{H}$ planes in Fig.~\ref{figww_add} and Fig.~\ref{figzz_add}, respectively. It was observed that inclusion of the $b\bar{b}H$ cross-sections furnish only a slight upward shift, and the current upper limits would still require an improvement of at least three orders of magnitude in order to probe the pMSSM parameter space of interest.

\subsubsection{Search for heavy Higgs decaying to $hh$ final state}

In SM, di-Higgs production cross-section is very small and production cross-section of single Higgs can be up to two orders of 
magnitude larger than the direct $hh$ production~\cite{CMS-PAS-HIG-13-032} (also see Table 1 of \cite{Bhattacherjee:2014bca}). In addition, measurement of $H\to hh$ decays can also exert non-trivial effects on the self-coupling measurement of the 125 GeV Higgs \cite{Bhattacherjee:2014bca,CMS-PAS-HIG-13-032,Adhikary:2017jtu}.
%
%
\begin{figure}[hbt]
\begin{center}
\includegraphics[width=\textwidth]{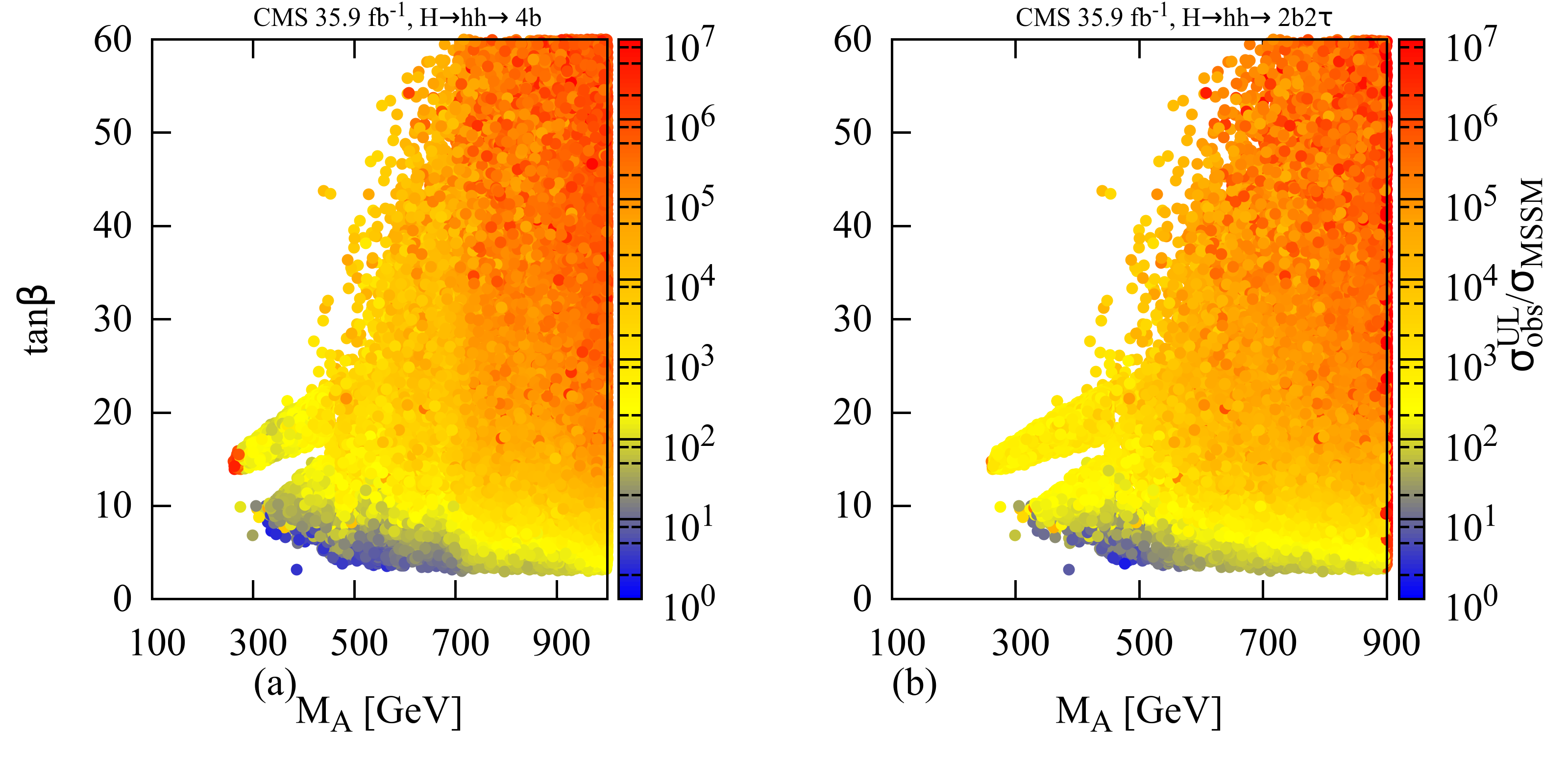}
\caption{In Fig.~\ref{H_hh}(a), we show the color palette plot of $\sigma_{Obs}^{UL}/\sigma_{MSSM}$ 
in the $M_{A} - \tan\beta$ plane. $\sigma_{Obs}^{UL}$ represents the upper limits
 on $\sigma_{H} \times \mathrm{Br}(H \rightarrow h h \rightarrow b \bar{b} b \bar{b})$ obtained by 
 CMS at 95$\%$ C.L. using LHC Run-II data with luminosity $\lum$ = 35.9 fb$^{-1}$~\cite{CMS-PAS-HIG-17-009}. Here, $\sigma_{H}$ corresponds to the production cross-section of $H$, while
$\sigma_{MSSM}$ represents the value of $\sigma_{ggH} \times \mathrm{Br}(H \rightarrow h h \rightarrow b \bar{b} b \bar{b})$ calculated for the allowed parameter space points in MSSM. In fig.~\ref{H_hh}(b), we present a similar plot with $\sigma_{Obs}^{UL}$ corresponding to the upper limits on $\sigma_{H} \times \mathrm{Br}(H \rightarrow h h \rightarrow b \bar{b} \tau^{+} \tau{-})$ obtained by 
 CMS at 95$\%$ C.L. using LHC Run-II data with luminosity $\lum$ = 35.9 fb$^{-1}$~\cite{Sirunyan:2017djm}, while $\sigma_{MSSM}$ corresponds to the value of $\sigma_{ggH} \times \mathrm{Br}(H \rightarrow h h \rightarrow b \bar{b} \tau^{+} \tau^{-})$ for the allowed parameter space points.}
\label{H_hh}
\end{center}
\end{figure}
In MSSM, branching fraction of $H \rightarrow hh$ is considerable only for low $\tan \beta$ and $M_A <$ 350 GeV, 
because at low $\tan \beta$ the branching ratio of $H \rightarrow t \bar t$ will take over as soon as the $t \bar t$ threshold is crossed.  
 The search for a BSM resonance decaying to a pair of 125 GeV Higgs boson has been 
 looked for by both CMS and ATLAS collaborations in the $b \bar b b \bar b $, $b \bar b \gamma \gamma$ and $b \bar{b} \tau^{+} \tau^{-}$  final states. We have also computed the
 production cross-section in the gluon fusion mode times branching ratios for the allowed parameter space and compared with the  
 corresponding upper limit obtained by CMS and ATLAS using  8 TeV and 13 TeV LHC data.

Fig.~\ref{H_hh}(a) represents the color palette plot of $\sigma_{Obs}^{UL}/\sigma_{MSSM}$ 
in the $M_{A} - \tan{\beta}$ plane, where $\sigma_{Obs}^{UL}$ represents the upper limits
 on $\sigma_{H} \times \mathrm{Br}(H \rightarrow h h \rightarrow b \bar{b} b \bar{b})$ derived by 
 CMS at 95$\%$ C.L. using Run-II data~\cite{CMS-PAS-HIG-17-009}, $\sigma_{H}$ corresponds 
  to the production cross-section of $H$ and $\sigma_{MSSM}$ represents the 
  value of $ \sigma_{ggH} \times Br(H \to hh \to b \bar{b} b \bar{b})$\footnote{The $ggH$ mode offers the most dominant contribution to MSSM heavy Higgs production at low $\tan\beta$ as one approaches the decoupling limit. The $H \to hh$ decay mode also attains an appreciable branching fraction in the low $\tan{\beta}$ region. Therefore, in this analysis, instead of taking a sum of all possible Higgs production channels, we consider only the gluon fusion mode of Higgs production.} 
  calculated for the allowed parameter space in MSSM. 
  A value of $\sigma_{Obs}^{UL}/\sigma_{MSSM}$ less than 1 would indicate that the current LHC data 
  have excluded the parameter space point.
  For the allowed parameter space points of our model we have $7.44~ \lesssim ~\sigma_{Obs}^{UL}/\sigma_{MSSM}~\lesssim~5.10\times 10^{9}$, showing that the existing bounds from this decay channel do not impose any constraints on our parameter space. 
  We present a similar plot for $ H \rightarrow h h \rightarrow b \bar{b} \tau^{+} \tau^{-} $ decay mode in Fig.~\ref{H_hh}(b), where we implement the upper limits obtained by CMS from Run-II data at a luminosity of $\lum$ = 35.9 fb$^{-1}$~\cite{Sirunyan:2017djm}. 
  In this case, for the allowed parameter space points we get $1.68~ \lesssim ~\sigma_{Obs}^{UL}/\sigma_{MSSM}~\lesssim~1.73\times 10^{10}$.

The future runs are expected to strengthen the existing upper limits on production cross-section times branching fraction, which would lead to a lower value of $\sigma_{Obs}^{UL}$. As a result, the ratio $\sigma_{Obs}^{UL}/\sigma_{MSSM}$ would get lowered  in both Fig.~\ref{H_hh}(a) and Fig.~\ref{H_hh}(b) and may be able to constrain the parameter space.
We expect the $ H \rightarrow hh$ decay mode to play a significant role in 
the search of heavy Higgs boson in low $\tan{\beta}$ and low $M_{A}$ region in high luminosity LHC (HL-LHC). 
It has been shown that the HL-LHC ($\sim$ 3000 fb$^{-1}$) will be able to probe the region $M_A \lesssim 500$ GeV for $\tan \beta \lesssim 8$~\cite{Bhattacherjee:2015sga}.

\subsubsection{Search for heavy Higgs decaying to $\gamma \gamma$  final states}

The diphoton invariant mass distribution has played a major role in the discovery of the 
125 GeV Higgs. But the branching fraction of this channel decreases as the Higgs mass 
increases. 
We have implemented the bounds on the fiducial cross-section times 
branching ratio for this channel from CMS and ATLAS for both 8 TeV and 13 TeV data on the allowed parameter space.

\begin{figure}[htb!]
\begin{center}
{\includegraphics[angle=0,width=0.48\textwidth]{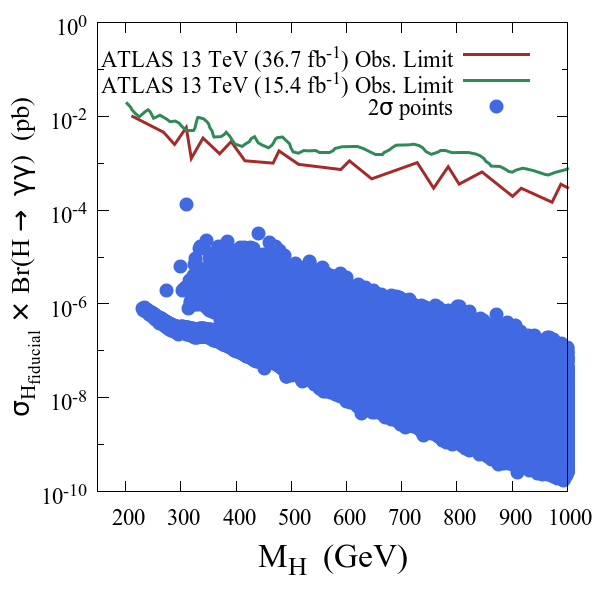}\includegraphics[angle=0,width=0.48\textwidth]{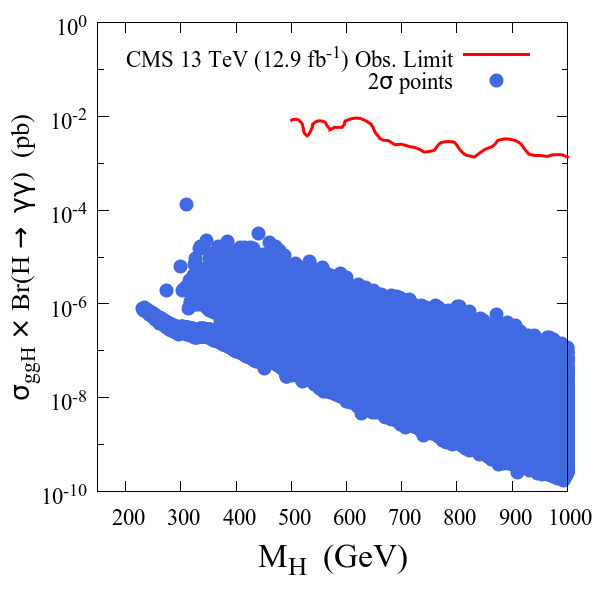}}
\caption{(a): Scatter plot showing $\sigma_{ggH} \times Br(H \to \gamma\gamma)$ for the allowed parameter space points. The green and brown solid lines represent 95\% C.L. upper limit on the product of fiducial cross-section of the heavy Higgs times its branching fraction into diphoton final state, derived by ATLAS using the 13 TeV \cite{ATLAS-CONF-2016-059,Aaboud:2017yyg} data. (b): Scatter plot in the $M_H-[\sigma_{ggH} \times$ Br$(H \to \gamma \gamma$)] plane for the allowed parameter 
space points. The red solid line represent the 95\% C.L. upper limit on the same quantity, derived by CMS using the 8 TeV + 13 TeV combined \cite{CMS-PAS-EXO-16-027} dataset.}
\label{figgamgam}
\end{center}
\end{figure}

In Fig.~\ref{figgamgam}, we present the $ggH$ production cross-section times $\mathrm{Br}(H \to \gamma \gamma)$ 
as a function of mass of the scalar resonance, for all allowed parameter space points. 
The green and brown solid curves in Fig.~\ref{figgamgam} (a) are the upper limit on the fiducial cross-section times $Br(H \to \gamma\gamma)$ from ATLAS 13 TeV results~\cite{ATLAS-CONF-2016-059,Aaboud:2017yyg}. In Fig.~\ref{figgamgam} (b), the red colored line is the upper limit on the inclusive Higgs production cross-section times  $Br(H \to \gamma\gamma)$
from the combined analysis of 8 TeV and 13 TeV data by CMS~\cite{CMS-PAS-EXO-16-027}. In this analysis, we compared the limits on fiducial cross-section against the total gluon fusion cross-section of the CP-even neutral heavy Higgs boson computed for our parameter space points. The fiducial cross-section is expected to be lesser than the total cross-section, and hence, for our parameter space points, the product of the fiducial cross-section times $Br(H \to \gamma\gamma)$, will attain a value lesser than that shown in Fig.~\ref{figgamgam}(a), rendering the upper limit more weaker. As a result, in the context of the current analysis, we have not evaluated the fiducial cross-sections to draw up the comparisons, as it would yield weaker results, and hence, will not exert any additional impact on our parameter space. At large values of $\tan\beta$, the $b\bar{b}H$ production mode overtakes the $ggH$ mode. Therefore, we also analyzed the effect of adding the $b\bar{b}H+ggH$ production cross-sections and the resulting upward shift was not very significant. The current upper limits would require an improvement of more than two orders of magnitude in order to be able to probe 
the region of our parameter space.

\subsubsection{Search for heavy Higgs decaying to $t \bar t$ final state}

\begin{figure}[htb]
\begin{center}
{\includegraphics[width=0.49\textwidth]{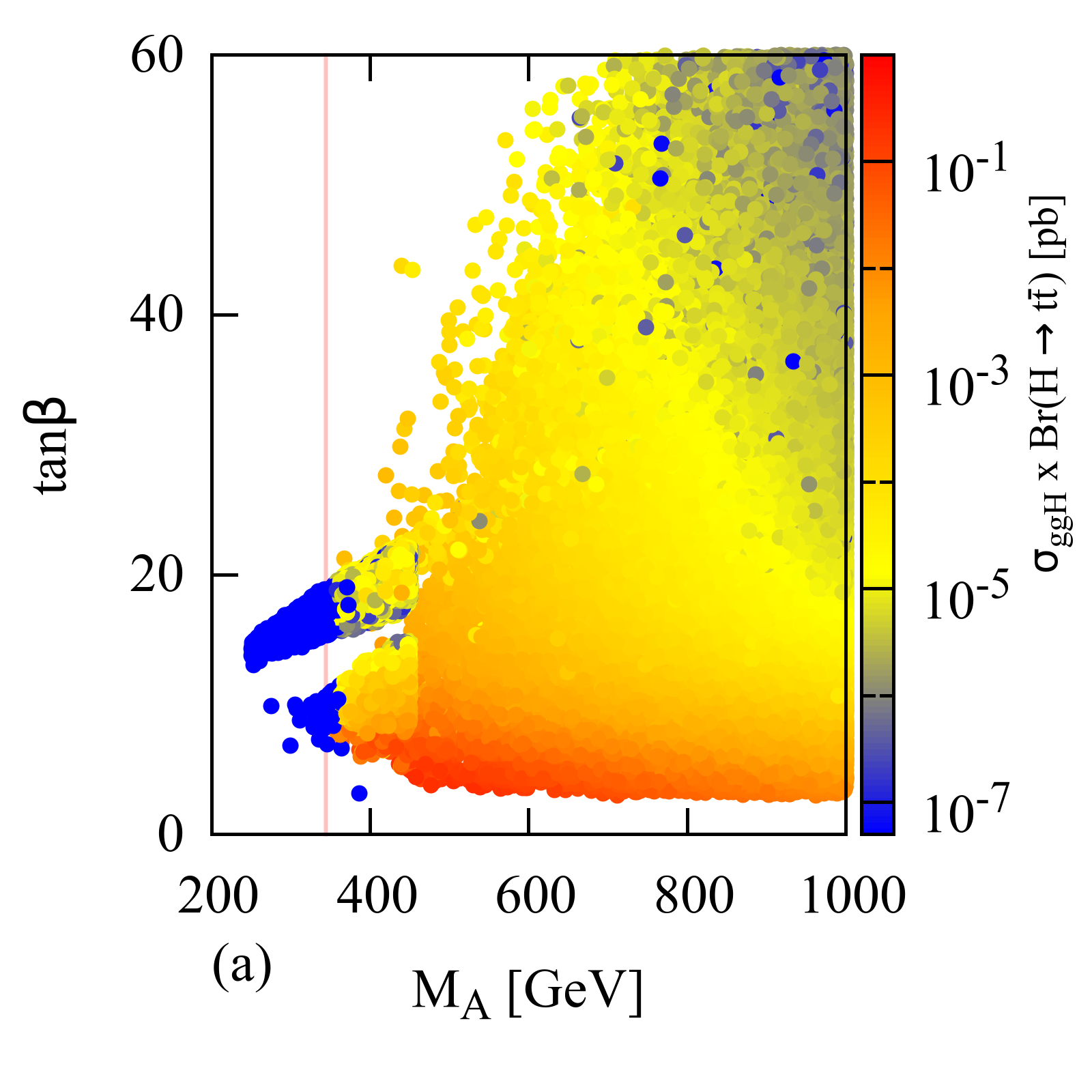}}
{\includegraphics[width=0.49\textwidth]{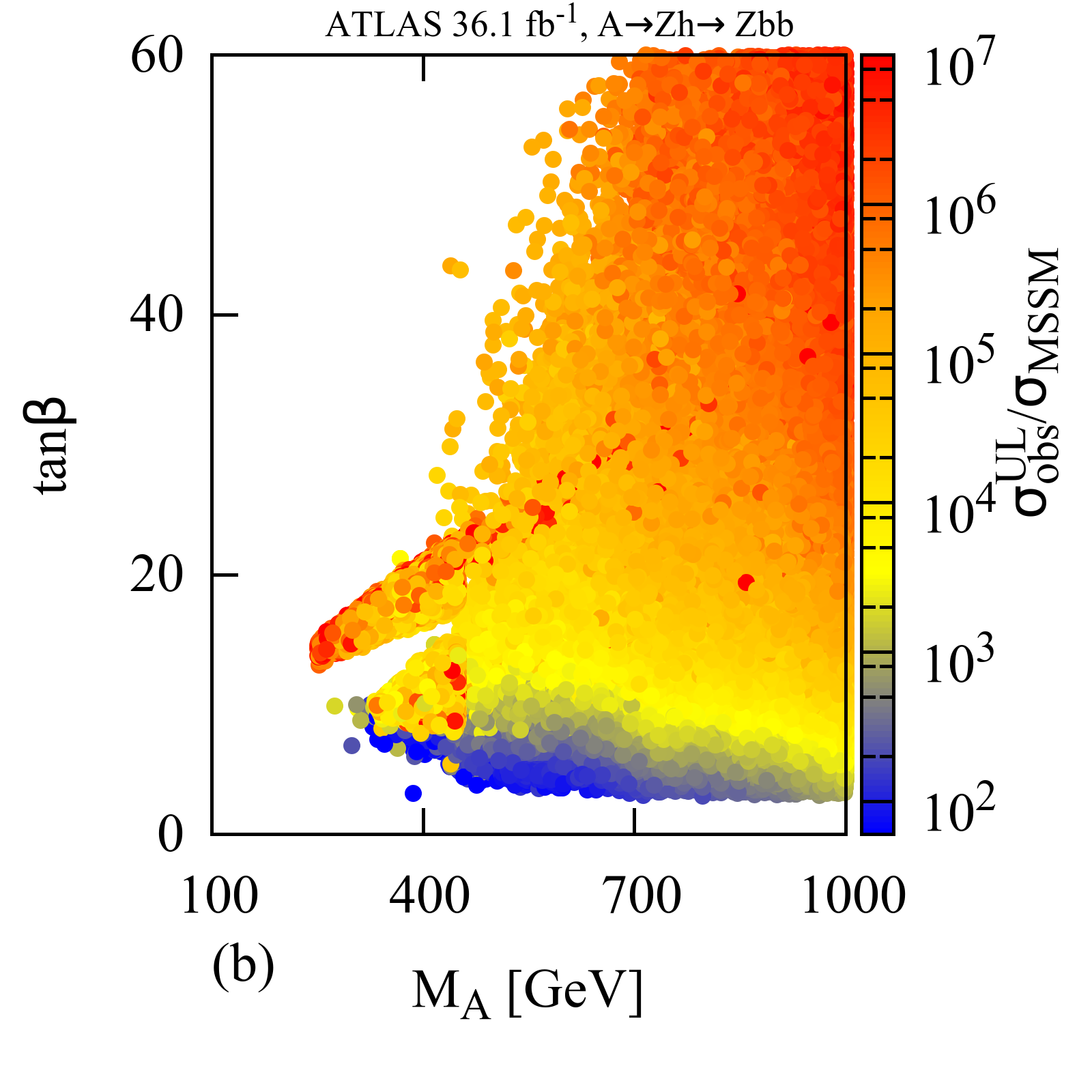}}
\caption{ 
(a): Scatter plot of $\sigma_{ggH} \times \mathrm{Br}(H \to t \bar t)$ in the 
$M_A - \tan \beta$ plane.  The magenta colored vertical line corresponds to $ 2 m_{t}$.
Fig.~\ref{A_zh_sigma_ratio}(b): Color palette plot of $\sigma_{Obs}^{UL}/\sigma_{MSSM}$ 
in the $M_{A} - \tan{\beta}$ plane. 
$\sigma_{Obs}^{UL}$ represents the upper limits on  $\sigma_{ggA} \times \mathrm{Br}(A \rightarrow Z h \rightarrow Z b \bar{b})$ derived by ATLAS at 95$\%$ C.L., using LHC Run-II data \cite{Aaboud:2017cxo}. Here, $\sigma_{ggA}$ corresponds to the production cross-section of the psedoscalar Higgs ($A$) through gluon fusion. 
 $\sigma_{MSSM}$ represents the corresponding values calculated for the allowed parameter space in MSSM.}
\label{A_zh_sigma_ratio}
\end{center}
\end{figure}

Branching ratio of heavy Higgs decaying into $t\bar{t}$ pair becomes large only in low $\tan \beta$ regime. 
At the same time, at low $\tan \beta$, the production cross- section of heavy Higgs through gluon-fusion also 
becomes large due to enhanced heavy Higgs to top Yukawa coupling. Therefore, near the $t \bar t$ kinematic 
threshold and in low $\tan \beta$ region, the production times branching ratio for this process is high and one 
would expect to see a resonance of heavy Higgs in the $t \bar t$ invariant mass distribution. 
But the $t \bar t$ invariant mass distribution is expected to be largely dominated by a continuous SM $t \bar t$ 
background and it might pose severe challenges, even for HL-LHC, to probe the parameter space of our interest through 
this channel~\cite{Bhattacherjee:2015sga}. Moreover, the interference effect between $t \bar t$ production 
through heavy Higgs mediation and SM-QCD production of $t \bar t$ plays an important role. The interference 
term in general distorts the invariant mass distribution of the $t \bar t$ pair and produces non-trivial peak-dip structure instead 
of a smooth Breit-Wigner resonance shape~\cite{Hespel:2016qaf,Carena:2016npr}. 
Projected sensitivities of future LHC to probe this channel have been discussed in Refs.~\cite{Hespel:2016qaf} and~\cite{Carena:2016npr}. 
In Fig~\ref{A_zh_sigma_ratio}(a) we show a scatter plot of cross-section times branching ratio for the process $gg \to H \to t \bar t$ 
through a color palette in the $M_A-\tan \beta$ plane. The pink colored vertical line corresponds 
to $2 m_{t}$ and represents the kinematic threshold for the on-shell 
decay of $H$ to a $t\bar{t}$ pair. The dark red points in 
Fig~\ref{A_zh_sigma_ratio}(a) correspond to 
$\sigma_{ggH} \times \mathrm{Br}(H \to t \bar t) \approx 0.5 - 1$ pb. 
An improvement in the background reduction is required to explore this channel further.

\subsubsection{Search for pseudoscalar heavy Higgs decaying to $Zh$ final state}

The pseudoscalar Higgs ($A$) can decay into a final state containing 125 GeV Higgs and $Z$ boson, when mass 
of the pseudoscalar Higgs ($M_A$) is greater than ($M_h+M_Z$). One should note that the branching ratio of this 
channel is considerable only with low values of $\tan \beta$ $(<10)$ and low $M_A$ $(<350$ GeV), 
because for low $\tan \beta$ and $M_A > 2m_{t}$, $A \rightarrow t \bar t$ is the dominant decay mode. 
CMS has searched for this resonance decay with two opposite sign leptons (from $Z$) and $b \bar b $ (from the decay of $h$) in 
the final state~\cite{Khachatryan:2015tha,Khachatryan:2015lba}. 
ATLAS has also given their exclusion limits for this process with both $b \bar b$ and $\tau^{+} \tau^{-}$ in the final state from $h$ 
decay~\cite{Aad:2015wra,Aad:2015wra,ATLAS-CONF-2016-015,Aaboud:2017cxo}. 
We computed the gluon fusion cross-section times branching ratio of this channel for the allowed parameter space points  and compared
with the corresponding  8 TeV and 13 TeV upper limits obtained by ATLAS and CMS.

We present the color palette plot for $\sigma_{Obs}^{UL}/\sigma_{MSSM}$ in 
the $M_{A} - \tan{\beta}$ plane in Fig.~\ref{A_zh_sigma_ratio}(b) where 
$\sigma_{Obs}^{UL}$ represents the upper limits on $\sigma_{ggA} \times \mathrm{Br}(A \rightarrow Z h \rightarrow Z b \bar{b})$ derived by ATLAS at 95$\%$ C.L. using LHC Run-II data~\cite{Aaboud:2017cxo}, with  $\lum=36.1 $ fb$^{-1}$.  
$\sigma_{MSSM}$ represents corresponding values calculated for our model. 
All of our allowed parameter space points have the value $\sigma_{Obs}^{UL}/\sigma_{MSSM} \gg 1$, and are in the range $79.4~ \lesssim ~\sigma_{Obs}^{UL}/\sigma_{MSSM}~\lesssim~6.66\times 10^{10}$. This indicates that our parameter space points elude 
the existing bounds from this decay channel. A $\sim 10^{2}$ times improvement 
in the measurement of these decay modes would enable this search channel to probe the psedudoscalar Higgs boson $A$ in the low $M_{A}$ and low $\tan{\beta}$ region.

%

\subsubsection{Search for heavy Higgs with $b\bar{b}$ and $\tau^{+} \tau^{-}$ final states}

The couplings of MSSM heavy Higgs $H$ and pseudoscalar Higgs $A$ with down 
type fermions ($b$-quark and $\tau$-lepton) are proportional to 
$\cos \alpha/ \cos \beta$ and $\tan \beta$  respectively. Therefore, 
for a fixed CP-even Higgs mixing angle $\alpha$, both the couplings 
increase with $\tan \beta$. So the production cross-section of 
heavy Higgs boson in association with $b$-quark is enhanced in the 
high $\tan\beta$ regime. For large values of $\tan \beta$ ($\geq$ 10), 
the dominant decay modes of both $H$ and $A$ are through $b \bar b$ ($\sim$ 90\%) 
and $\tau^{+} \tau^{-}$ ($\sim$ 10 \%) channels. 
CMS and ATLAS have presented their results on both $\tau^{+} \tau^{-}$ 
\cite{Aad:2014vgg,CMS-PAS-HIG-14-029,ATLAS-CONF-2015-061,Aaboud:2017sjh,CMS-PAS-HIG-16-006,Aad:2014vgg,ATLAS-CONF-2016-085,CMS-PAS-HIG-16-037,CMS:2017epy} and $b \bar b ~$\cite{CMS-PAS-HIG-16-025,Khachatryan:2015tra} final states of heavy Higgs boson decay, for production through gluon-fusion and in association with $b$-quark, with 8 TeV and 13 TeV LHC data. 
\begin{figure}[!tb]
\begin{center}
{\includegraphics[width=0.48\textwidth]{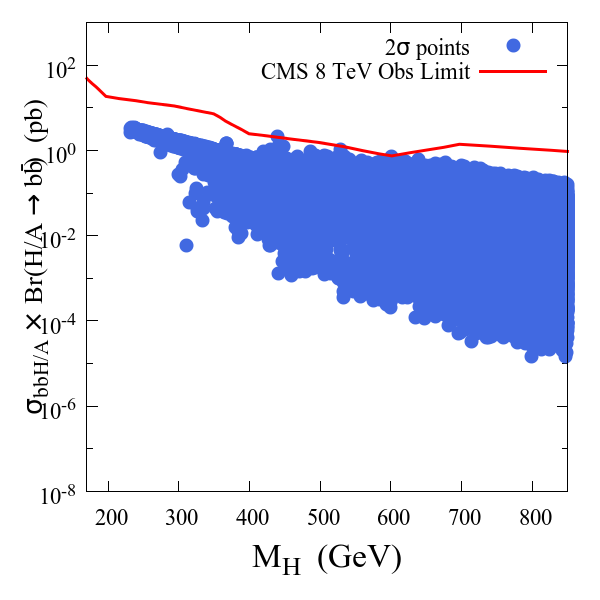}}
\caption{Scatter plot of $\sigma_{bbH/A} \times \mathrm{Br}(H/A \rightarrow b \bar{b})- M_A$ plane, for the allowed parameter space points. The red colored line represents the upper limit on $\sigma_{bbH} \times \mathrm{Br}(H \rightarrow b \bar{b})$ derived by CMS~\cite{Khachatryan:2015tra} from 8 TeV LHC data.  }
\label{figbb}
\end{center}
\end{figure}

\begin{figure}[!tb]
\begin{center}
{\includegraphics[angle=0,width=0.49\textwidth]{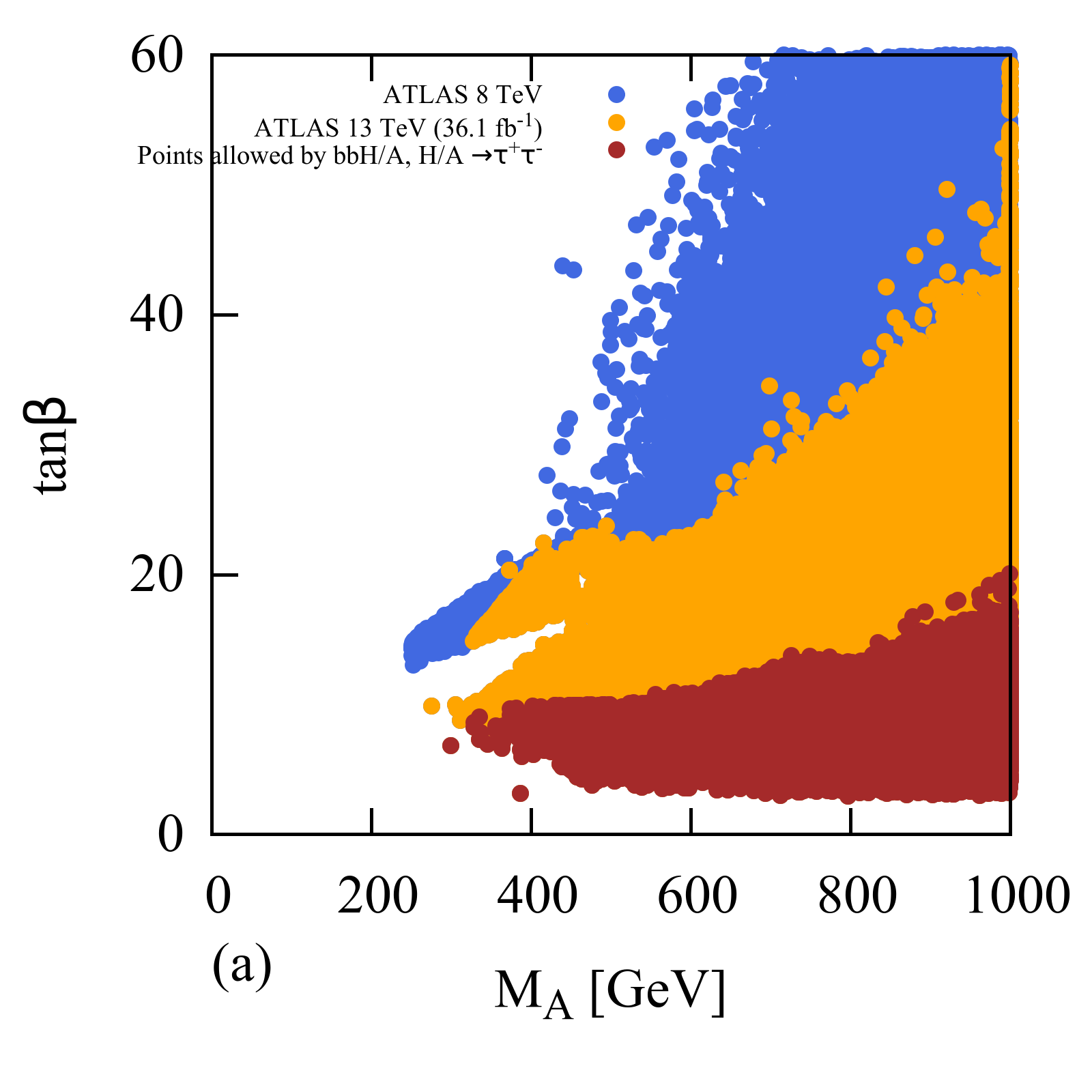}}{\includegraphics[angle=0,width=0.49\textwidth]{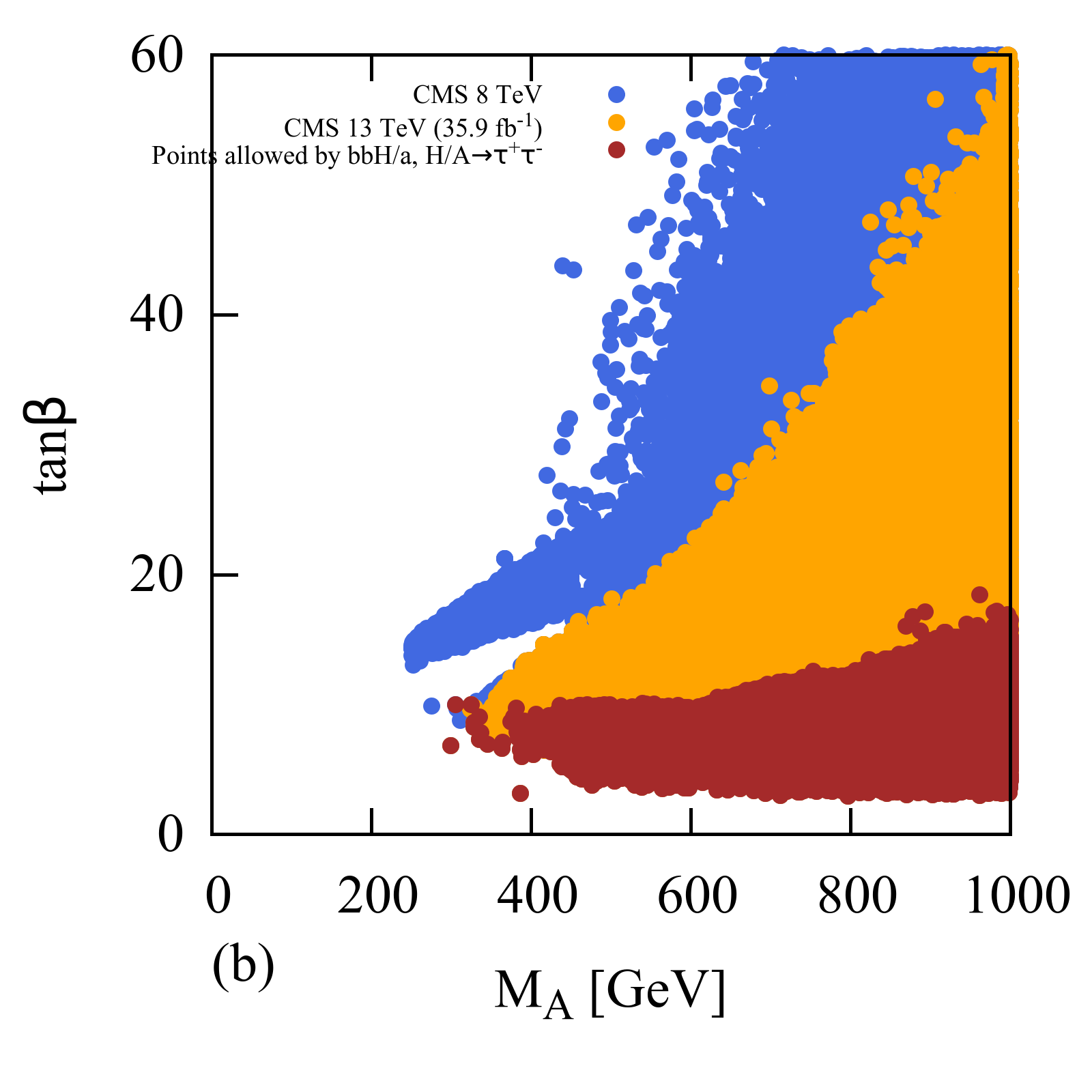}} \\
{\includegraphics[angle=0,width=0.60\textwidth]{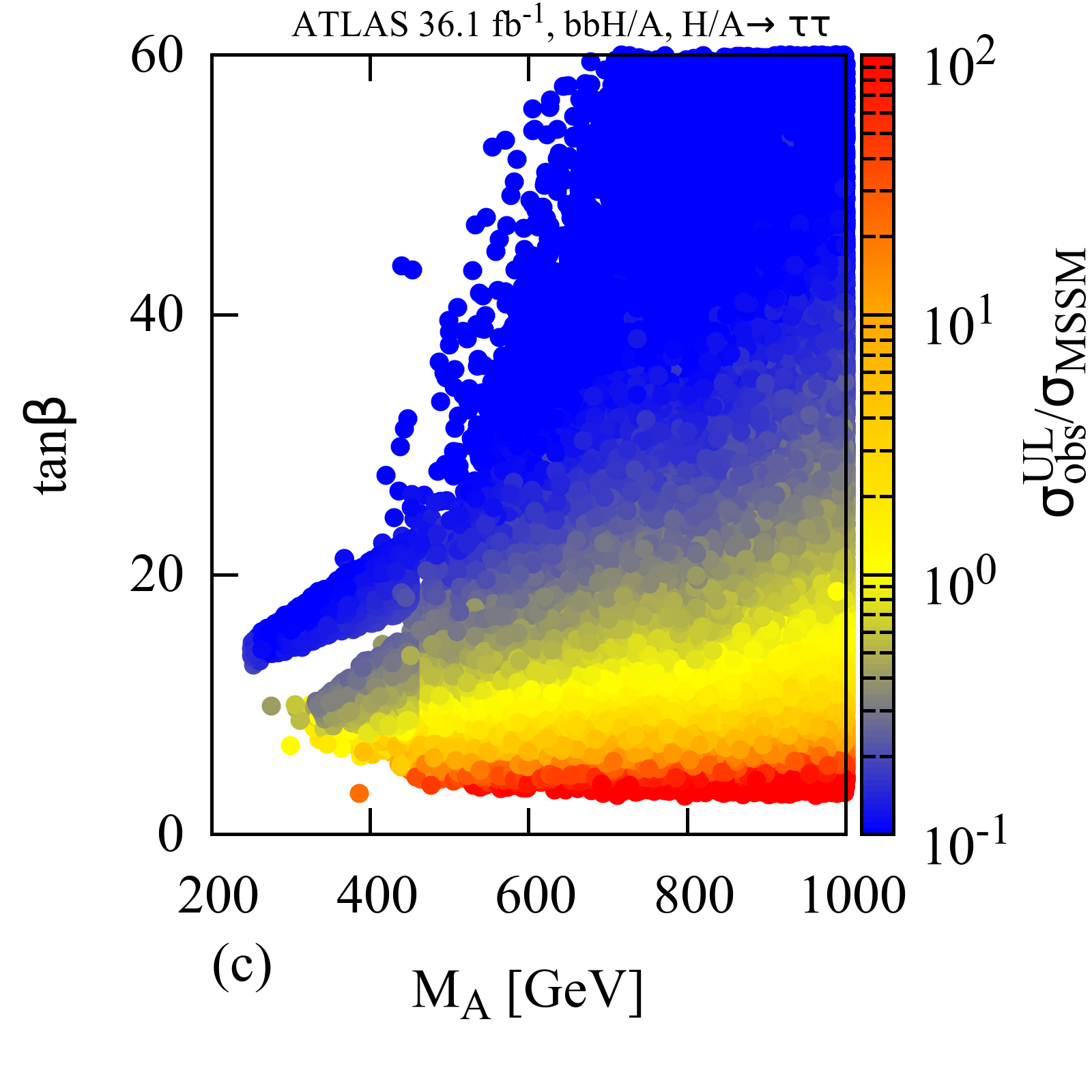}}
\caption{(a): Scatter plot in the $M_A - \tan \beta$ plane 
assuming associated production of $H/A$ with $b$ quarks. Brown colored points represent those allowed parameter space points which are also allowed by the most updated Heavy Higgs search results from ATLAS corresponding to $H/A \to \tau^{+}\tau^{-}$ channel. The blue colored points correspond to those allowed parameter space points which are excluded by ATLAS 8 TeV data in the direct search of $H/A \to \tau^{+} \tau^{-}$ at 95\% C.L~\cite{Aad:2014vgg}. The orange colored points are excluded by ATLAS 13 TeV data corresponding to $36.1~$fb$^{-1}$~\cite{Aaboud:2017sjh} of integrated luminosity, in the same channel. Fig.~\ref{figtau}(b): Blue colored points are excluded by the CMS 8 TeV data in the direct search of $H/A \to \tau^{+} \tau^{-}$ at 95\% C.L~\cite{CMS-PAS-HIG-14-029}, while the orange colored points are excluded by the CMS 13 TeV data corresponding to $35.9~$fb$^{-1}$~\cite{CMS:2017epy}. Fig.~\ref{figtau}(c): Color palette plot in the $M_A - \tan \beta$ plane with $\sigma^{UL}_{Obs}/\sigma_{MSSM}$ as the third axis. Here, $\sigma^{UL}_{Obs}$ corresponds to the upper limit on $\sigma_{bbH/A}\times Br(H/A \to \tau^{+} \tau^{-})$ derived by ATLAS using 13 TeV LHC data at an integrated luminosity of $36.1~$fb$^{-1}$~\cite{Aaboud:2017sjh}.}
\label{figtau}
\end{center}
\end{figure}

We have calculated the $b\bar{b}H/A$ cross-section times branching ratio for both $H/A \to b \bar{b}$ and $H/A \to \tau^{+} \tau^{-}$ final 
states for the allowed parameter space and the results are shown in Fig.~\ref{figbb} and Fig.~\ref{figtau}.
The blue points in Fig.~\ref{figbb} correspond to $\sigma_{b\bar{b}H/A} \times \mathrm{Br}(H/A \rightarrow b \bar{b})$ for 
the allowed parameter space points, while the red line shows the limit obtained by CMS at 8 TeV~\cite{Khachatryan:2015tra}. 
We observe that the upper limit derived by CMS glances over the parameter space at $M_{A} \approx 600~{\rm GeV}$.

In Fig.~\ref{figtau}(a) and Fig.~\ref{figtau}(b) we show the impact of direct search in $H/{A} \to \tau^{+}\tau^{-}$  
channel in the $M_A - \tan \beta$ plane, on the allowed parameter space points.
The blue colored points correspond to the fraction of allowed parameter space points which gets disallowed upon the application of 8 TeV ATLAS (Fig.~\ref{figtau}(a))~\cite{Aad:2014vgg} and CMS (Fig.~\ref{figtau}(b))~\cite{CMS-PAS-HIG-14-029} upper limits, and we are eventually left with the orange colored points. Finally, implementation of the 13 TeV constraints from ATLAS ($36.1~$fb$^{-1}$~\cite{Aaboud:2017sjh}) and CMS ($35.9~$fb$^{-1}$~\cite{CMS:2017epy}) leave us with the brown colored points in Fig.~\ref{figtau}(a) and Fig.~\ref{figtau}(b), respectively.

%
From Fig.~\ref{figtau} it is evident that $H/A \to \tau^{+} \tau^{-}$ channel puts the strongest constraint on our choice of MSSM parameter space among all  possible decay channels of heavy Higgs bosons. 
%
%
As both production and decay get enhanced with $\tan \beta$, the LHC constraint becomes more stringent for low to moderately large values of $M_A$. We would also like to mention that there is no significant difference between the latest 13 TeV limits from ATLAS (36.1~fb$^{-1}$) and CMS (35.9~fb$^{-1}$), and roughly speaking, both exclude the same parameter region as is evident from Fig.~\ref{figtau}(a) and Fig.~\ref{figtau}(b).

In Fig.~\ref{figtau}(c) we plot the ratio between the observed upper limit (by ATLAS at 13 TeV~\cite{Aaboud:2017sjh}) and the value of $b\bar{b}H/A$ cross-section times branching ratio of the heavy Higgs decaying to $\tau^{+} \tau^{-}$ channel computed in MSSM, in the $M_A - \tan \beta$ plane as a color palette. Points for which the quantity $\sigma^{UL}_{Obs}/\sigma_{MSSM} < 1$ are excluded by this search. In this analysis, the allowed parameter space points have the value of this ratio in the range $0.03~ \lesssim ~\sigma_{Obs}^{UL}/\sigma_{MSSM}~\lesssim~1.0\times 10^{3}$, which indicates that this channel is highly promising in order to probe the low and moderate $\tan\beta$ regions, which will be feasible in the future HL-LHC. We would also like to mention that a $10$ times ($30$ times) improvement in the sensitivity of this channel will be able to exclude parameter space points with $\tan\beta \gtrsim 13 $ ($\tan\beta \gtrsim 6 $) for $M_{A} \sim 1~{\rm TeV}$.

\subsubsection{Search for $H^{\pm}$ with $\tau \nu$ final states}

The observation of a charged Higgs boson will be a clear hint of BSM physics. 
The bounds on charged Higgs masses from the direct searches at the 
Large Electron-Positron (LEP) collider is $M_H^{\pm} > 78.6$ GeV~\cite{Abbiendi:2013hk}. 
The production and decay of the charged Higgs mostly depend on the 
charged Higgs mass. If the charged Higgs is lighter than the top-quark, 
i.e. $M_H^{\pm}< (m_t - m_b)$, then the charged Higgs is produced mostly 
from $t \bar t $ process and  decays to $\tau \nu$ final state. A light 
enough charged Higgs can also decay into a $c \bar s$ final state. 
Both ATLAS and CMS have looked for the charged Higgs state decaying 
into $\tau \nu$~\cite{Aad:2014kga,Aaboud:2016dig,Khachatryan:2015qxa,ATLAS-CONF-2016-088} and $ t \bar{b}$~\cite{Khachatryan:2015qxa,ATLAS-CONF-2016-089,Aad:2015typ} final states. In a large fraction of our 
parameter space $M_H^{\pm}> (m_t - m_b)$ and therefore $t \rightarrow H^{\pm} b$
 channel is forbidden. Thus charged Higgs will mostly be produced 
 through the associated production with top and bottom quarks. 
 We apply the bounds on the MSSM parameter space from both CMS and ATLAS 
 coming from 8 TeV and 13 TeV data.

\begin{figure}[!tb]
\begin{center}
{\includegraphics[angle=0,width=0.48\textwidth]{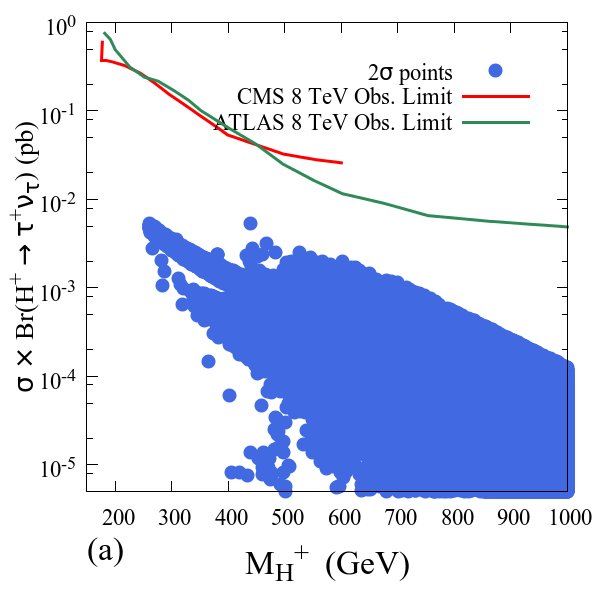}}
{\includegraphics[angle=0,width=0.48\textwidth]{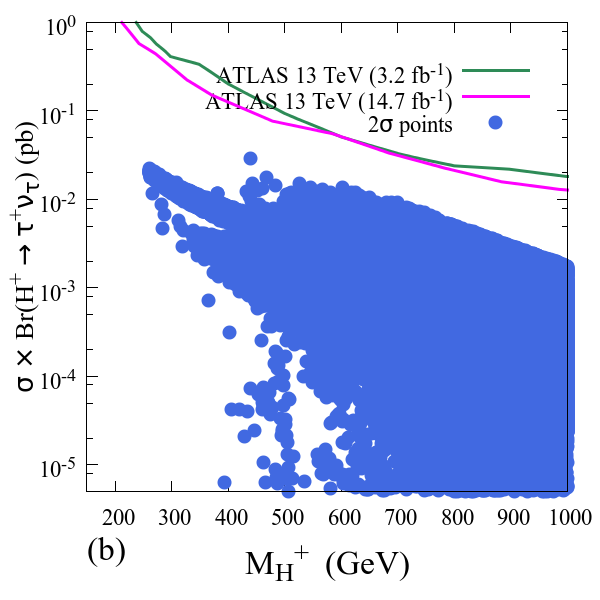}}
\caption{Scatter plot in the $M_H-[\sigma \times$ Br$(H^{\pm} \to \tau \nu$)] plane 
for the allowed parameter space for 8 TeV (Fig.~\ref{figtaunu}(a)) and  13 TeV (Fig.~\ref{figtaunu}(b)). 
The green dashed (red solid) line in Fig.~\ref{figtaunu}(a) denotes the 95\% C.L. upper limit on $\sigma \times$ Br$(H^{\pm} \to \tau \nu)$   given by ATLAS~\cite{Aad:2014kga} (CMS~\cite{Khachatryan:2015qxa}) at 8 TeV. The green dashed (purple dashed) points in Fig.~\ref{figtaunu}(b) denotes the pre-ICHEP 2016~\cite{Aaboud:2016dig} (post-ICHEP 2016~\cite{ATLAS-CONF-2016-088}) upper limits at 13 TeV.}
\label{figtaunu}
\end{center}
\end{figure}

We have calculated the cross-section times branching ratio for the allowed parameter space points
for the channel $pp \to H^{\pm} t b \to (\tau \nu) t b$. Fig.~\ref{figtaunu}(a) and Fig.~\ref{figtaunu}(b) 
show $\sigma \times$ Br$(H^{\pm} \to \tau \nu)$ for the allowed parameter space points, for 8 TeV and 13 TeV respectively.
In Fig.~\ref{figtaunu}(a) the red solid and green dashed lines 
represent the upper limits on $\sigma \times$ Br$(H^{\pm} \to \tau \nu)$ at 95\% C.L. for 8 TeV data from CMS~\cite{Khachatryan:2015qxa} and ATLAS~\cite{Aad:2014kga} respectively.  
The green dashed line in Fig.~\ref{figtaunu}(b) represents the corresponding  upper limit 
from ATLAS 13 TeV data at an integrated luminosity of 3.2$~$fb$^{-1}$~\cite{Aaboud:2016dig}, while the purple dashed line corresponds 
to a higher integrated luminosity of 14.7$~$fb$^{-1}$~\cite{ATLAS-CONF-2016-088}.  From the 13 TeV results it is clear that at least an order of magnitude improvement in these cross-section limits might make this channel sensitive enough to probe certain fractions of the allowed parameter space.

\subsubsection{Search for $H^{\pm}$ with $t \bar b$ final states}

We have already mentioned that when the charged Higgs is heavy i.e. $M_H^{\pm}> (m_t + m_b)$, it is primarily produced in the process 
 $gg \rightarrow tbH^{\pm}$ and its major decay channel would be $H^{\pm} \rightarrow t \bar b$. Both ATLAS and CMS have looked for this 
 channel and have derived upper limits on  production cross-section times branching ratio for the process $gg \rightarrow tbH^{\pm}$ and 
 $H^{\pm} \rightarrow t \bar b$~\cite{Aad:2015typ,Khachatryan:2015qxa,ATLAS-CONF-2016-089}.

\begin{figure}[!tb]
\begin{center}
{\includegraphics[angle=0,width=0.48\textwidth]{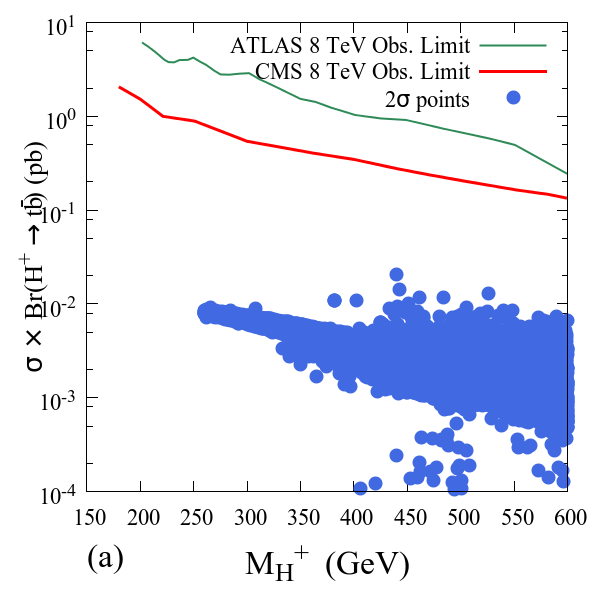}}
{\includegraphics[angle=0,width=0.48\textwidth]{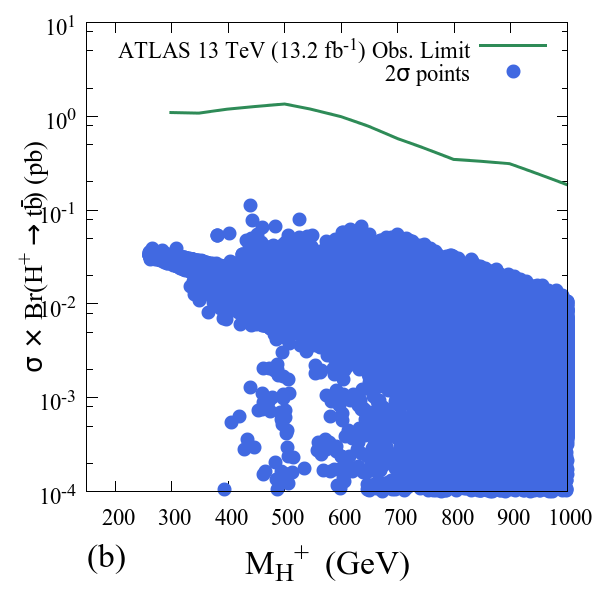}}
\caption{Scatter plot in the $M_H-[\sigma \times$ Br$(H^{\pm} \to t \bar b$)] plane of the allowed parameter space points for 8 TeV (Fig.~\ref{figtb}(a)) and 13 TeV (Fig.~\ref{figtb}(b)).  
The green dashed (red solid) line in Fig.~\ref{figtb}(a) denotes the 95\% C.L. upper limit on $\sigma \times$ Br$(H^{\pm} \to t \bar b)$ from ATLAS~\cite{Aad:2015typ} (CMS~\cite{Khachatryan:2015qxa}) 8 TeV data. Green dashed line in Fig.~\ref{figtb}(b) denotes the 95\% C.L. upper limit derived by ATLAS 13 TeV with 13.2 fb$^{-1}$ data~\cite{ATLAS-CONF-2016-089}.}
\label{figtb}
\end{center}
\end{figure}

We show the quantity $\sigma \times$ Br$(H^{\pm} \to t \bar b$) as a function of the mass of the 
charged Higgs in Fig.~\ref{figtb}(a) and Fig.~\ref{figtb}(b).
We superimpose the upper limits on $\sigma \times$ Br$(H^{\pm} \to t \bar b$) derived by both ATLAS~\cite{Aad:2015typ} and CMS~\cite{Khachatryan:2015qxa} at 8 TeV in Fig.~\ref{figtb}(a), as represented by green dashed and red solid lines respectively. 
In Fig.~\ref{figtb}(b), we superimpose the upper limits on $\sigma \times$ Br$(H^{\pm} \to t \bar b$) from ATLAS 13 TeV data 
with 13.2 fb$^{-1}$ data~\cite{ATLAS-CONF-2016-089}.
We can see from the figures that the CMS bound is much stronger\footnote{The limits on $\sigma \times Br(H^+ \to \tau^+ \nu_{\tau})$ 
 obtained by CMS collaboration are much stronger than the limits 
 derived by ATLAS collaboration at 8 TeV. The possible reasons are: 
 a) The jet substructure technique adopted by ATLAS to reconstruct 
 the top quark is highly efficient in the highly boosted region and 
 reconstruction efficiency falls in the low $M_{H^{+}}$ region, 
 b) In the hadronic tau analysis, CMS has proposed an observable $R_{\tau}$, 
 which is sensitive to different $\tau$ polarizations and is quite efficient 
 in suppressing the background events with $W \to \tau \nu_{\tau}$ and hence 
 improves signal by background ratio.}, 
  although all the allowed parameter points 
 of our scan are well below the reach of these bounds.
 We observe that an improvement of at least one order of magnitude in the cross-section 
 measurement is required to probe the allowed parameter space through the direct charged Higgs search in this channel.

\section{Favored parameter space after 13 TeV LHC data}
\label{sec:III}

In previous sections we have discussed the constraints on the MSSM parameter space in detail. We have explored the impact of light Higgs mass constraint and a  global $\chi^{2}$ analysis is performed by combining Higgs signal strength constraints and flavor physics observables. We have also analysed the impact of various heavy Higgs searches and studied their implications on the allowed parameter space. We have considered both charged and neutral Higgses and imposed an upper limit on cross-section times branching ratio in different possible decay channels, using both 8 TeV and 13 TeV data published by both ATLAS and CMS collaborations. We have observed that of all possible decay channels of the heavy Higgs bosons, $H/A \to \tau^{+} \tau^{-}$ imposes the strongest constraint and we showed its effect on the allowed parameter space, in the $M_A - \tan \beta$ plane. 

Now that we have discussed all these constraints on the MSSM Higgs sector individually, it would be comprehensive and conclusive, if we discuss the status of the parameter space and the correlation between various signal strengths when all these constraints are put together.

\begin{figure}[!htb]
\begin{center}
{\includegraphics[angle=0,width=0.58\textwidth]{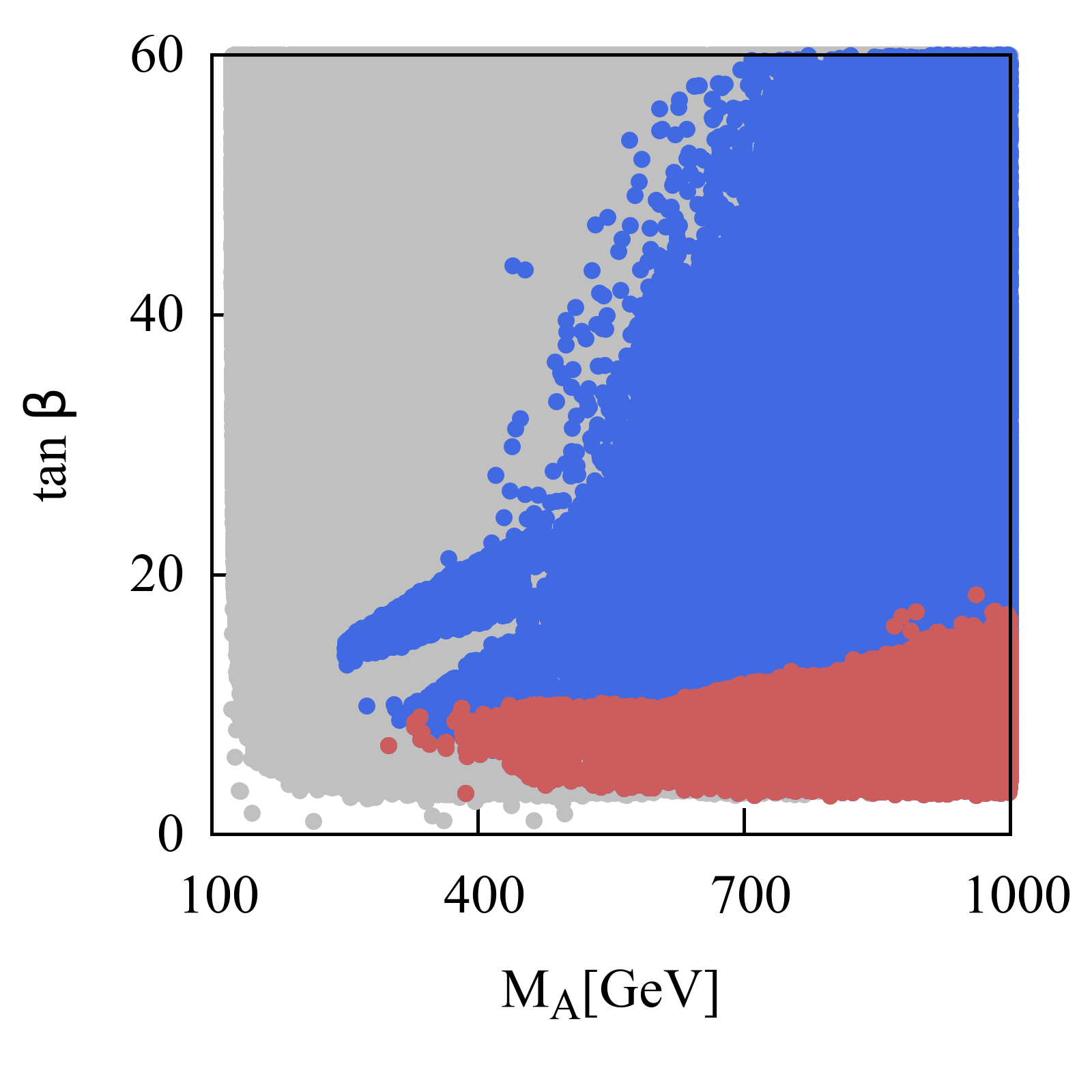}}
\caption{Scatter plot in the $M_A - \tan \beta$ plane. Grey points represent the region allowed by light Higgs mass constraint. Blue points are allowed by the global $\chi^{2}$ analysis performed by combining the signal strength constraints tabulated in Table~\ref{table-mu8} and Table~\ref{table-mu13}, along with the flavor physics constraints. Brown points are allowed by the most stringent heavy Higgs search limits derived by both ATLAS and CMS from the $ H \to \tau^{+} \tau^{-}$ decay mode using LHC Run-II data~\cite{Aaboud:2017sjh,CMS:2017epy}, along with all earlier constraints.}
\label{tb_mA_final}
\end{center}
\end{figure}

\subsection{The $M_A - \tan \beta$ plane}

We first discuss the effect of all the constraints on $M_A - \tan \beta$ plane. In Fig.~\ref{tb_mA_final}, grey points denote the region allowed by the light Higgs mass constraints, while the blue points correspond to the allowed parameter space obtained from global $\chi^{2}$ analysis taking into account the Higgs signal strength constraints and the flavor physics observables. 
Brown points are those which are also allowed by the most stringent heavy Higgs search limit put by $H/A \to \tau^{+} \tau^{-}$ channel, derived by ATLAS and CMS, from LHC Run-II data at $\lum \sim 36~$fb$^{-1}$~\cite{Aaboud:2017sjh,CMS:2017epy}, along with all the earlier constraints.


\subsection{Correlations between $\alpha$, $\beta$ and $M_A$}

\begin{figure}[!tb]
\begin{center}
{\includegraphics[angle=0,width=1\textwidth]{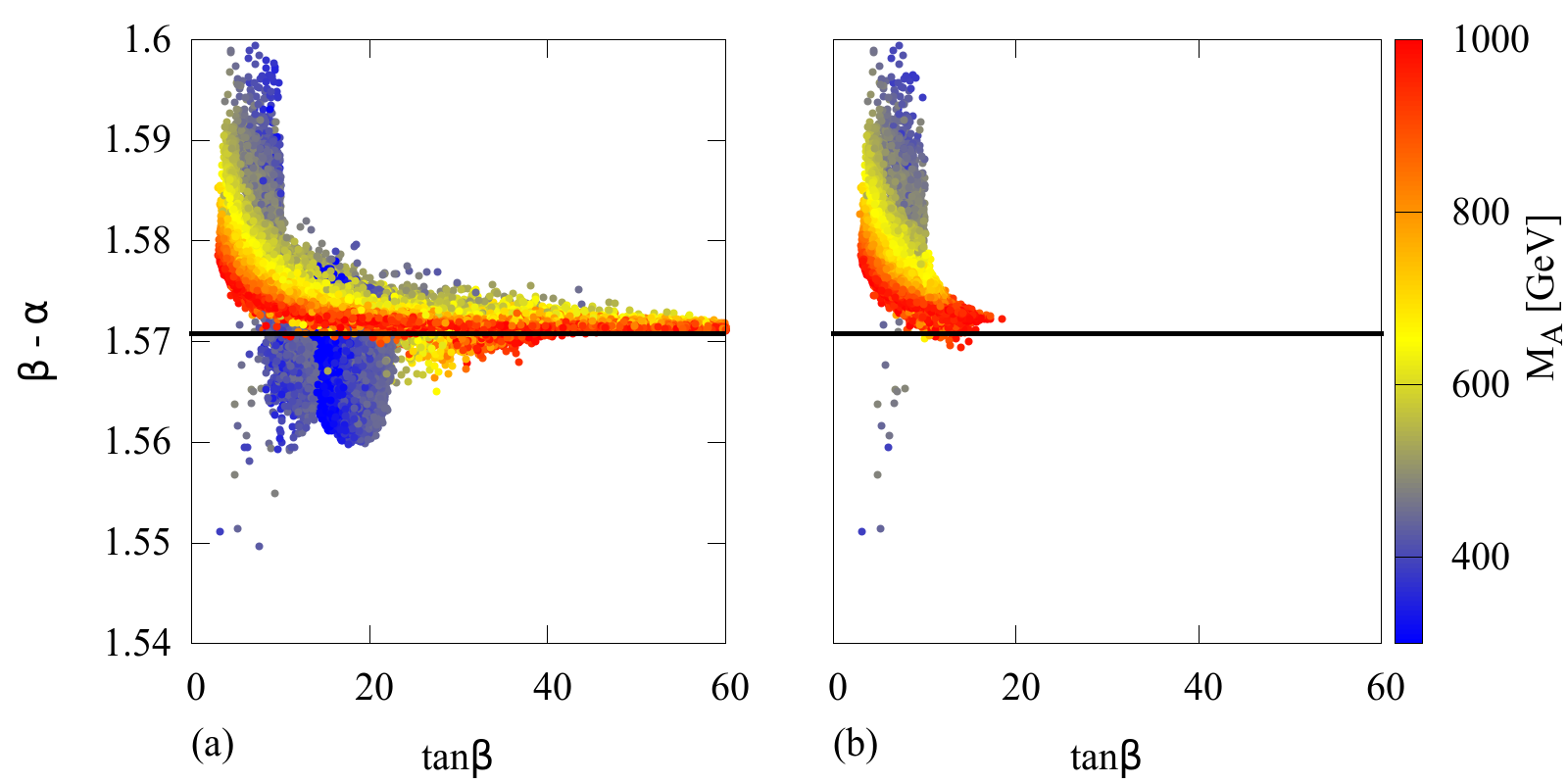}}
\caption{Scatter plot in the $\tan \beta - (\beta-\alpha)$ plane: Fig.~\ref{tb_alpha_beta}(a) represents parameter space points within $2\sigma$ interval of $\chi^{2}_{min}$, obtained by combining the Higgs signal strength constraints tabulated in Table~\ref{table-mu8}, Table~\ref{table-mu13} and the flavor physics observables. Fig.~\ref{tb_alpha_beta}(b) corresponds to those parameter space points of Fig.~\ref{tb_alpha_beta}(a) which are also allowed by the 13 TeV heavy Higgs direct search limits. All points satisfy the light Higgs mass constraint. The black horizontal line corresponds to $(\beta-\alpha) = \pi/2$.}
\label{tb_alpha_beta}
\end{center}
\end{figure}

In this subsection, we discuss the effect of all the constraints, discussed earlier, on the Higgs mixing angle. We begin by discussing the `alignment limit' ~\cite{Gunion:2002zf,Delgado:2013zfa,Craig:2013hca,Carena:2013ooa,Carena:2014nza,Bechtle:2016kui,Bechtle:2015pma} briefly. In MSSM the couplings of the gauge bosons with the neutral CP-even Higgs bosons are given by the following equations:
\\
\begin{equation}
g_{hVV}=\sin(\beta - \alpha) g_V \; ,
\label{align1}
\end{equation}
\begin{equation}
g_{HVV}=\cos(\beta-\alpha) g_V \; ,
\label{align2}
\end{equation}
\\
where $\alpha$ is the mixing angle between the neutral CP-even Higgs bosons and $g_V$ is the SM gauge boson coupling 
($g_V = 2iM_V^2/v$). Alignment limit is the limiting case when one of the CP-even neutral
 Higgs bosons, $h$ and $H$, mimics the behavior of the SM Higgs boson. In this work, we have assumed that the lighter one, i.e. $h$, resembles the SM Higgs boson. 
  It can be achieved when $\sin(\beta - \alpha)\sim1$  in Eq.~\eqref{align1} or $\cos(\beta-\alpha)\sim0$ in Eq.~\eqref{align2}, which imply
   $(\beta - \alpha) \sim \pi/2$. This condition is known as the `alignment criteria'.

 In Fig.~\ref{tb_alpha_beta} we show a scatter plot in the $\tan \beta$ vs. $(\beta-\alpha)$ plane, with $M_A$ shown through the color palette.    
In Fig.~\ref{tb_alpha_beta}(a) we show the parameter space points which lie within $\chi^{2}<\chi^{2}_{min}+6.18$, where $\chi^{2}_{min}$ has been obtained by combining the signal strength constraints tabulated in Table~\ref{table-mu8}, Table~\ref{table-mu13} and the flavor physics observables, 
and satisfy the light Higgs mass constraint as well.
In Fig.~\ref{tb_alpha_beta}(b) we present those of Fig.~\ref{tb_alpha_beta}(a) which are also allowed by the 13 TeV heavy Higgs direct search limits. 

The figures show that we are indeed very close to the alignment limit. 
We should mention here that even with low $M_A$, in the range 200 -- 400 GeV, it is possible to be in the alignment region. In the MSSM the alignment limit can be realised independently of the decoupling of the heavier Higgs states through a cancellation between tree-level and higher-order  contributions in the Higgs sector. This cancellation can occur at relatively large values of $\tan \beta$ (increasing $\mu A_t/M_S$ it is possible to achieve alignment limit even with low $\tan \beta$)~\cite{Carena:2014nza,Bechtle:2016kui,Bechtle:2015pma}. However, large values of $\tan \beta$ are disfavored by the direct search of heavy Higgs in the $\tau^{+} \tau^{-}$ channel. So the points in Fig.~\ref{tb_alpha_beta}(a) and Fig.~\ref{tb_alpha_beta}(b) which are close to the alignment limit actually possess the largest possible $\tan \beta$ values still allowed by the direct search of heavy Higgs. Detailed discussion on the `alignment without decoupling' scenario can be found in the Refs.~\cite{Carena:2014nza,Bechtle:2016kui,Bechtle:2015pma}. Further analysis in this direction is beyond the scope of this paper and will be addressed in~\cite{futurework}.

\subsection{Signal strength correlations}

\begin{figure}[!t]
\begin{center}
{\includegraphics[angle=0,width=1.0\textwidth]{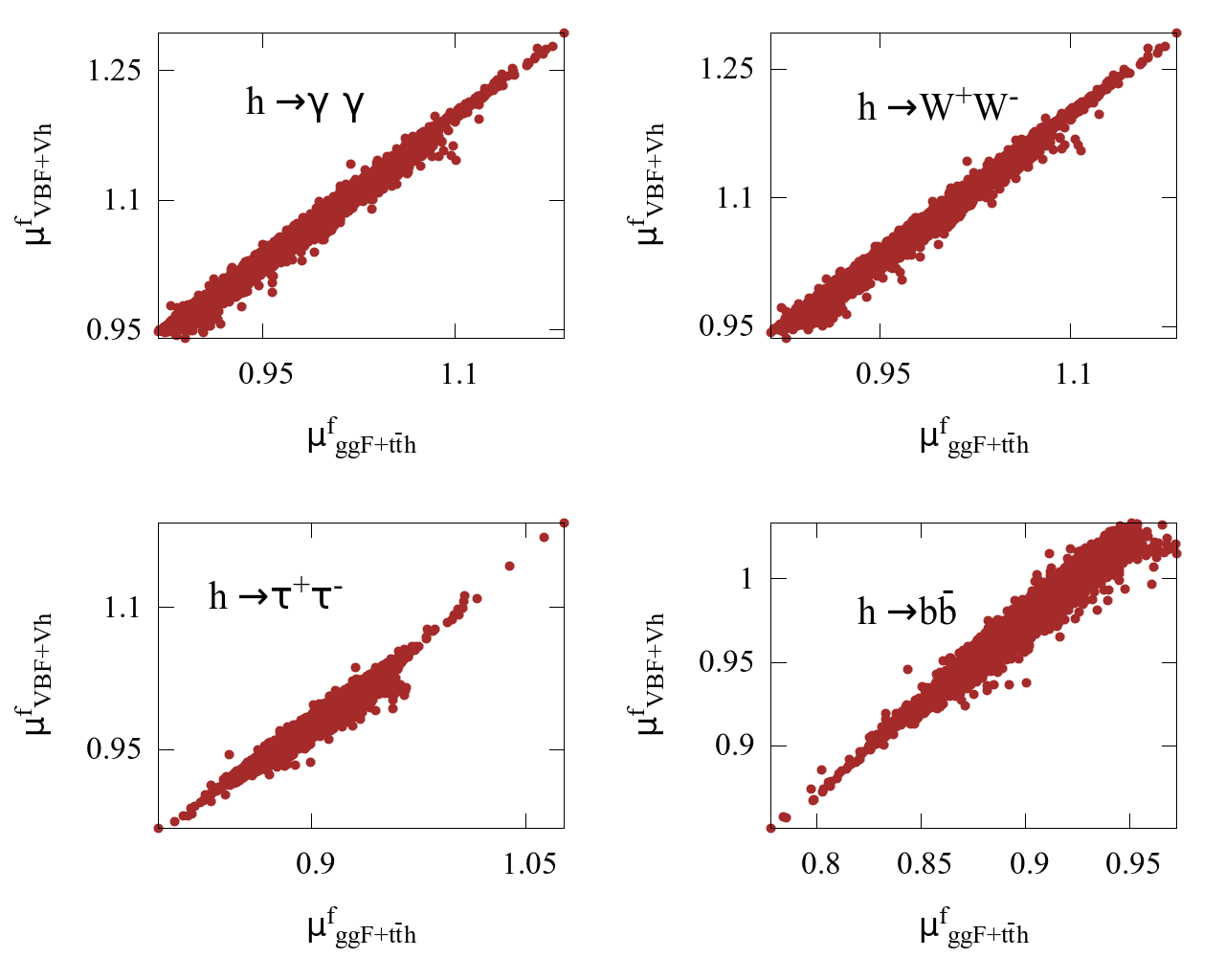}}
\caption{Correlations between signal strengths $\mu_{\text{ggF}+t\bar{t}h}$ and $\mu_{VBF+Vh}$ for the final states $\gamma \gamma$, $W^{+}W^{-}$, $\tau^{+}\tau^{-}$ and $b\bar{b}$. The brown colored points correspond to those which are allowed by the global $\chi^{2}$ analysis and also evade the heavy Higgs direct search limits (tabulated in Table.~\ref{tab:heavysearches} and Table.~\ref{tab:heavysearchesichep}).}
\label{corr_pre_post_ichep}
\end{center}
\end{figure}

In Fig.~\ref{corr_pre_post_ichep} we plot the correlations between signal strength variables $\mu_{ggF+t\bar{t}h}$ and $\mu_{VBF+Vh}$ for $\gamma \gamma$, $W^{+}W^{-}$, $\tau^{+} \tau^{-}$ and $b\bar{b}$ final states. All the parameter points shown in Fig.~\ref{corr_pre_post_ichep} satisfy the light Higgs mass constraint and lie within $2\sigma$ interval of $\chi^{2}_{min}$ in the global $\chi^2$ analysis. 
These points are also allowed by the heavy Higgs direct search results tabulated in Table.~\ref{tab:heavysearches} and Table.~\ref{tab:heavysearchesichep}.


\subsection{$t \bar t h$ correlations}

\begin{figure}[!t]
\begin{center}
{\includegraphics[angle=0,width=0.48\textwidth]{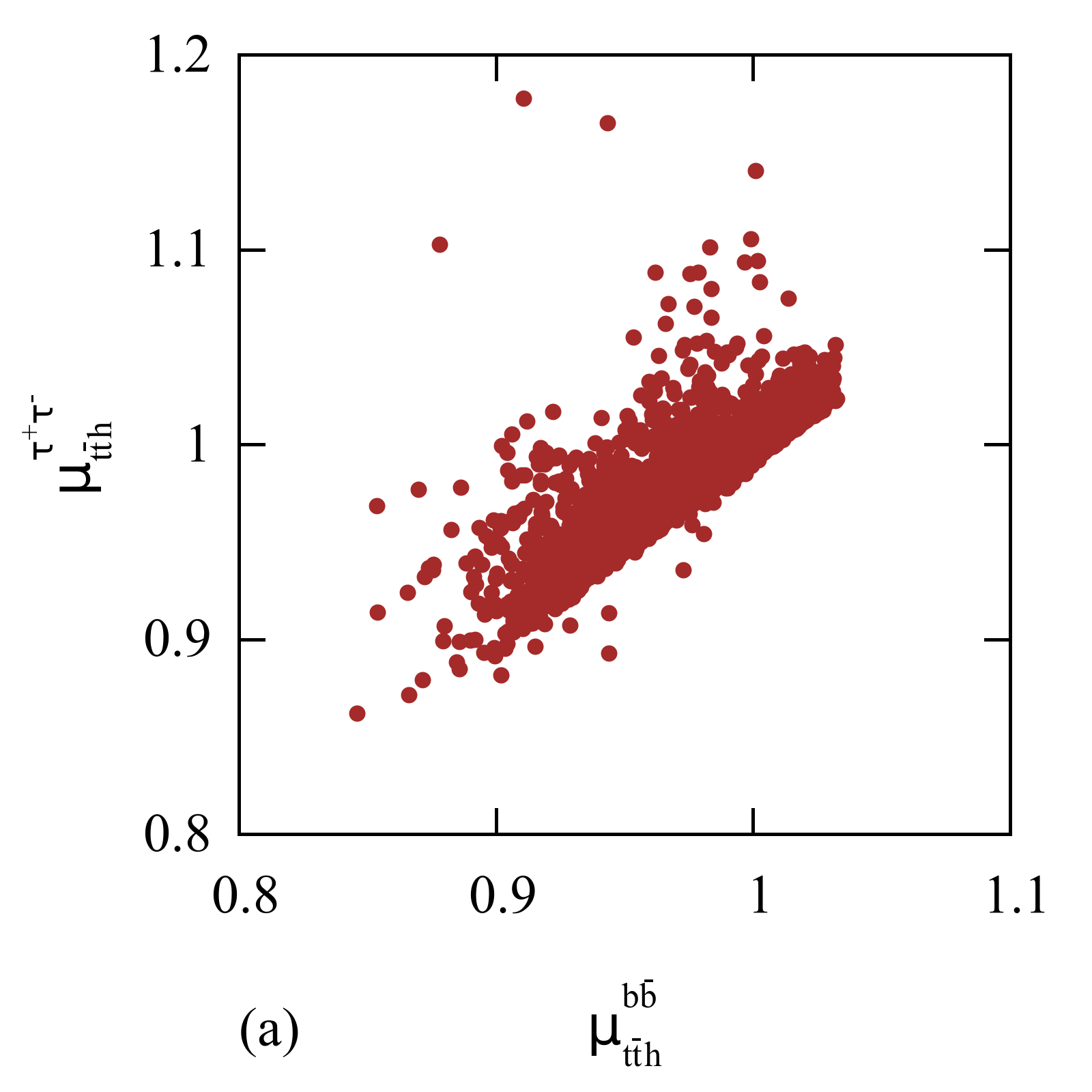}}
{\includegraphics[angle=0,width=0.48\textwidth]{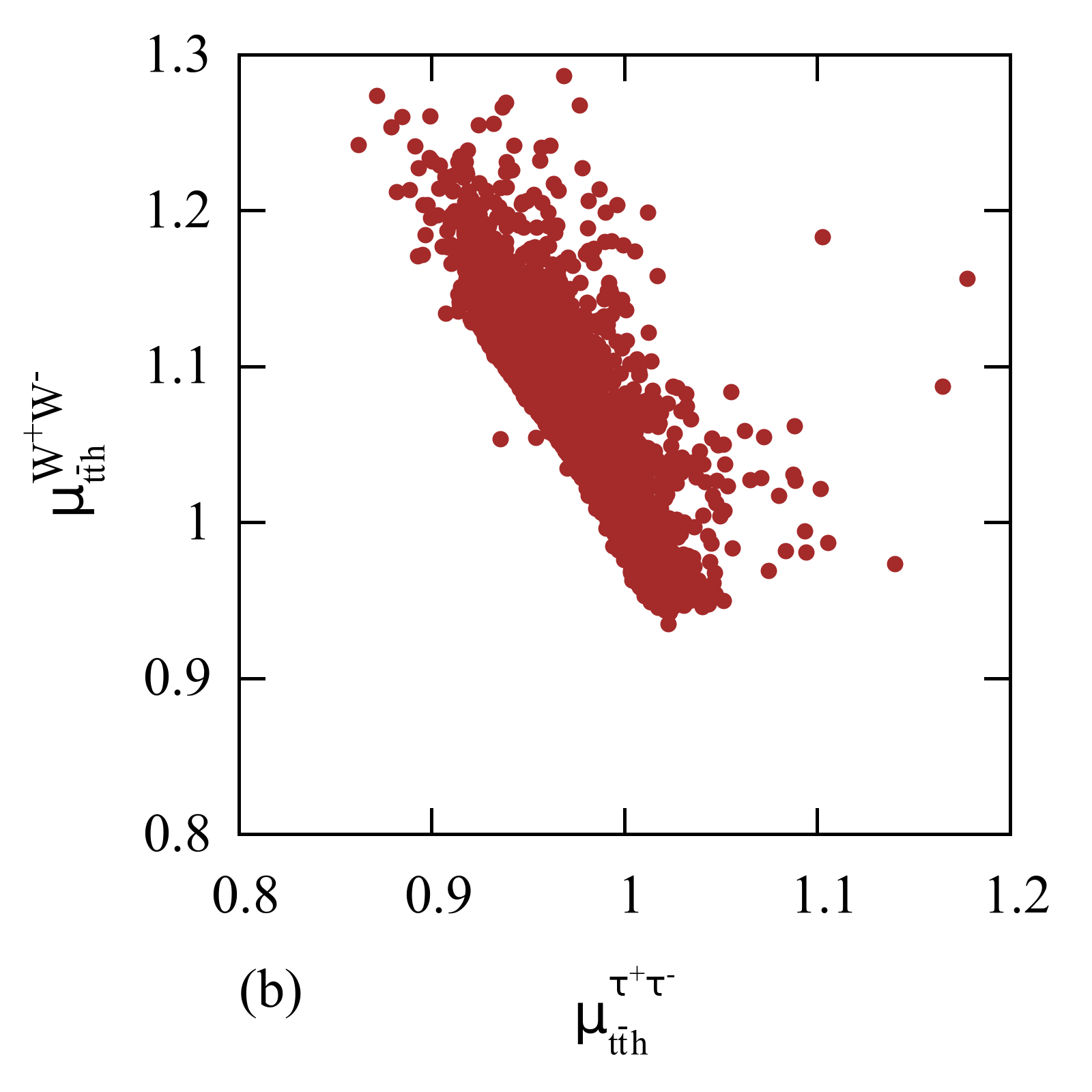}}
\caption{Correlations between signal strength variables: Fig.~\ref{corr_tth_ichep}(a) shows correlation between $\mu_{t\bar{t}h}$ in $b\bar{b}$ and $\tau^{+} \tau^{-}$ final states, while Fig.~\ref{corr_tth_ichep}(b) shows correlation between $\mu_{t\bar{t}h}$ in $W^{+}W^{-}$ and $\tau^{+} \tau^{-}$ final states, for the parameter space points allowed by global $\chi^{2}$ analysis and heavy Higgs direct search limits.}
\label{corr_tth_ichep}
\end{center}
\end{figure}

In Fig.~\ref{corr_tth_ichep}(a) we plot the scatter diagram for correlation between $\mu_{t\bar{t}h}$ in the $b\bar{b}$ and $\tau^{+} \tau^{-}$ final states. In Fig.~\ref{corr_tth_ichep}(b) we plot the scatter diagram for correlation between $\mu_{t\bar{t}h}$ in the $\tau^{+} \tau^{-}$ and $W^{+}W^{-}$ final states. 
The points are allowed at $2\sigma$ by the global $\chi^{2}$ analysis performed by combining 8 TeV and 13 TeV signal strength data and flavor physics constraints and also evade the heavy Higgs direct search limits.
Here again, we observe the anti-correlation between the $h \rightarrow \tau^+ \tau^-$ and $h \rightarrow WW$ and correlation between $h \rightarrow b \bar b$ and $h \rightarrow \tau^+ \tau^-$. 



\subsection{Higgs coupling with the bottom quark}

Precise measurement of the Higgs signal strength can be a probe for Higgs coupling measurements. We have studied the details of bottom Yukawa coupling in this regard. The loop corrections involving various SUSY particles can modify the bottom Yukawa coupling significantly. 
To qualitatively understand this, let us consider the effective two-Higgs doublet model Lagrangian of the MSSM, which contains the following couplings of the bottom-quark to the CP-even neutral Higgs bosons
\begin{equation}
{\mathcal{L}}_{eff} = Y_b H^0_d b \bar b + \Delta Y_b H^0_u b \bar b \ .
\end{equation}
In MSSM, the tree level $H^0_u b \bar b $ coupling does not exist, as $H^0_u$ couples only to up-type quarks at tree level, but a non-vanishing $\Delta Y_b$ can be generated dynamically at one loop level. Although $\Delta Y_b$  is loop suppressed, once the Higgs fields $H^0_u$ and $H^0_d$ acquire vacuum expectation values, a small $\Delta Y_b$ shift can introduce a large modification to the tree level relation between the bottom mass and its Yukawa coupling as it is enhanced by $\tan\beta$:
\begin{align*}
m_{b} = Y_{b} v_{b} \longrightarrow m_{b} = v_{b} (Y_{b} + \Delta Y_{b} \tan\beta) = Y_{b} v_{b} (1 + \Delta m_{b})\ ,
\end{align*}
where $\Delta m_{b} = (\Delta Y_{b}/Y_{b}) \tan\beta$~\cite{Carena:1999py}. 
\begin{figure}[htb]
\begin{center}
\includegraphics[width=0.9\textwidth]{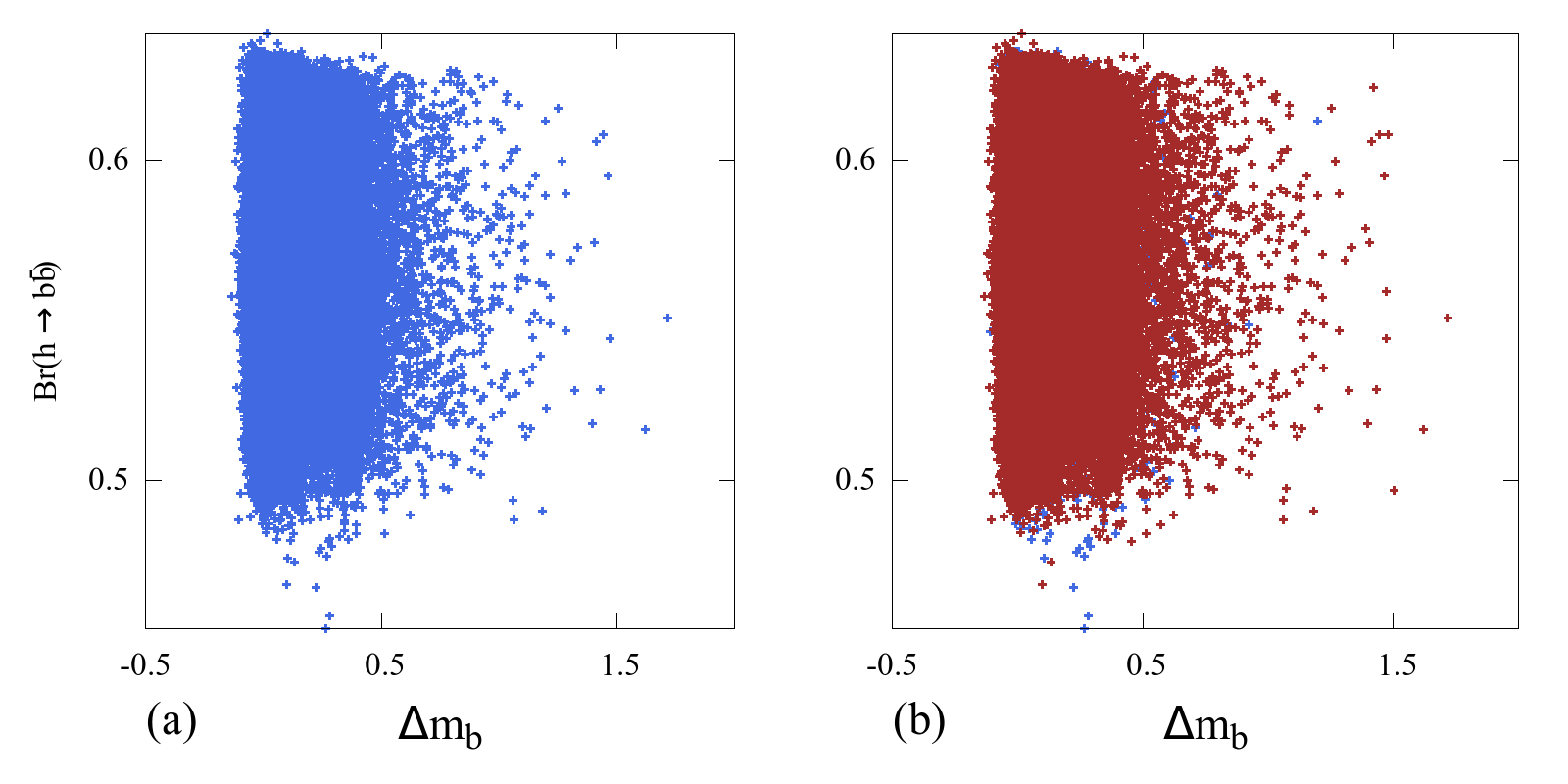}
\caption{Scatter plot in $\Delta m_b$ and Br($h \to b \bar b)$ plane: (a) Parameter points within $2\sigma$ interval of $\chi^{2}_{min}$, (b) Parameter points of Fig.~\ref{deltamb}(a) which are allowed by 13 TeV heavy Higgs direct search limits.}
\label{deltamb}
\end{center}
\end{figure}
In MSSM the SUSY particles contribute to the threshold corrections at loop level and that would imply a shift in the bottom Yukawa coupling~\cite{Carena:1999py,Hall:1993gn,Guasch:2003cv,Dawson:2011pe}. We have performed a scan of the parameter space and plotted the Higgs branching ratio for $h \to b \bar b$ decay as a function of $\Delta m_b$, calculated by FeynHiggs 2.12.0, in Fig.~\ref{deltamb}. In Fig.~\ref{deltamb}(a) we see the Br($h \to b \bar b$) as a function of
 $\Delta m_b$ for the scanned blue points that are allowed by light Higgs mass, flavor data and 8 TeV and 13 TeV combined Higgs signal strength data. In Fig.~\ref{deltamb}(b) we show those parameter points which are also allowed by the 13 TeV heavy Higgs direct search limits, along with the earlier blue points.  
 
 %

\section{Heavy Higgs decay to SUSY states}
\label{sec:IV}

In this section we discuss the scenarios where 
the heavy Higgs bosons can decay to sparticles like third generation 
squarks/sleptons or to electroweakinos. 
To probe these scenarios we scan the parameter space by decoupling 
the gluinos and first two generations of squarks and sleptons by fixing their masses at $2~{\rm TeV}$.  
Hence apart from the SM final states, the heavy Higgs bosons can decay into a 
pair of stops, staus, sbottoms. 
We will also consider the possibility where the heavy Higgs bosons decay into invisible,  
semi-invisible or visible electroweakinos in the final states i.e., 
$H/A$ decaying to $\lspone \lsptwo, \lspone \lspthree, \lsptwo \lspthree$, $\lspone \lspone, \lsptwo \lsptwo, \lspthree \lspthree$ etc.

\subsection{Heavy Higgs decaying to Electroweakinos }

\begin{figure}[t!]
\begin{center}
\includegraphics[width=0.48\textwidth]{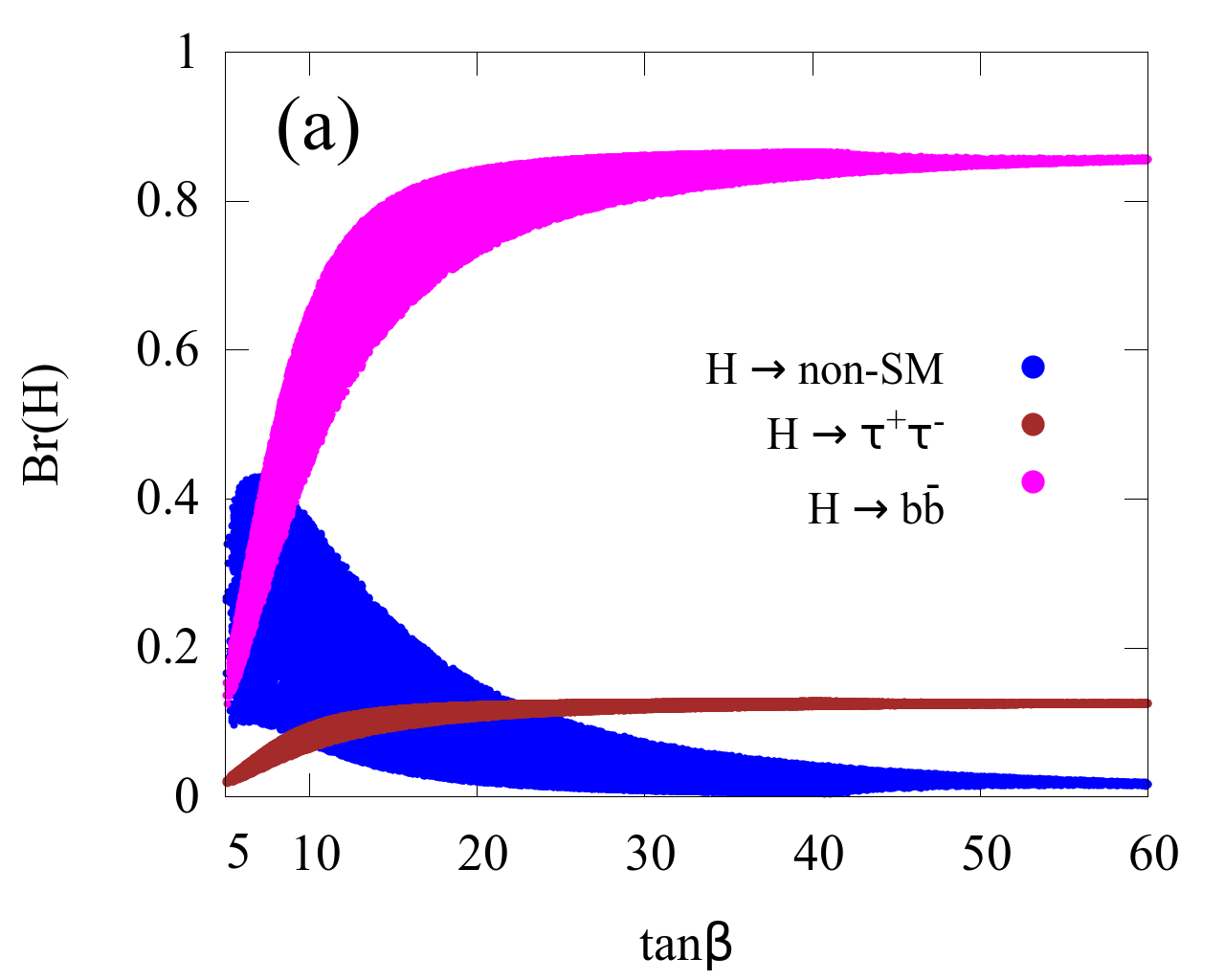}
\includegraphics[width=0.48\textwidth]{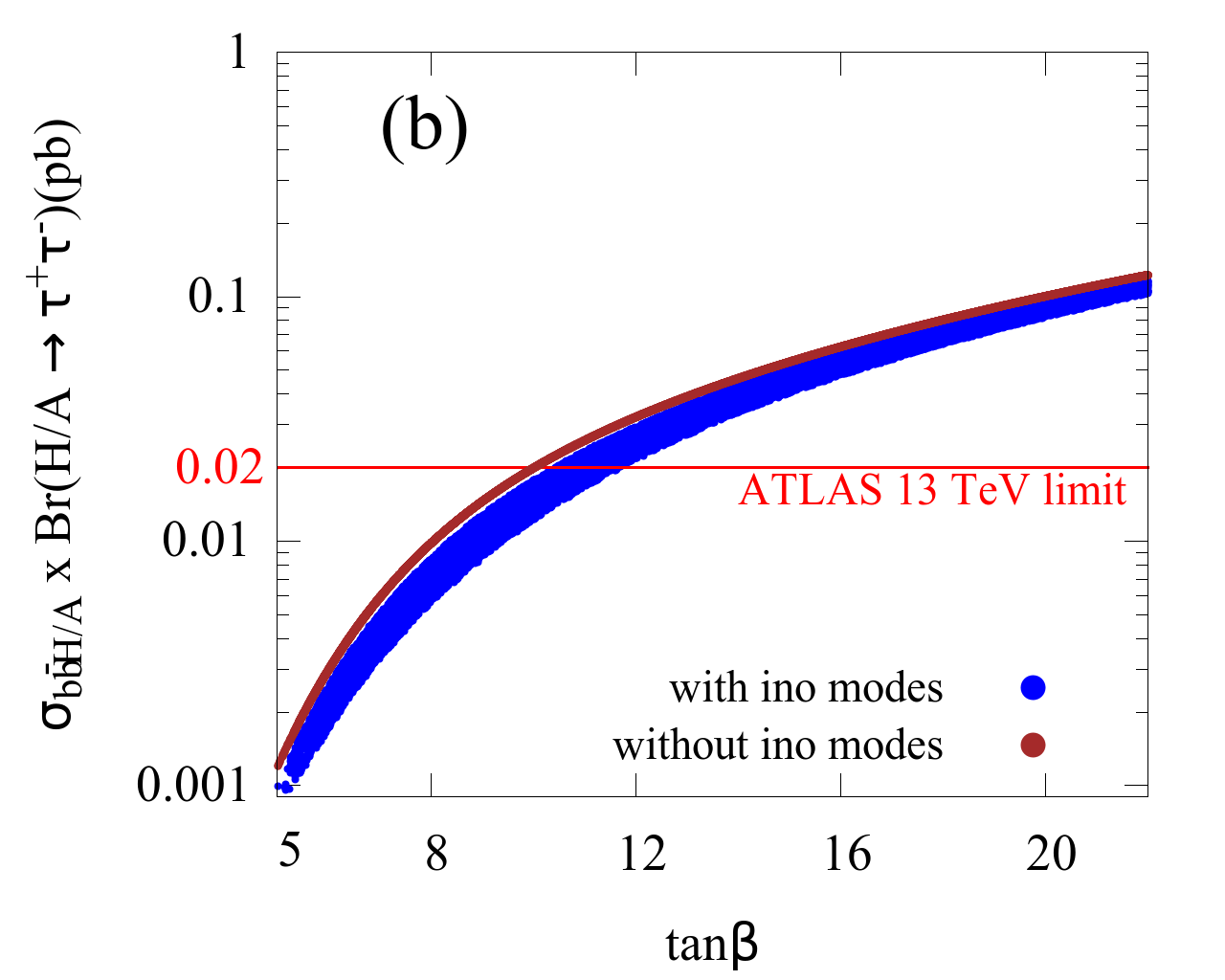}
\caption{In Fig.~\ref{fig:m1_mu}(a) we show the branching fractions of the scalar heavy Higgs boson: $H \to \tau^{+} \tau^{-} $ (brown dots), $ b \bar{b}$ (magenta dots) and `electroweakino' or `ino' pairs (blue dots). Here $H \to$ (non-SM) has been obtained by summing over the branching fractions of all possible `ino' decay modes. In Fig.~\ref{fig:m1_mu}(b) we show a scatter plot in the $\tan{\beta} - [\sigma_{b\bar{b}H/A} \times \mathrm{Br}(H/A \to \tau^{+} \tau^{-})]$ plane. The blue (brown) points signify the value of $\sigma_{b\bar{b}H/A} \times \mathrm{Br}(H/A \to \tau^{+} \tau^{-})$ in the presence (absence) of possible heavy Higgs to `ino' decay modes. Here the `inos' are admixtures of binos and higgsinos. The red horizontal line in Fig.~\ref{fig:m1_mu}(b) represents the 95$\%$ C.L. upper limit on the quantity $\sigma_{b\bar{b}H/A} \times \mathrm{Br}(H/A \to \tau^{+} \tau^{-})$ given by the ATLAS 13 TeV data ($ 36.1~{\rm fb^{-1}}$) for $M_A = 600$ GeV. All parameter points here lie within $2\sigma$ interval of the $\chi_{min}^{2}$, computed earlier.
}
\label{fig:m1_mu}
\end{center}
\end{figure}

In the MSSM parameter space, there exist certain regions with intermediate $\tan\beta$ 
($\sim$~5 - 15) where the heavy Higgs coupling to SM particles become very 
small~\cite{Barman:2016kgt} and non-SM decays are appreciable.  
In the presence of light electroweakinos, and if kinematically allowed, these heavy 
Higgs bosons can decay to charginos and neutralinos with a significantly high 
branching fraction\footnote{For detailed study on heavy Higgs decay to electroweakinos see Refs.~\cite{Ananthanarayan:2015fwa,Barman:2016kgt,Gori:2018pmk}.}. 
These `ino' decay modes crucially depend on the gaugino-higgsino mixing, or 
precisely on the composition of these electroweakino states. 
The heavy Higgs bosons will couple with the charginos and neutralinos  if and only if 
the electroweakinos are admixture of the higgsinos and gauginos (bino or wino).
For higgsino-dominated or gaugino-dominated scenarios these couplings are highly 
suppressed. It may be noted that direct electroweakino searches at LHC \cite{Aad:2014vma,Aad:2015eda,Aad:2014yka,Aad:2015jqa} have obtained strong bounds on electroweakino masses, especially searches in the chargino-neutralino pair production channel, which are  the most stringent ones. For a neutralino LSP of mass $\sim100-150$ GeV and for a degenerate wino-like $\chi_{2}^{0}$ and $\chi_{1}^{\pm}$, the lower limit on the masses of these NLSPs is $\sim300$ GeV. These mass bounds disfavour heavy Higgs mass below $\sim400-450$ GeV to produce a visible `ino' final state due to kinematic reasons. However, in certain regions of parameter space, it is possible to obtain a nearly-degenerate LSP and NLSP, and within such scenarios the above mentioned limit does not apply.

The couplings of the heavy Higgs bosons with the charginos and neutralinos are parametrized by the wino, bino and higgsino mass parameters. We perform a random scan varying the gaugino and higgsino mass parameters and $\tan{\beta}$, while keeping the other parameters fixed, to study the heavy Higgs to `ino' decay modes. We fix $M_A$ at 600 GeV, while the slepton and the squark masses are fixed at a much higher value such that the decay of the heavy Higgs bosons to the sleptons and squarks are kinematically forbidden. 

\begin{figure}[tb!]
\begin{center}
\includegraphics[width=0.48\textwidth]{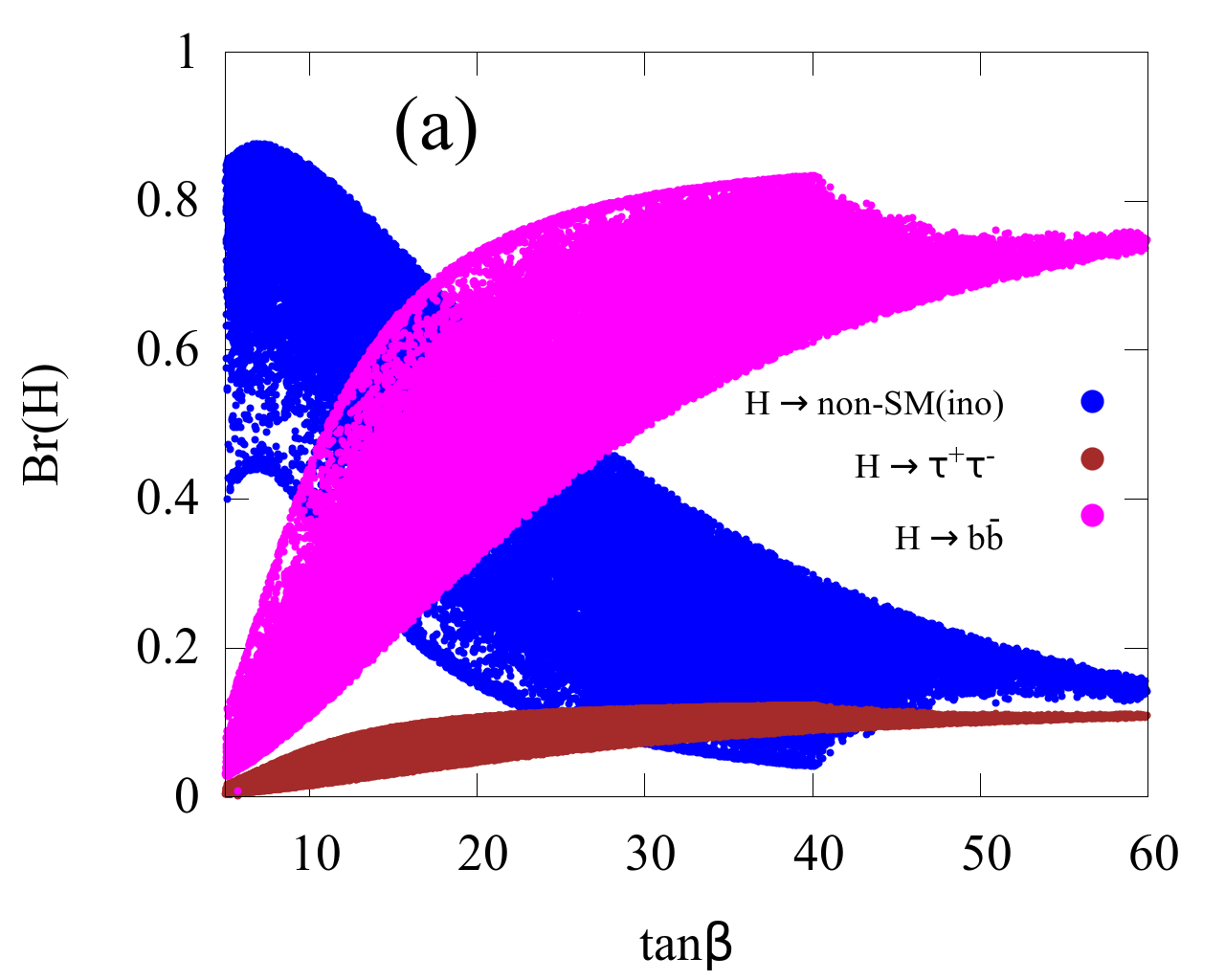}
\includegraphics[width=0.48\textwidth]{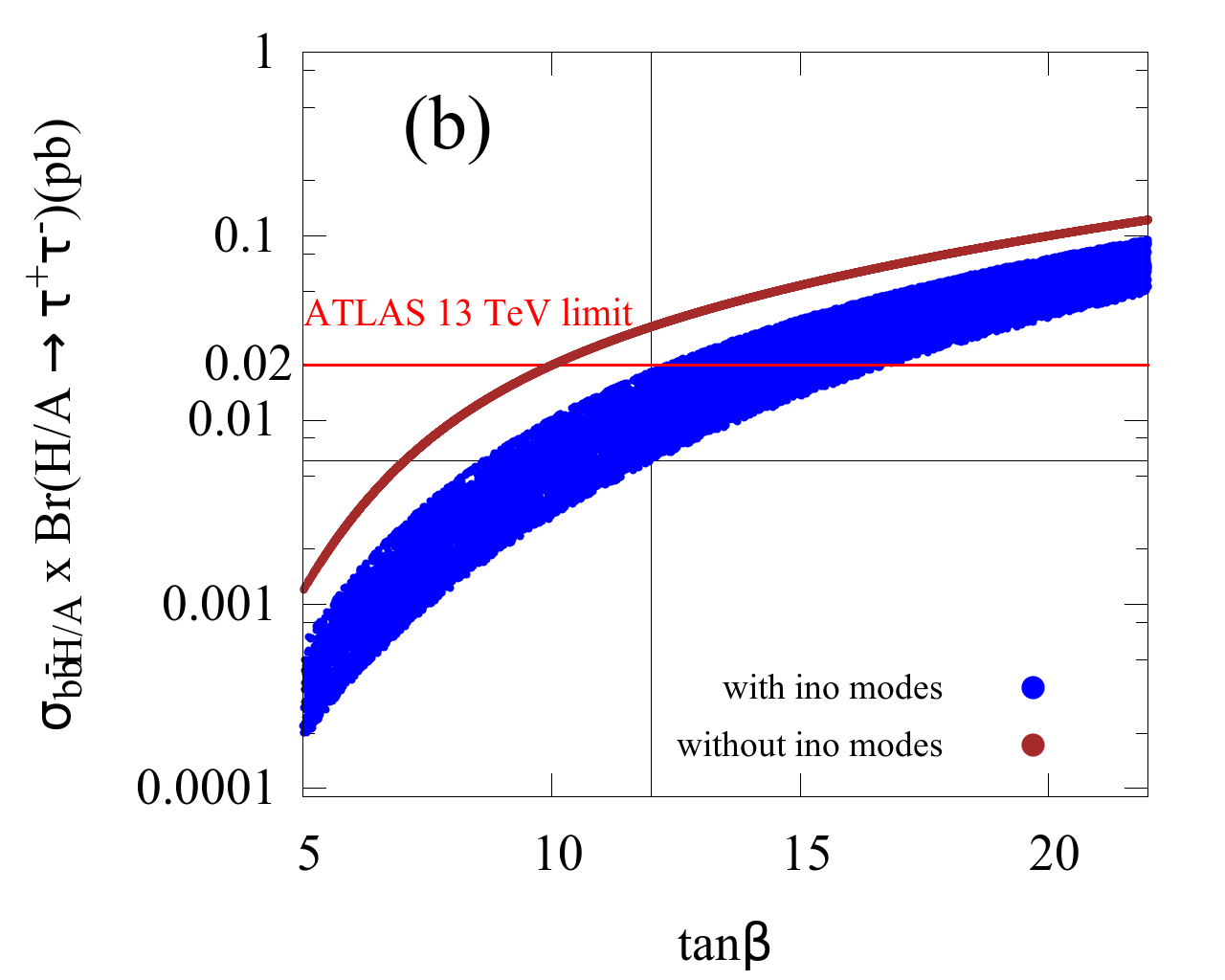}
\caption{We show the branching fractions of heavy scalar Higgs boson: $H \to \tau^{+} \tau^{-} $ (brown dots), $ b \bar{b}$ (magenta dots) and `ino' pairs (blue dots), in Fig.~\ref{ino:m2_mu}(a). Here $H \to$ (non-SM) has been obtained by summing over the branching fractions of all possible `ino' decay modes. In Fig.~\ref{ino:m2_mu}(b) we show the scatter plot in the $\tan{\beta} - [\sigma_{b\bar{b}H/A} \times \mathrm{Br}(H/A \rightarrow \tau^{+} \tau^{-})]$ plane. The blue (brown) points signify the value of $\sigma_{b\bar{b}H/A} \times \mathrm{Br}(H/A \to \tau^{+} \tau^{-})$ in the presence (absence) of possible heavy Higgs to `ino' decay modes. Here the `ino's are admixtures of winos and higgsinos.
The red horizontal line in Fig.~\ref{ino:m2_mu}(b) represents the 95$\%$ C.L. upper limit on the quantity $\sigma_{b\bar{b}H/A} \times \mathrm{Br}(H/A \to \tau^{+} \tau^{-})$ given by the ATLAS 13 TeV data ($36.1 ~{\rm fb^{-1}}$) for $M_A = 600$ GeV. All parameter points here lie within $2\sigma$ interval of the $\chi_{min}^{2}$, computed earlier.
}
\label{ino:m2_mu}
\end{center}
\end{figure}

We consider two different scenarios here. In the first scenario, we choose the light `inos'\footnote{By light `inos', we refer to those neutralinos or charginos into which the decay of the heavy Higgs bosons is kinematically allowed.} to be admixtures of bino and higgsino components. Here we vary $M_{1}$ and $\mu$ in such a way that $M_{1} + \mu < M_{A}$ and both $M_{1},\mu > 100$ GeV. In addition, $ |M_{\chi_{1}^{0}} - M_{\chi_{3}^{0},\chi_{1}^{\pm}}|$ is always less than the mass of W boson. In such cases, the Higgs to `ino' branching fraction can reach up to 40$\%$ in the moderately low $\tan{\beta}$ region, as presented in  Fig.~\ref{fig:m1_mu}(a). The `ino' branching fraction has been calculated by summing over all possible `Higgs-to-ino' decay modes. We investigate the effect of the presence of heavy Higgs bosons to `ino' decay modes on the $\rm H \to \tau^{+} \tau^{-}$ channel, since heavy Higgs searches in this channel provide the strongest constraints on the MSSM parameter space. Blue points in 
Fig.~\ref{fig:m1_mu}(b) represent the $\sigma_{b\bar{b}H/A} \times \mathrm{Br}(H/A \rightarrow \tau^{+} \tau^{-})$ values in the presence of `ino' decay modes, with the `ino's being mixtures of binos and higgsinos, while the brown line represents the corresponding cross-section times branching ratio in the absence of light `ino's.

In the second scenario we choose the `ino' mass parameters in such a way that the light `ino's are compounded from wino and higgsino mixing. We choose  $M_{2} + \mu < M_{A}$, with both $M_{2}, \mu > 100$ GeV, and  $|M_{\chi_{1}^{0}} - M_{\chi_{3}^{0},\chi_{2}^{\pm}}|$ less than the W boson mass. Here the Higgs to `ino' branching fraction can go as high as 80$\%$ in the moderate $\tan{\beta}$ region, as shown in  Fig.~\ref{ino:m2_mu}(a). We present the $\sigma_{b\bar{b}H/A} \times \mathrm{Br}(H/A \to \tau^{+} \tau^{-})$ values in Fig.~\ref{ino:m2_mu}(b) with blue points. These points reflect the effect of heavy Higgs to `ino' decay modes on the $H \to \tau^{+} \tau^{-}$ branching fraction. 
Let us focus on a particular value of $\tan{\beta}$, say $\tan{\beta}=12$. We observe that the latest and the most stringent upper limit on $\sigma_{b\bar{b}H/A} \times \mathrm{Br}(H/A \to \tau^{+} \tau^{-}) \approx 0.02~$pb, derived by ATLAS at 95$\%$ C.L., would be able to rule out this point for the case when there are no possible `ino' decay modes in the model. However, because of the presence of these `ino' decay modes, the upper limit  bound on $\sigma_{b\bar{b}H/A} \times \mathrm{Br}(H/A \rightarrow \tau^{+} \tau^{-})$ is required to be approximately one-third of the existing value in order to rule out the same $\tan{\beta}>12$ region. This would in turn require $\sim 9$ times increase in luminosity to maintain the same sensitivity. 

\subsection{Heavy Higgs to sfermions}

In this section we explore the possibility of 
heavy Higgs decay to third generation sfermions. 
It may be noted that for first two generation squarks, 
the couplings to heavy Higgs is almost zero as they are 
proportional to the corresponding fermion masses. 
Due to the large third generation fermion masses and large 
mixing in the third generation squark-sector, heavy Higgs decaying 
to $\lstop \lstop$ or $\lsbot \lsbot$ could be the dominant 
decay mode. For large $A_\tau$ and large $\tan \beta$, the Higgs coupling to a pair of staus will be large. But, in such scenarios, the coupling to $b \bar b$ is also enhanced and $H \to b \bar b$ always dominates. We perform a dedicated scan for light ${\widetilde{\tau}_1}$ scenarios and obtain that Br($ H \to {\widetilde{\tau}_1} {\widetilde{\tau_1}}$ ) is very small (typically $< 1\%$).

\begin{table}[tb!]
\begin{center}
\begin{tabular}{|c|c||c||c|c|} \hline
 Benchmark & Parameters (GeV) & Mass (GeV) & Processes & Branching \\ 
 Points& & & & Fraction\\ \hline\hline
 	& $ M_{A} = 950, \quad M_{1} = 300,$ &  $ M_H = 784 $  & 
 	$ H \rightarrow {\widetilde{t_1}}{\widetilde{t_1}}$ & $  96\% $ \\
BP-1 	& $ M _{2} = 1500, \quad \mu = 8000, $ & $M_{\widetilde{t_1}} = 402 $ & 
$ H \rightarrow b \bar b$ & $  3\% $ \\
& $ \tan{\beta} = 20,\quad A_{t} = -1400, $ & $M_{\widetilde{t_2}} = 1524 $ & 
$ H \rightarrow \tau \bar \tau$ & $  1\% $ \\
	 & $ m_{\tilde{Q}_{3L}} = 425, \quad m_{\tilde{t}_{R}} = 1500, $ & $M_{\widetilde{b_1}} = 407 $ & $ A \rightarrow b \bar b$ & $  73\% $ \\
 	& $ m_{\tilde{b}_{R}} = 5000, A_{b} = -3000,A_{\tau} = 0 $ & $M_{\widetilde{b_2}} = 5001 $ & 
$ A \rightarrow \tau \bar \tau$ & $  27\% $ \\
	 & $ M_{3} = 3000, M_{\widetilde{q}_{1,2}} = M_{\widetilde{L}} = 4000$ &$M_{\chi_{1}^{0}}$ = 300    &  & \\  	 \hline	 \hline  

	& $ M_{A} =700, \quad M_{1} = 290,$ &  $ M_H = 700 $  & 
 	$ H \rightarrow {\widetilde{b_1}}{\widetilde{b_1}}$ & $  27\% $ \\
BP-2 	& $ M _{2} = 1500, \quad \mu = 3000, $ & $M_{\widetilde{t_1}} = 1295 $ & 
$ H \rightarrow b \bar b$ & $  42\% $ \\
& $ \tan{\beta} = 10,\quad A_{t} = -2800, $ & $M_{\widetilde{t_2}} = 3011 $ & 
$ H \rightarrow \tau \bar \tau$ & $  11\% $ \\
	 & $ m_{\tilde{Q}_{3L}} = 1300, \quad m_{\tilde{t}_{R}} = 3000, $ & $M_{\widetilde{b_1}} = 295 $ & $ A \rightarrow b \bar b$ & $  54\% $ \\
 	& $ m_{\tilde{b}_{R}} = 325, A_{b} = -12000,A_{\tau} = 0 $ & $M_{\widetilde{b_2}} = 1305 $ & 
$ A \rightarrow \tau \bar \tau$ & $  15\% $ \\
	 & $ M_{3} = 3000, M_{\widetilde{q}_{1,2}} = M_{\widetilde{L}} = 4000$ &$M_{\chi_{1}^{0}}$ = 290    & $ A \rightarrow t \bar t$  &  31$\%$\\   
	 \hline
	 \hline
& $ M_{A} =1000, \quad M_{1} = 275,$ &  $ M_H = 1000 $  & 
	$ H \rightarrow {\widetilde{b_1}}{\widetilde{b_1}}$ & $  53\% $ \\
BP-3 	& $ M _{2} = 1500, \quad \mu = 3000, $ & $M_{\widetilde{t_1}} = 1303 $ & 
$ H \rightarrow b \bar b$ & $  31\% $ \\
& $ \tan{\beta} = 20,\quad A_{t} = -2800, $ & $M_{\widetilde{t_2}} = 4005 $ & 
$ H \rightarrow \tau \bar \tau$ & $  13\% $ \\
	 & $ m_{\tilde{Q}_{3L}} = 1300, \quad m_{\tilde{t}_{R}} = 4000, $ & $M_{\widetilde{b_1}} = 313 $ & $ A \rightarrow b \bar b$ & $  64\% $ \\
 	& $ m_{\tilde{b}_{R}} = 400, A_{b} = -15000,A_{\tau} = 0 $ & $M_{\widetilde{b_2}} = 1311 $ & 
$ A \rightarrow \tau \bar \tau$ & $  28\% $ \\
	 & $ M_{3} = M_{\widetilde{q}_{1,2}} = M_{\widetilde{L}} = 4000$ &$M_{\chi_{1}^{0}}$ = 290    & $ A \rightarrow t \bar t$  &  7$\%$\\  
	 \hline
	 \hline

\end{tabular}
\end{center}

\caption{Input parameters, output masses of heavy Higgs bosons and the third generation squarks 
and relevant branching fractions of  heavy Higgs bosons for selected benchmark points. Here 
all the input mass parameters and output masses are in GeV.  All the benchmark points are within the range of $\chi^2_{min} + 6.18$.}
\label{tab:bmp}
\end{table}

\subsubsection{Heavy Higgs to stops}

Due to a large mixing term driven by the top-quark mass $m_t$ in the stop mass-matrix, 
the lighter mass eigenvalues ($\mlstop$) can be much lighter than the masses of all other 
squarks. This will make the decay of the heavy Higgs to a pair of stops kinematically 
allowed, even when the Higgs is not so heavy.  
For not too heavy Higgs and small $\tan \beta$ or for 
intermediate $\tan \beta$, or heavy Higgs with large $\mu$ and $A_t$, the partial
 decay width into stop squarks can be very large and can dominate over the $t \bar t$ 
 and electroweakino final states. The heavy Higgs decay into light stops has been discussed in detail in \cite{Liebler:2015ddv}.
 
To probe such scenarios we scan the MSSM parameter space by fixing the 
masses of first two generation squarks and all three generation sleptons 
at 4 TeV. It should also be noted that the LHC Run-I data and recent 13 TeV 
data for direct stop searches (see \cite{ATLAS-CONF-2016-077,ATLAS-CONF-2016-050,ATLAS-CONF-2016-076,Aaboud:2016tnv,Aad:2015pfx}) have severely constrained 
the parameter space depending upon the decay modes of stop.
 
For $\widetilde{t_1} \to t \lspone $ mode,   
stop mass below 800 GeV is excluded from 13 TeV 13.2 fb$^{-1}$ LHC data for $\mlspone <$ 240 GeV
~\cite{ATLAS-CONF-2016-050,ATLAS-CONF-2016-077}. 
Even for $\widetilde{t_1} \to c \lspone $   or  $b W \lspone $ or $b f f^{\prime} \lspone $ 
 modes, $\mlstop$ approximately below 350 GeV is already ruled for 
  $\mlspone <$ 225 GeV \cite{atlas-stop-summary}. 
Hence to look for $H \to \lstop \lstop$ decay modes we basically 
concentrate in the region with $M_H > $ 700 GeV.  
In Table~\ref{tab:bmp}  we present a benchmark point BP-1 which 
is allowed by 125 GeV Higgs data and other flavor physics 
constraints. The relevant input parameters and output masses, 
branching ratios are also summarized in the Table~\ref{tab:bmp} for BP-1. 
In BP-1, the dominant decay mode of $H$ is $H \to \lstop \lstop$ ($\sim 96\%$).

\subsubsection{Heavy Higgs to sbottoms}

\begin{figure}[!htb]
\begin{center}
{\includegraphics[angle=0,width=0.48\textwidth]{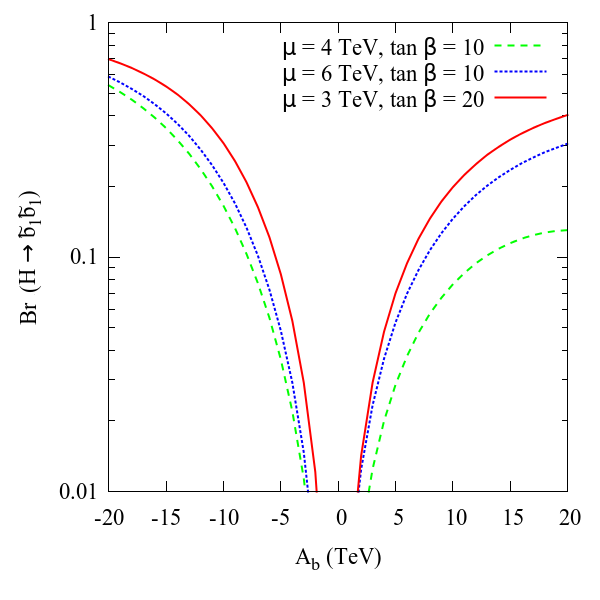}}
\caption{Branching fraction of heavy Higgs boson $H \to \lsbot \lsbot $ with respect to the variation of $A_b$. The green, blue and red lines correspond to  ($\mu$, $\tan\beta$) = (4 TeV, 10), (6 TeV, 10) and (3 TeV, 20), respectively. Other parameters are exactly same as BP3 (see Table~\ref{tab:bmp}).}
\label{sbot_brs}
\end{center}
\end{figure}

Similar to the stop scenarios, light sbottom is also tightly constrained 
from Run-I and Run-II data  \cite{Aad:2015pfx,Khachatryan:2015wza,Aaboud:2016nwl}. 
$\mlsbot <$ 800 GeV is excluded for $\mlspone \lesssim 250 $ GeV. 
Apart from the degenerate regions,  this limits hold up to $\mlspone \sim 400 $ GeV. 
Hence the LHC limits only allow the scenarios where a heavy Higgs 
(typically below 1 TeV) can decay to a sbottom pair ($\mlsbot > 300$ GeV) 
and the lightest sbottom is almost degenerate to LSP.

In case of sbottoms, the heavy Higgs couplings will be large with larger values of both $A_b$ and $\tan\beta$. Moreover, with large $A_b$ the mixing will be large in sbottom sector 
and the one of the sbottom masses will be lighter which may kinematically allow the decay of heavy Higgs  into a pair of sbottoms. 
We present two benchmark points, BP-2 and BP-3, in Table~\ref{tab:bmp} with relevant input 
parameters and output masses and branching ratios. In BP-2, Br($ H \to {\widetilde{b_1}}{\widetilde{b_1}}$) is about 27\%, but due to enhancement in $\tan\beta$ and $A_b$, this branching ratio increases to 53\% in BP-3. 
To illustrate the effect of $\tan\beta,~\mu$ and $A_b$ 
we present the branching ratios of heavy Higgs boson $H \to \lsbot \lsbot $ decay 
with respect to the variation of $A_b$ in Fig.~\ref{sbot_brs}. 
The green, blue and red lines correspond to  ($\mu, \tan\beta$) = (4 TeV, 10), 
(6 TeV, 10) and (3 TeV, 20) respectively. Other parameters are exactly same as BP3 
(see Table~\ref{tab:bmp}). 
The mixing in the sbottom sector depends 
on $\widetilde{A}_{b} \equiv A_{b} - \mu \tan\beta$ and hence the branching ratios 
may be large particularly in the large $\tan\beta$ scenarios (red solid line in 
Fig.~\ref{sbot_brs}). 
We also perform a general scan by varying the masses of third generation 
squarks, trilinear couplings, $\mu$  and $\tan\beta$ and notice that 
 Br($H \to \lsbot \lsbot$ or $\lstop\lstop$) can be as large as $~95\%$. 
In such scenarios the direct LHC bounds from stop/sbottom production 
mostly allow the parameter space where stop/sbottom is nearly 
degenerate to LSP and the search for heavy Higgs bosons need special 
attention \cite{futurework}.

\section{Summary and outlook}
\label{sec:V}

Since the discovery of 125 GeV Higgs boson, the CMS and ATLAS collaborations at LHC have performed numerous studies to decipher the properties of the observed resonance. Studies performed on the production, as well as the decay of the 125 GeV Higgs boson to SM particles have been presented in terms of the signal strength variables in association with the uncertainties in the measurements. These measured values are used to constraint the models which fall under the purview of ``beyond the SM" physics.
The ATLAS and CMS collaborations have also performed numerous searches for the heavy Higgs bosons through their decay to Standard Model particles. However, none of these searches have been able to observe any clear signature of the additional Higgs bosons. As a result, upper limits have been derived on the production cross-section times branching fraction of the respective search channels at $95\%$ C.L. Our objective in this work is to understand the effect of the latest bounds from Higgs signal strength measurements and heavy Higgs searches on the MSSM parameter space.

We scan over a wide range of MSSM parameter space. The allowed parameter space is required to have the light Higgs mass in the range 122 GeV to 128 GeV.
We perform a global $\chi^{2}$ analysis by combining the Higgs signal strength constraints derived by ATLAS and CMS corresponding to the $8$ TeV and $13$ TeV runs of LHC (see Table.~\ref{table-mu8} and Table.~\ref{table-mu13}), and the flavor physics constraints, derived on the branching fraction of rare $B$-decays, $\mathrm{Br}( B_{s} \rightarrow X_{s}\gamma)$, $\mathrm{Br}( B_{s} \rightarrow \mu^{+} \mu^{-})$ and $\mathrm{Br}(B^+ \to \tau^+ \nu_{\tau})$ (see Sec.~\ref{sec:II}). The allowed parameter space, thus obtained, is then subjected to consistency checks with respect to the existing bounds from numerous heavy Higgs searches. 

Key findings of this work are the following: 
\begin{itemize}

\item  As the signal strength measurements are in favor of the alignment/decoupling 
limit, the direct searches with $H \to W^{+}W^{-},ZZ$ final states are not effective to 
probe the relevant parameter space of our interest. 

\item Upper bounds derived on $ H/A \rightarrow \tau^{+}\tau^{-}$ are found to impose the strongest constraints on the parameter space and rules out a significant region of parameter space in the high $\tan\beta$ and low $M_{A}$ region. Compared to Run-I data, the recent 13 TeV data is more stringent in the region where $\tan\beta>$ 10. 

\item One requires an improvement of around two orders of magnitude in the observed upper limits in order to make the channels like $ H \rightarrow hh,~ A \rightarrow Z h$ sensitive to heavy Higgs searches. These searches are important because they will be able to probe a region of the $M_A - \tan \beta$ parameter space which is complementary to the region sensitive to the $H \rightarrow \tau^+ \tau^-$ search. For an order of magnitude improvement in observed upper limits in the $H^{\pm} \to \tau \nu$ and $H^{\pm} \to t \bar{b}$ channels, the allowed parameter space might become sensitive to charged Higgs searches as well. 

\item It is observed that presence of heavy Higgs to ino decay modes severely affects the constraints imposed on the parameter space from heavy Higgs searches. We, in particular, study it's effect on the constraints imposed by the upper bounds on $ H/A \rightarrow \tau^{+} \tau^{-}$, which, as discussed earlier, provides the strongest constraints on our parameter space and it is observed that these non-SM decay modes can significantly modify the exclusion limits on $\tan\beta$ derived from the heavy Higgs direct searches, as discussed in detail in Sec.~\ref{sec:IV}. In case its kinematically possible for the heavy Higgs bosons decay to SUSY particles, the Higgs to non-SM decay modes can receive significant branching fractions.
~For example, we observe that heavy Higgs to ino decay modes can reach up to 80$\%$ in the moderately low $\tan\beta$ region, when the inos are admixtures of wino and higgsino components. Furthermore, we have observed that for specific regions of the parameter space, decay of the heavy Higgs to top- and bottom-squarks can be enhanced. It is observed that the constraints on the parameter space get much weaker in presence of these additional decay modes and much improved measurements of the observables would be required in order to rule out those regions of MSSM parameter space, which had been excluded previously assuming the absence of those decay modes.
\end{itemize}

\begin{acknowledgments}
We acknowledge useful communication with Thomas Hahn and Sven Heinemeyer regarding FeynHiggs package. Work of B.B. is supported by Department of Science and Technology, Government of INDIA under the Grant Agreement numbers IFA13-PH-75 (INSPIRE Faculty Award). The research leading to these results has received funding from the European Research Council under the European Union's Seventh Framework Programme (FP/2007-2013)/ERC Grant Agreement n.~279972. Work of S.R. is supported by the Department of Science and Technology, Government of India through the INSPIRE Faculty Fellowship (Grant agreement number IFA12-PH-41). R.K.B. would like to thank Amit Adhikary for helpful discussions and comments. B. and D.C. would like to thank the Centro de Ciencias de Benasque Pedro Pascual for its hospitality during the initial stages of this work. S.R. would like to thank Prof.~Yuval Grossman, LEPP, Cornell University for hosting the visit as a part of INSPIRE Faculty Fellowship. The work of A.C.  is supported by the Lancaster-Manchester-Sheffield Consortium for Fundamental Physics under STFC Grant No. ST/L000520/1.	   
\end{acknowledgments}
\providecommand{\href}[2]{#2}\begingroup\raggedright\endgroup

\section{Appendix: A}
\begin{figure}[htb!]
\begin{center}
{\includegraphics[angle=0,width=0.40\textwidth]{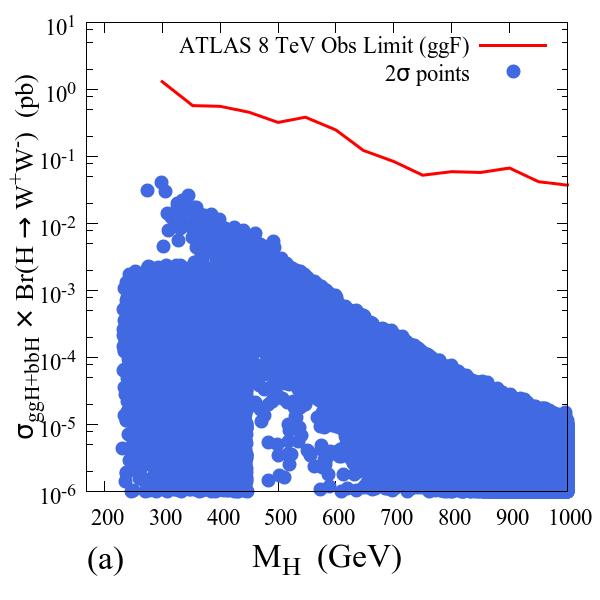}}
{\includegraphics[angle=0,width=0.40\textwidth]{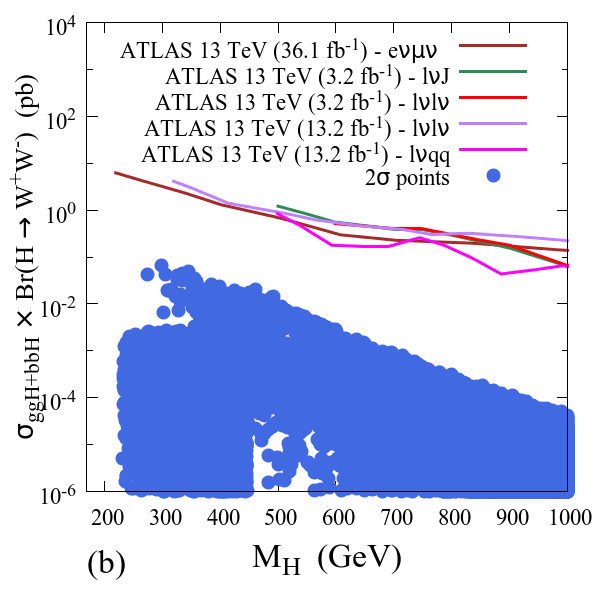}}
\caption{Scatter plot in the $M_H-[\sigma_{ggH+b\bar{b}H} \times$ Br$(H \to W^{+}W^{-}$)] plane, for the allowed parameter space (blue colored points in fig.~\ref{fig:ma_tb2}(d)). The $ggH+b\bar{b}H$ production cross-section times
branching ratio for 8 (13) TeV are presented in fig.~\ref{figww_add}(a) (fig.~\ref{figww_add}(b)). The red solid line in fig.~\ref{figww_add}(a) denotes the 95\% C.L. upper limit derived by ATLAS for LHC 8 TeV data~\cite{Aad:2015agg}. The colored lines of fig.~\ref{figww_add}(b) represent the upper limit obtained by ATLAS in various final states using 13 TeV data~\cite{ATLAS-CONF-2016-074,ATLAS-CONF-2016-062,ATLAS-CONF-2016-021,Aaboud:2017gsl}. }
\label{figww_add}
\end{center}
\end{figure}

\begin{figure}[htb!]
\begin{center}
{\includegraphics[angle=0,width=0.40\textwidth]{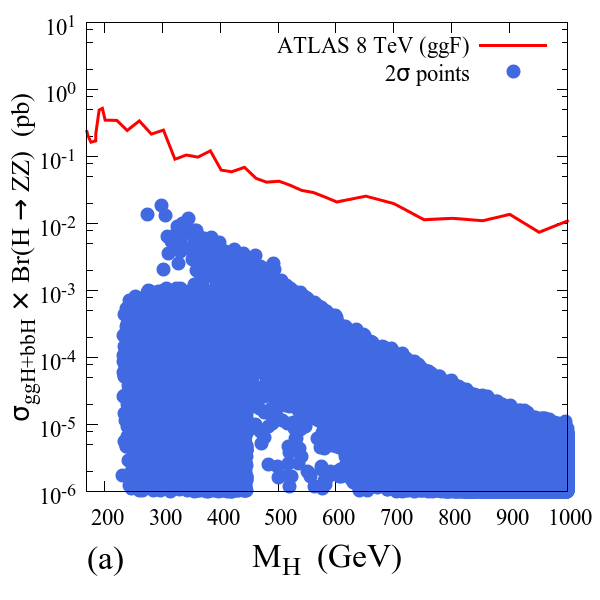}}
{\includegraphics[angle=0,width=0.40\textwidth]{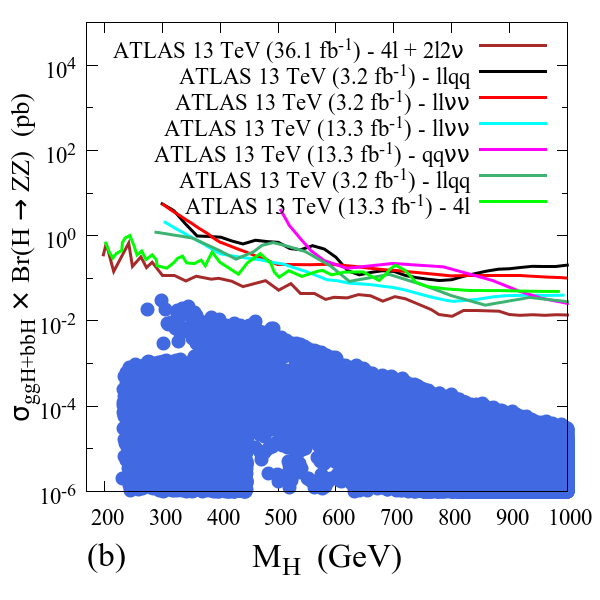}}
\caption {Scatter plot in the $M_H-[\sigma_{ggH+b\bar{b}H} \times$ Br$(H \to ZZ$)] plane, for the allowed parameter space. The $ggH+b\bar{b}H$ cross-section times branching ratios for 8 (13) TeV are presented in fig.~\ref{figzz_add}(a) (fig.~\ref{figzz_add}(b)). The red solid line in fig.~\ref{figzz_add}(a) represents the 95\% C.L. upper limit on the $ggH$ production cross-section times branching ratio given by ATLAS at 8 TeV~\cite{Aad:2015kna}. The colored lines in fig.~\ref{figzz_add}(b) represent the 13 TeV~\cite{ATLAS-CONF-2016-012,ATLAS-CONF-2016-016,ATLAS-CONF-2016-056,ATLAS-CONF-2016-082,ATLAS-CONF-2016-079,Aaboud:2017rel} ATLAS upper limit for different final states.}
\label{figzz_add}
\end{center}
\end{figure}

The allowed parameter space points (blue colored points in figs. 2 (d)) have been shown in the $\sigma_{ggH + b\bar{b}H} \times Br(H \to WW) - M_{H}$ and $\sigma_{ggH + b\bar{b}H} \times Br(H \to ZZ) - M_{H}$ planes in Fig.~\ref{figww_add} and Fig.~\ref{figzz_add}, respectively.

\end{document}